\documentclass{article}
\PassOptionsToPackage{authoryear,round}{natbib}
\usepackage{conference,times}

\usepackage[utf8]{inputenc}
\usepackage[T1]{fontenc}
\usepackage{hyperref}
\usepackage{url}
\usepackage{booktabs}
\usepackage{amsfonts}
\usepackage{nicefrac}
\usepackage{microtype}
\usepackage{graphicx}
\usepackage{subcaption}
\usepackage{amsmath}
\usepackage{amssymb}
\usepackage{mathtools}
\usepackage{amsthm}
\usepackage{makecell}
\usepackage[table]{xcolor}
\usepackage{multirow}
\usepackage{placeins}
\usepackage[capitalize,noabbrev]{cleveref}
\usepackage{tabularx}
\usepackage{array}
\usepackage{etoc}

\theoremstyle{plain}
\newtheorem{theorem}{Theorem}[section]

\theoremstyle{definition}

\theoremstyle{remark}

\title{Scalable Quantum Machine Learning via Multi-layer Fully-Connected Variational Quantum Circuits}

\author{
Howard Su$^{1}$\thanks{Corresponding author: \texttt{h.su24@imperial.ac.uk}}, Chen-Yu Liu$^{2}$, Samuel Yen-Chi Chen$^{3}$,
Kuan-Cheng Chen$^{1}$, Huan-Hsin Tseng$^{3}$\\[0.5em]
\small $^{1}$Imperial College London, United Kingdom\\
\small $^{2}$National Taiwan University, Taiwan\\
\small $^{3}$Brookhaven National Laboratory, Upton, NY, USA
}

\makeatletter
\iclrfinaltrue
\def\@maketitle{\vbox{\hsize\textwidth
{\LARGE\sc \@title\par}
\def\And{\end{tabular}\hfil\linebreak[0]\hfil
\begin{tabular}[t]{l}\bf\rule{\z@}{24pt}\ignorespaces}%
\def\AND{\end{tabular}\hfil\linebreak[4]\hfil
\begin{tabular}[t]{l}\bf\rule{\z@}{24pt}\ignorespaces}%
\begin{tabular}[t]{l}\bf\rule{\z@}{24pt}\@author\end{tabular}%
\vskip 0.3in minus 0.1in}}
\makeatother

\begin{document}

\maketitle
\fancyhead{}
\etocdepthtag.toc{main}

\begin{abstract}
Variational quantum circuits (VQCs) face an expressivity--trainability
dilemma and scalability challenges. We propose Multi-Layer
Fully-Connected Variational Quantum Circuits (FC-VQC), a
general-purpose quantum machine learning framework that connects
local VQC blocks through measurement, deterministic parameter-free
routing, and re-encoding. All trainable model parameters reside
within the quantum blocks, without trainable classical neural
components. We study fully connected, sliding-window,
and parallel block mixing and analyze computational costs,
conditional error propagation, and block information exchange.
Using classical simulation, we evaluate tabular regression and
classification, with our main comparison addressing spatio-temporal
function approximation for the Black--Scholes, Burgers, and
time-dependent oscillatory PDEs with up to $144$ spatial dimensions.
Comparisons include neural-network, explicit angle-feature,
tensor-network, and gradient-boosted-tree baselines. The results
show that FC-VQC achieves the lowest trajectory relative MAE
for all PDEs at the higher dimensions $d\in\{81,144\}$.
Gradient-dynamics and depolarizing-noise experiments provide
complementary empirical diagnostics of trainability and
preliminary noise sensitivity.
\end{abstract}
\section{Introduction}
\label{sec:introduction}

Quantum machine learning (QML) with variational quantum circuits (VQCs)
faces two key challenges: the \emph{expressivity--trainability dilemma}
and \emph{scalability}. Small circuits may lack expressive capacity,
while wider or deeper circuits can suffer from vanishing
gradients~\citep{mcclean2018barren,cerezo2021cost}. Scalability is
constrained by monolithic circuit width: with one qubit per feature,
a $d$-dimensional input requires $d$ qubits and $2^d$ state-vector
amplitudes for direct classical simulation. We ask:
\emph{Can VQC-based models achieve competitive high-dimensional
performance without a trainable classical neural encoder?}

Related approaches include federated
QML~\citep{chen2021federated,chehimi2022_QFL},
quantum convolutional neural networks
(QCNNs)~\citep{cong2019qcnn,chen2022qcnn}, tensor-network
methods~\citep{stoudenmire2016supervised,rieser2023tensor},
and multi-chip quantum
models~\citep{park2025_Multi_Chip_QNN,chen2025_Multi_Chip_QLSTM}.
Circuit knitting decomposes a prescribed large circuit into local
subcircuits and reconstructs its observables through quasiprobability
sampling, incurring cut-dependent sampling
overhead~\citep{piveteau2023circuit}. In contrast, FC-VQC uses measured
local outputs directly as hidden features rather than reconstructing
a prescribed circuit.

We introduce Multi-Layer Fully-Connected Variational Quantum Circuits
(FC-VQC), a general-purpose QML framework supporting regression,
classification, and high-dimensional function approximation.
FC-VQC builds trainable models from local quantum blocks connected
through measurement, deterministic parameter-free routing, and
re-encoding. All trainable model parameters reside within the quantum
blocks, without trainable classical neural components. We study three
block-mixing variants: fully-connected (FC), sliding-window (SW),
and parallel (PL).

Our contributions are threefold. First, we develop the FC-VQC
framework with three mixing variants and characterize their
computational costs. Second, we analyze conditional error
propagation, block receptive fields, and support-induced approximation lower bounds. Third, we evaluate tabular regression and
classification and compare FC-VQC with neural-network,
explicit angle-feature, tensor-network, and gradient-boosted-tree
baselines on spatio-temporal function approximation for the
Black--Scholes, Burgers, and time-dependent oscillatory PDEs
with up to $144$ spatial dimensions. FC-VQC shows stronger relative performance at higher
tested dimensions, with FC-VQC-FC generally the strongest variant.
Additional experiments examine empirical trainability and
preliminary noise robustness.
\section{Scalable Variational Quantum Architectures}
\label{sec:architectures}

\begin{figure}[h]
    \centering
    \includegraphics[width=\textwidth]{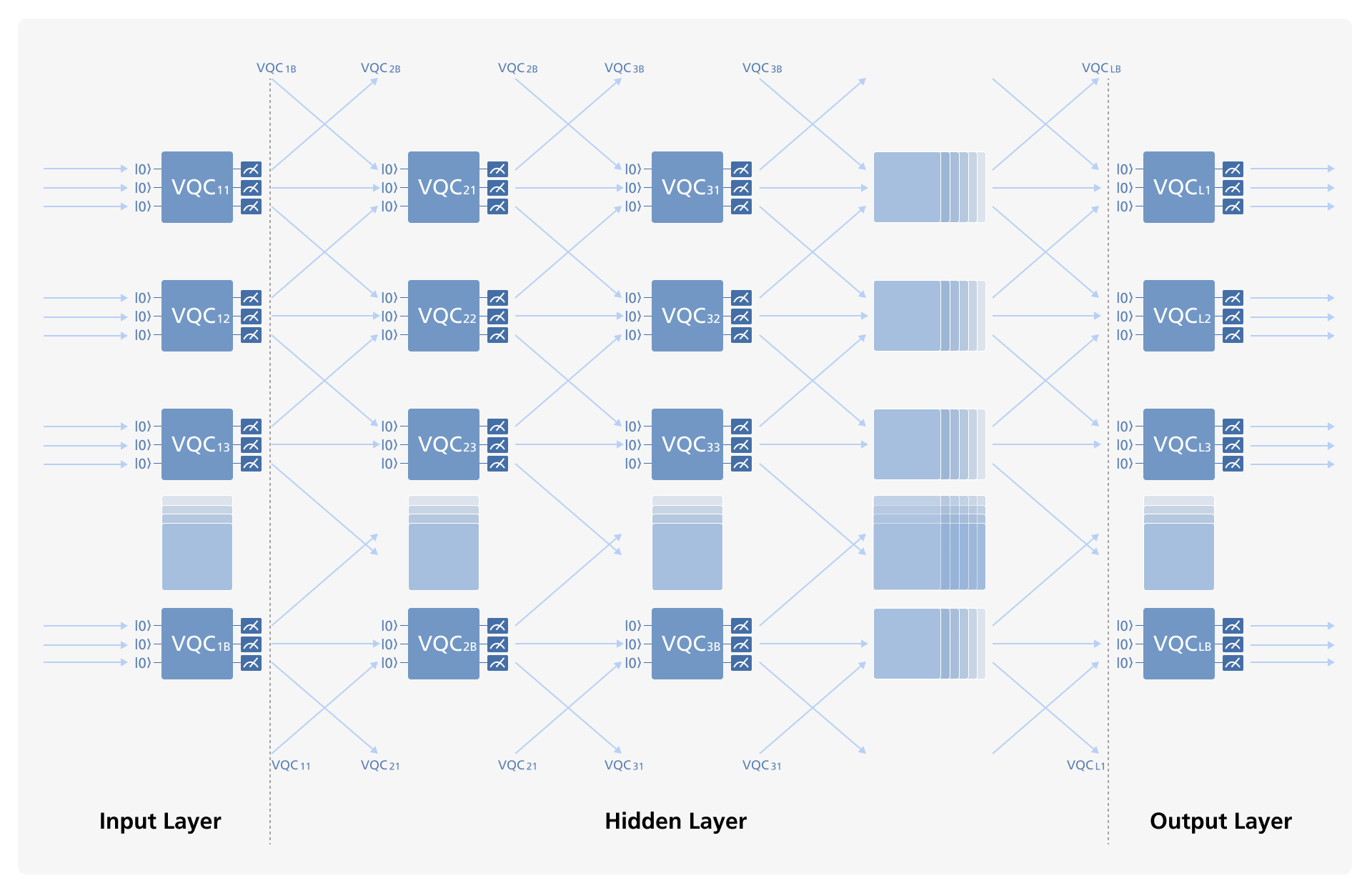}
    \caption{Overview of the FC-VQC architecture.}
    \label{fig:architecture}
\end{figure}

FC-VQC is a modular architecture that composes fixed-size local VQC blocks through deterministic, parameter-free block mixing. This keeps each quantum computation local while allowing the number of trainable quantum parameters to grow with the input dimension. Figure~\ref{fig:architecture} illustrates the FC-VQC architecture used as the core scalable model in this work.

\vspace{-2mm}

\subsection{VQC Block and Architecture Types}
\label{sec:type1}

The basic building block is a $q$-qubit VQC map $f_{\Theta}:\mathbb{R}^q\to\mathbb{R}^{n_{\rm out}}$. Given an input block $h=(h_1,\ldots,h_q)$, we apply rotation encoding,
\begin{equation}
    |\psi_{\rm enc}(h)\rangle = \bigotimes_{j=1}^{q} R_{\alpha}(h_j)|0\rangle,
    \label{eq:encoding}
\end{equation}
followed by $K$ StronglyEntanglingLayers~\cite{schuld2020circuit}. Each layer applies general single-qubit Euler rotations and an entangling CNOT pattern, giving $3qK$ trainable parameters per block. The block output is obtained from Pauli-$Z$ expectation values,
\begin{equation}
    f_{\Theta}(h)_j =
    \langle \psi_{\rm enc}(h)|U^{\dagger}(\Theta)O_jU(\Theta)|\psi_{\rm enc}(h)\rangle,
    \qquad j=1,\ldots,n_{\rm out}.
    \label{eq:vqc_readout}
\end{equation}

We use four architecture types. \textbf{Type~1} is the standard monolithic VQC: it sets $q=d$ and processes all input features in a single circuit. \textbf{Type~2} stacks $L$ monolithic VQC blocks with measure-and-re-encode interfaces,
$h^{(l)} = f_{\Theta^{(l)}}(h^{(l-1)})$, with $h^{(0)}=x$, but each layer remains a $d$-qubit circuit. \textbf{Type~3} is the main FC-VQC architecture: it partitions the input into local VQC blocks connected across layers through deterministic routing. \textbf{Type~4} extends Type~3 by applying a deterministic feature expansion before block partitioning, which is useful for low-dimensional tabular tasks where the raw input dimension provides too few VQC blocks.

\subsection{Input Layer}
\label{sec:input}

For Type~3, let $d=Bq$, where $q$ is the fixed number of qubits per local VQC block and $B$ is the number of blocks. The input is partitioned into local blocks,
\[
    x=[x^{(1)},\ldots,x^{(B)}],
    \qquad x^{(b)}\in\mathbb{R}^q .
\]
Each block is processed by an independent local VQC. If the input dimension is not divisible by $q$, we zero-pad the input to match the block structure.

Type~4 uses the same block partitioning after deterministic feature expansion. We increase the input dimension by repeating the input or concatenating fixed nonlinear transformations, such as polynomial, root, or logarithmic features. This introduces no trainable classical encoder, but provides more local VQC blocks and hence more trainable quantum parameters for low-dimensional tasks.

\subsection{Hidden Layer}
\label{subsec:hidden_layer}

\begin{figure}[htbp]
    \centering
    \includegraphics[width=\textwidth]{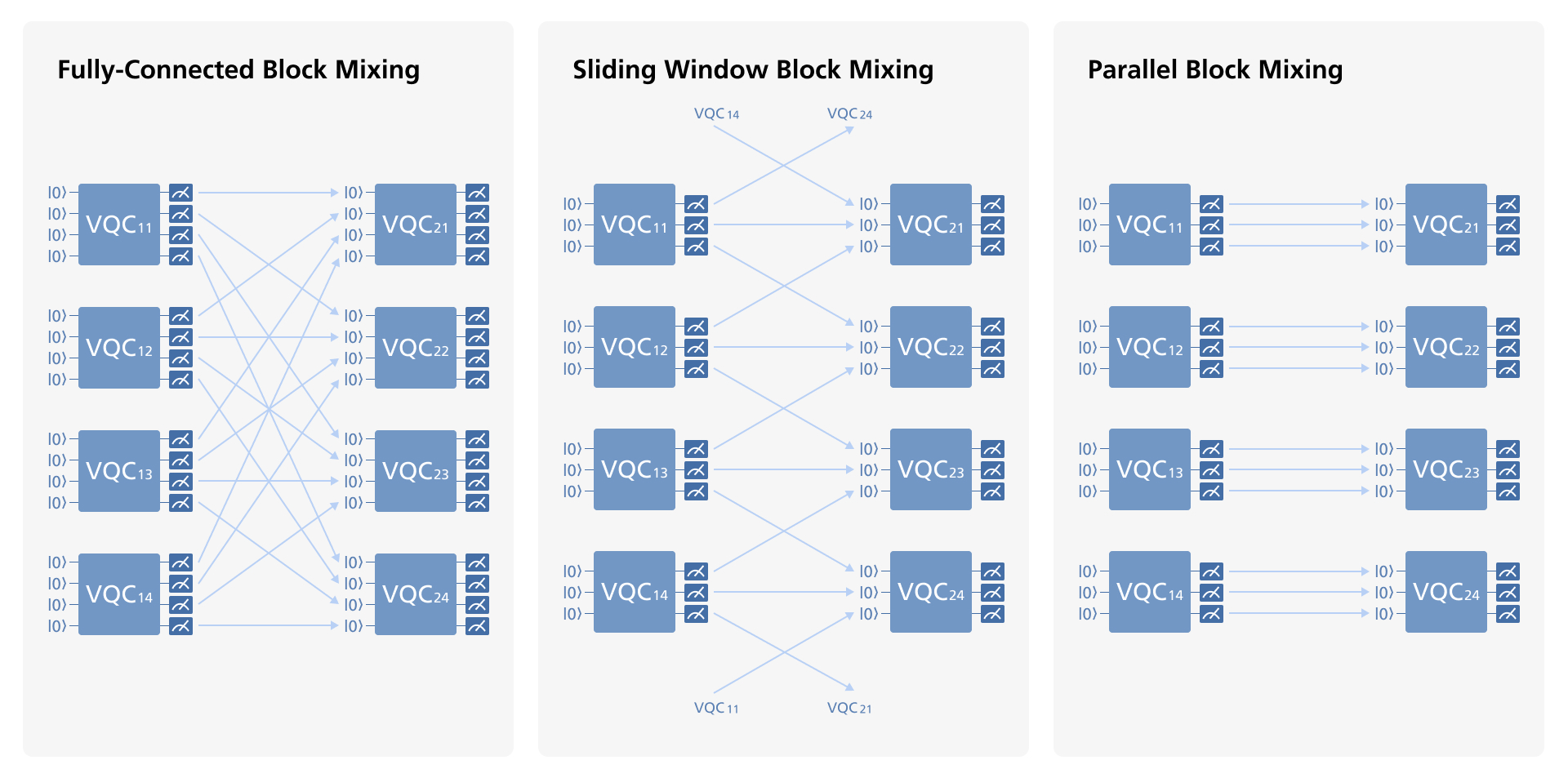}
    \caption{Block mixing: fully-connected (FC, left), sliding-window
    (SW, middle), and parallel (PL, right). Arrows between blocks denote
    classical routing of measured outputs for re-encoding.}
    \label{fig:block_mixing}
\end{figure}

Each hidden layer routes measured block outputs through a deterministic,
parameter-free map $g_b^{(l)}$, then re-encodes them into a local VQC:
\begin{equation}
h^{(l+1,b)} = f_{\Theta_b^{(l)}}\!\left(
 g_b^{(l)}\!\left(h^{(l,1)},\ldots,h^{(l,B)}\right)\right),
\quad b=1,\ldots,B.
\label{eq:fcvqc_hidden_update}
\end{equation}
Figure~\ref{fig:block_mixing} shows three block-mixing rules.
\textbf{Fully connected (FC)} routes components from every
preceding-layer block into each next-layer block, using
$q=\lceil\sqrt{d}\rceil$ qubits per block.
\textbf{Sliding-window (SW)} routes components from neighbouring
blocks on a ring, spreading information across blocks over
successive layers.
\textbf{Parallel (PL)} passes each block's output to the
corresponding block in the next layer, without cross-block exchange.
SW and PL keep $q$ fixed as $d$ increases.
All inter-block connections are classical; trainable parameters
reside within the VQC blocks.

\subsection{Output Layer}
\label{sec:output}

FC-VQC supports both dimension-preserving and dimension-reducing outputs. For BSDE/PDE solvers, the model uses a dimension-preserving map $d\mapsto d$, so that the output has the same spatial dimension as the input. This is required when approximating a full gradient vector or state-dependent solution component at each time step.

For scalar or low-dimensional prediction, such as tabular regression and classification, FC-VQC uses staged dimensionality reduction. This is implemented by measuring fewer observables per VQC block, so that each stage reduces the feature dimension. We denote a topology by $d_{\rm in}t d_1t\cdots t d_{\rm out}$. For example, $27$t$9$t$3$t$1$ represents a three-stage reduction 
$27\to9\to3\to1$, implemented by successive output stages. An illustrative example is provided in Appendix~\ref{sec:appendix_experimental_setup}, Figure~\ref{fig:appendix_dim_reduction}.

\subsection{Computational Complexity on Classical Simulators}
\label{subsec:simulator_complexity}

We consider dense state-vector simulation of one forward evaluation
of a dimension-preserving model. Let $L$ be the number of network
layers, $K$ the number of circuit layers per block, and
$B=\lceil d/q\rceil$ the number of local blocks.
Each block uses $O(Kq)$ gates and costs $O(Kq2^q)$ to simulate.
Including $O(LBq)$ work for deterministic routing gives $C_{\mathrm{mod}}=O(LBKq2^q+LBq)=O(LKd2^q)$,
where $Bq=O(d)$ for $q\le d$, including padding.
The resulting costs are
\begin{equation}
C(d)=
\begin{cases}
O(LKd2^d),
& \text{full-width VQCs},\\
O(LKd),
& \text{FC-VQC-SW/PL with fixed }q,\\
O\!\left(LKd2^{\lceil\sqrt{d}\rceil}\right),
& \text{FC-VQC-FC}.
\end{cases}
\end{equation}
Thus, SW and PL scale linearly with $d$ when $q$, $L$, and $K$
are fixed. FC uses $q=\lceil\sqrt{d}\rceil$, reducing the
exponential state-vector factor from $2^d$ to
$2^{\lceil\sqrt{d}\rceil}$.
Each local block stores $2^q$ complex amplitudes, compared
with $2^d$ for a full-width circuit.

These bounds describe direct state-vector simulation of one
forward evaluation, not end-to-end quantum-hardware training.
Under the local-Jacobian training scheme considered here,
hardware training additionally requires repeated state
preparation, measurement, parameter-shift evaluations, and
input-angle derivatives through re-encoding.
Appendix~\ref{app:hardware_cost} accounts for the corresponding
circuit settings and measurement shots.

\section{Experimental Results}
\label{sec:experiments}

We evaluate FC-VQC on Concrete Strength regression and Wine Quality
classification, with our main evaluation focusing on high-dimensional
function approximation for the Black--Scholes, Burgers, and
time-dependent oscillatory PDEs. We compare block-mixing variants
with classical baselines and monolithic VQCs, examining predictive
performance, scalability, empirical trainability, and preliminary
noise robustness. Experimental protocols are provided in
Appendix~\ref{sec:appendix_experimental_setup}.

\subsection{Tabular Predictive Performance}
\label{subsec:predictive_performance}

We evaluate FC-VQC on Concrete Strength regression ($d=8$)
and Wine Quality classification ($d=11$).
Table~\ref{tab:predictive_summary} compares FC-VQC with a
standard monolithic VQC and a fully connected deep neural
network (DNN). FC-VQC achieves the highest mean test performance
among the evaluated models on both tasks, with $R^2=0.8928$
on Concrete Strength and $63.6\%$ accuracy on Wine Quality.

\begin{table}[htbp]
\centering
\caption{Tabular predictive performance (mean $\pm$ standard deviation). Higher is better.}
\label{tab:predictive_summary}
\footnotesize
\setlength{\tabcolsep}{3pt}
\renewcommand{\arraystretch}{1.05}
\begin{tabular*}{\linewidth}{@{\extracolsep{\fill}}llccc@{}}
\toprule
Task & Metric & Monolithic VQC & DNN & FC-VQC \\
\midrule
Concrete Strength & Test $R^2$ & $0.6768\pm0.0218$ & $0.8486\pm0.0291$ & $\mathbf{0.8928\pm0.0189}$ \\
Wine Quality & Accuracy & $57.2\%\pm1.6\%$ & $58.4\%\pm3.1\%$ & $\mathbf{63.6\%\pm1.1\%}$ \\
\bottomrule
\end{tabular*}
\end{table}

\subsection{High-Dimensional Approximation Performance}
\label{subsec:pde_results}

We evaluate FC-VQC on the Black--Scholes, Burgers, and
time-dependent oscillatory PDEs at spatial dimensions
$d\in\{9,36,81,144\}$. We use a backward stochastic differential
equation (BSDE) reformulation with $N=10$ time steps and measure
accuracy using trajectory relative mean absolute error (RelMAE)
against analytical solution trajectories; lower values are better.
The formulations and training protocols are provided in
Appendix~\ref{sec:appendix_experimental_setup}.

\paragraph{Compared models.}
Classical baselines include a DNN with hidden width $64$,
a parameter-matched DNN, FC-DNN, FC-Angle-Linear, a matrix
product state (MPS) model, XGBoost, and CatBoost.
The parameter-matched DNN uses a trainable parameter count
comparable to FC-VQC. FC-DNN and FC-Angle-Linear retain the
three-coordinate block partition and sliding-window routing
of FC-VQC-SW, replacing each local VQC with a small DNN or a
linear map over explicit product-angle features, respectively.
MPS is a tensor-network baseline, while XGBoost and CatBoost
are gradient-boosted-tree models.

For modular VQCs (Type~3), we compare FC-VQC-FC, FC-VQC-SW,
and FC-VQC-PL, corresponding to global, local, and no cross-block
exchange. Monolithic VQC (Type~1) and stacked full-width VQC
(Type~2) are evaluated only at $d=9$,
with $L=1$ and $L\in\{3,5\}$, respectively.

\paragraph{Reporting.}
Tables~\ref{tab:pde_black_scholes}--\ref{tab:pde_oscillatory}
report the best tested configuration for each model, PDE,
and dimension, selected by the lowest mean trajectory RelMAE
across five random seeds. Each entry gives the mean and
standard deviation for that configuration. Boldface identifies
the lowest displayed mean in each column, and dashes indicate
unreported results. At $d=9$, FC-VQC-SW is not reported separately from
FC-VQC-FC; the FC-VQC-FC result is shown in the main tables.
Full configuration results and trajectory-level error plots
are provided in Appendices~\ref{app:bsde_pde_benchmarks}
and~\ref{app:trajectory_plots}, respectively.

\paragraph{Predictive performance.}
Classical DNN baselines achieve the lowest reported mean errors
at $d=9$. At $d=36$, FC-VQC-SW leads on Black--Scholes, while
MPS leads on Burgers and the oscillatory PDE.
At $d\in\{81,144\}$, an FC-VQC variant achieves the lowest
five-seed mean trajectory RelMAE among the evaluated models
for each PDE, with FC-VQC-FC leading in five of the six settings.
These results demonstrate competitive high-dimensional performance
within our simulation budget, whereas direct state-vector simulation
and training of full-width monolithic VQCs at these dimensions
exceed the resources of our experimental setup.

At $d=9$, FC-VQC-FC achieves lower mean errors than both
monolithic and stacked full-width VQCs on all three PDEs.
Type~2 improves on Type~1 for Burgers and the oscillatory PDEs.

Among the modular variants, FC generally performs best,
while SW improves on PL in many settings. However, FC also
uses wider local circuits, so differences involving FC cannot
be attributed solely to routing.

\begin{table}[h]
\centering
\caption{Black--Scholes PDE: trajectory RelMAE (mean $\pm$ standard deviation). Lower is better.}
\label{tab:pde_black_scholes}
\footnotesize
\setlength{\tabcolsep}{2pt}
\renewcommand{\arraystretch}{1.08}
\begin{tabular*}{\linewidth}{@{\extracolsep{\fill}}lcccc@{}}
\toprule
Model & $d=9$ & $d=36$ & $d=81$ & $d=144$ \\
\midrule
DNN & $\mathbf{0.0374\pm0.0010}$ & $0.0306\pm0.0024$ & $0.0239\pm0.0003$ & $0.0230\pm0.0004$ \\ 
Parameter-matched DNN & $0.0421\pm0.0011$ & $0.0303\pm0.0028$ & $0.0237\pm0.0006$ & $0.0229\pm0.0007$ \\ 
FC-DNN & $0.1612\pm0.0093$ & $0.0799\pm0.0038$ & $0.0534\pm0.0021$ & $0.0457\pm0.0049$ \\ 
FC-Angle-Linear & $0.0408\pm0.0015$ & $0.0249\pm0.0012$ & $0.0252\pm0.0007$ & $0.0252\pm0.0021$ \\ 
\midrule
MPS & $0.0439\pm0.0017$ & $0.0235\pm0.0006$ & $0.0275\pm0.0012$ & $0.0231\pm0.0008$ \\ 
XGBoost & $0.0553\pm0.0011$ & $0.0466\pm0.0049$ & $0.0366\pm0.0026$ & $0.0293\pm0.0017$ \\ 
CatBoost & $0.0471\pm0.0018$ & $0.0458\pm0.0011$ & $0.0377\pm0.0006$ & $0.0303\pm0.0005$ \\ 
\midrule
Monolithic VQC (Type~1) & $0.0471\pm0.0013$ & --- & --- & --- \\ 
FC-VQC-Type2 & $0.0478\pm0.0015$ & --- & --- & --- \\ 
FC-VQC-PL & $0.0387\pm0.0011$ & $0.0225\pm0.0005$ & $0.0179\pm0.0003$ & $0.0174\pm0.0005$ \\ 
FC-VQC-SW & --- & $\mathbf{0.0224\pm0.0004}$ & $0.0178\pm0.0004$ & $0.0175\pm0.0003$ \\ 
FC-VQC-FC & $0.0391\pm0.0018$ & $0.0232\pm0.0003$ & $\mathbf{0.0165\pm0.0007}$ & $\mathbf{0.0133\pm0.0003}$ \\ 
\bottomrule
\end{tabular*}
\end{table}

\begin{table}[h]
\centering
\caption{Burgers PDE: trajectory RelMAE (mean $\pm$ standard deviation). Lower is better.}
\label{tab:pde_burgers}
\footnotesize
\setlength{\tabcolsep}{2pt}
\renewcommand{\arraystretch}{1.08}
\begin{tabular*}{\linewidth}{@{\extracolsep{\fill}}lcccc@{}}
\toprule
Model & $d=9$ & $d=36$ & $d=81$ & $d=144$ \\
\midrule
DNN & $0.1040\pm0.0050$ & $0.3609\pm0.0246$ & $0.4410\pm0.0185$ & $0.5752\pm0.0246$ \\ 
Parameter-matched DNN & $\mathbf{0.1032\pm0.0067}$ & $0.3601\pm0.0283$ & $0.4364\pm0.0144$ & $0.5705\pm0.0311$ \\ 
FC-DNN & $0.3492\pm0.0230$ & $0.4518\pm0.0174$ & $0.5017\pm0.0155$ & $0.6500\pm0.0115$ \\ 
FC-Angle-Linear & $0.1156\pm0.0070$ & $0.4122\pm0.0161$ & $0.5733\pm0.0050$ & $0.6913\pm0.0362$ \\ 
\midrule
MPS & $0.1360\pm0.0032$ & $\mathbf{0.2776\pm0.0098}$ & $0.4236\pm0.0159$ & $0.6347\pm0.0394$ \\ 
XGBoost & $0.1407\pm0.0178$ & $0.2964\pm0.0289$ & $0.4184\pm0.0159$ & $0.4926\pm0.0259$ \\ 
CatBoost & $0.1311\pm0.0100$ & $0.3121\pm0.0085$ & $0.4348\pm0.0173$ & $0.5394\pm0.0166$ \\ 
\midrule
Monolithic VQC (Type~1) & $0.1313\pm0.0080$ & --- & --- & --- \\ 
FC-VQC-Type2 & $0.1244\pm0.0091$ & --- & --- & --- \\ 
FC-VQC-PL & $0.1462\pm0.0039$ & $0.3528\pm0.0141$ & $0.4233\pm0.0076$ & $0.5074\pm0.0118$ \\ 
FC-VQC-SW & --- & $0.3260\pm0.0103$ & $\mathbf{0.4136\pm0.0090}$ & $0.4957\pm0.0076$ \\ 
FC-VQC-FC & $0.1155\pm0.0084$ & $0.3245\pm0.0050$ & $0.4275\pm0.0123$ & $\mathbf{0.4850\pm0.0121}$ \\ 
\bottomrule
\end{tabular*}
\end{table}

\begin{table}[h]
\centering
\caption{Time-dependent oscillatory PDE: trajectory RelMAE (mean $\pm$ standard deviation). Lower is better.}
\label{tab:pde_oscillatory}
\footnotesize
\setlength{\tabcolsep}{2pt}
\renewcommand{\arraystretch}{1.08}
\begin{tabular*}{\linewidth}{@{\extracolsep{\fill}}lcccc@{}}
\toprule
Model & $d=9$ & $d=36$ & $d=81$ & $d=144$ \\
\midrule
DNN & $0.0234\pm0.0012$ & $0.0749\pm0.0107$ & $0.1445\pm0.0227$ & $0.2487\pm0.0060$ \\ 
Parameter-matched DNN & $\mathbf{0.0219\pm0.0007}$ & $0.0636\pm0.0018$ & $0.1339\pm0.0226$ & $0.2523\pm0.0140$ \\ 
FC-DNN & $0.2051\pm0.0075$ & $0.1850\pm0.0112$ & $0.1809\pm0.0091$ & $0.2506\pm0.0084$ \\ 
FC-Angle-Linear & $0.0299\pm0.0042$ & $0.1641\pm0.0271$ & $0.3143\pm0.0155$ & $0.3982\pm0.1492$ \\ 
\midrule
MPS & $0.0296\pm0.0020$ & $\mathbf{0.0459\pm0.0008}$ & $0.1239\pm0.0143$ & $0.2505\pm0.0064$ \\ 
XGBoost & $0.0643\pm0.0091$ & $0.1008\pm0.0130$ & $0.1141\pm0.0008$ & $0.1303\pm0.0114$ \\ 
CatBoost & $0.0569\pm0.0037$ & $0.1013\pm0.0023$ & $0.1304\pm0.0132$ & $0.1369\pm0.0027$ \\ 
\midrule
Monolithic VQC (Type~1) & $0.0446\pm0.0038$ & --- & --- & --- \\ 
FC-VQC-Type2 & $0.0291\pm0.0012$ & --- & --- & --- \\ 
FC-VQC-PL & $0.0510\pm0.0019$ & $0.0629\pm0.0056$ & $0.0944\pm0.0055$ & $0.1461\pm0.0061$ \\ 
FC-VQC-SW & --- & $0.0543\pm0.0022$ & $0.0909\pm0.0043$ & $0.1477\pm0.0035$ \\ 
FC-VQC-FC & $0.0252\pm0.0015$ & $0.0553\pm0.0022$ & $\mathbf{0.0816\pm0.0066}$ & $\mathbf{0.1004\pm0.0049}$ \\ 
\bottomrule
\end{tabular*}
\end{table}

\subsection{Scalability}
\label{subsec:scalability}

\paragraph{Circuit resources.}
Full-width VQCs (Types~1 and~2) require $d$ qubits and
$2^d$ complex amplitudes for dense state-vector simulation.
We evaluate these baselines only at $d=9$, as their simulation
and training at $d\ge36$ exceed our experimental resources.
The modular FC-VQC variants are evaluated at all four
dimensions, up to $d=144$.

FC-VQC-SW and FC-VQC-PL accommodate larger inputs by adding
blocks, with $B=\lceil d/q\rceil$, while keeping $q$ fixed.
Their forward-evaluation cost therefore grows linearly with
$d$ at fixed $q$, $L$, and $K$.
FC-VQC-FC instead uses $q=\lceil\sqrt{d}\rceil$, corresponding
to $q\in\{3,6,9,12\}$ for the tested dimensions.
At $d=144$, it uses 12-qubit local circuits rather than a
single 144-qubit circuit. Its per-block state-vector size
still grows as $2^{\lceil\sqrt{d}\rceil}$, so square-root
circuit-width scaling does not imply square-root runtime.
The simulation-cost derivations are given in
Section~\ref{subsec:simulator_complexity}.

\paragraph{Information-exchange trade-off.}
PL keeps blocks independent, while SW propagates information
across neighbouring blocks. With the unmixed first stage,
Theorem~\ref{thm:rf_sliding_window_main} gives
$|\mathcal{R}^{(L)}(b)|\le\min\{B,2(L-1)r+1\}$,
with $r=1$ in the evaluated SW architecture.
Theorem~\ref{thm:support_lower_bound} shows that broader nested
access cannot increase the support-induced approximation floor,
without guaranteeing lower trained error.
FC provides global block-level access after its first fully
connected routing step but uses wider local circuits.
Thus, the variants trade circuit width against the depth required
for cross-block exchange.

\paragraph{Measured computational cost.}
Appendix~\ref{app:runtime_memory} reports representative
single-run training times and peak GPU memory for the
oscillatory PDE at $d=36$. In this benchmark, the FC-VQC
variants require more runtime than the evaluated classical
baselines. These fixed-configuration measurements complement
the simulation-cost analysis but do not establish an empirical
scaling law. Reduced local circuit width should therefore
not be interpreted as a demonstrated runtime or memory
advantage over the classical baselines.

\subsection{Trainability and Gradient Dynamics}
\label{subsec:trainability}

We examine Types~1--4 on Concrete Strength through predictive
performance and training-gradient dynamics.

\paragraph{Architectural progression.}
Table~\ref{tab:architecture_evolution_concrete} shows increasing
test $R^2$ across the selected configurations, from $0.6768$
for Type~1 to $0.8928$ for Type~4. Parameter counts increase
from $216$ to $4{,}887$ alongside stacking, modularity, and
feature expansion. This comparison evaluates these architectural
changes jointly; it does not isolate the effect of parameter count.

\begin{table}[htbp]
\centering
\caption{Architecture comparison on Concrete Strength.}
\label{tab:architecture_evolution_concrete}
\small
\setlength{\tabcolsep}{6pt}
\renewcommand{\arraystretch}{1.02}
\begin{tabular}{@{}llcccc@{}}
\toprule
Type & Architecture & Layers & Depth & Params. & Test $R^2$ \\
\midrule
Type~1 & $8$t$1$         & $1$ & $9$ & $216$     & $0.6768$ \\
Type~2 & $8$t$1$         & $3$ & $9$ & $648$     & $0.7360$ \\
Type~3 & $8$t$3$t$1$     & $3$ & $5$ & $720$     & $0.8141$ \\
Type~4 & $32$t$11$t$4$t$1$ & $3$ & $9$ & $4{,}887$ & $0.8928$ \\
\bottomrule
\end{tabular}
\end{table}

\paragraph{Gradient dynamics.}
Appendix~\ref{app:gradients_concrete} reports gradient variance
across parameter components during training.
For Type~1, this statistic declines substantially, reaching
approximately $10^{-14}$ at $K=3$ and falling below $10^{-8}$
at $K=9$, with recurring spikes in the latter setting.
Type~2 also exhibits decay in configurations with few layers
and shallow circuits.
Many Type~3 configurations with larger $L$ and $K$ instead
show persistent fluctuations after the initial training phase,
with similar behaviour in Type~4. Shallow modular configurations
can nevertheless exhibit decay.

\paragraph{Interpretation and limitations.}
These curves are empirical diagnostics, not evidence of
barren-plateau immunity: declining gradient variance can also
reflect convergence. Locality-based trainability results for
QCNNs~\citep{pesah2021absence} rely on architecture-specific
assumptions and do not directly extend to FC-VQC.
Gradients may still be suppressed within local blocks or
attenuated through downstream Jacobians across successive stages.
Fixed-width FC-VQC-SW/PL also permit efficient direct classical
simulation, illustrating the trainability--simulability trade-off
discussed by \citet{cerezo2025simulability}.
FC-VQC-FC instead uses $q=\lceil\sqrt{d}\rceil$ and is not
covered by the fixed-width scaling argument; this does not
establish classical intractability.
Appendix~\ref{app:locality_trainability} provides the supporting
discussion and gradient-propagation analysis.

\subsection{Depolarizing-Noise Sensitivity}
\label{subsec:nisq_robustness}

To complement the noiseless experiments, we evaluate FC-VQC
on Concrete Strength under depolarizing noise with
$p\in\{0.001,0.01\}$. Due to computational constraints,
we use the representative Type~4 architecture
$32$t$11$t$4$t$1$, testing circuit depths
$K\in\{3,5,7,9\}$ across five random seeds.

\paragraph{Results.}
Table~\ref{tab:nisq_noisy_concrete} compares noiseless and noisy
test $R^2$. Relative to the noiseless setting, changes in mean
$R^2$ range from $-0.0200$ to $+0.0042$.
The response is therefore small and non-monotonic, with
increased between-seed variability in several configurations.
The slight increase at $K=7$ and $p=0.001$ does not establish
a systematic improvement from noise.

\begin{table}[h]
\centering
\caption{Test $R^2$ under noiseless and depolarizing-noise
simulation on Concrete Strength ($32t11t4t1$).
Values are mean $\pm$ standard deviation across five seeds.}
\label{tab:nisq_noisy_concrete}
\small
\setlength{\tabcolsep}{5pt}
\renewcommand{\arraystretch}{1.02}
\begin{tabular}{@{}lcccc@{}}
\toprule
Setting & $K=3$ & $K=5$ & $K=7$ & $K=9$ \\
\midrule
Noiseless
& $0.8868 \pm 0.0092$
& $0.8773 \pm 0.0348$
& $0.8791 \pm 0.0261$
& $0.8928 \pm 0.0189$ \\

Noisy, $p=0.001$
& $0.8804 \pm 0.0376$
& $0.8762 \pm 0.0381$
& $0.8833 \pm 0.0414$
& $0.8758 \pm 0.0281$ \\

Noisy, $p=0.01$
& $0.8668 \pm 0.0349$
& $0.8759 \pm 0.0389$
& $0.8698 \pm 0.0378$
& $0.8753 \pm 0.0265$ \\
\bottomrule
\end{tabular}
\vspace{-2mm}
\end{table}

\paragraph{Scope.}
These results provide a preliminary sensitivity assessment
for one architecture and task, rather than a comprehensive
hardware-noise study. The simplified model does not capture
hardware-specific connectivity, finite-shot effects,
calibration drift, or correlated errors. More realistic
noise models and hardware experiments remain future work.

\section{Theoretical Results}
\label{sec:theory_main}

We summarize three theoretical components that motivate our architectural design choices.
Full assumptions, proof details, and extended discussions are deferred to Appendix~\ref{sec:appendix_theory}.

\subsection{Layerwise Error Propagation}
\label{subsec:layerwise_error_propagation}

We bound how implementation bias and finite-shot errors propagate through $L$ VQC stages connected by deterministic routing.

\begin{theorem}[Layerwise error propagation]
\label{thm:layerwise_error_propagation}
For fixed circuit parameters and a common initial input, let $H^{(L)}$ and $\widetilde H^{(L)}$ denote the ideal and implemented final outputs. At stage $l$, let $z^{(l)}:\mathcal D_{l-1}\to\mathbb R^{m_l}$ be the ideal readout and $W^{(l)}$ the fixed linear outgoing routing map. Assume $z^{(l)}$ is $L_{z,l}$-Lipschitz in $\ell_2$. Conditional on preceding randomness, suppose the routed local bias has norm at most $\beta_l$, and each implemented readout coordinate before routing has variance at most $1/S_l$, uniformly over reachable inputs and histories. Setting $\rho_l=\|W^{(l)}\|_2L_{z,l}$ gives
\begin{equation}
\mathbb E\|\widetilde H^{(L)}-H^{(L)}\|_2
\le
\sum_{l=1}^{L}
\left(\prod_{j=l+1}^{L}\rho_j\right)
\left(\beta_l+\|W^{(l)}\|_2\sqrt{\frac{m_l}{S_l}}\right),
\label{eq:layerwise_error_bound}
\end{equation}
where an empty product equals one. For a direct final readout $W^{(L)}=I$, and for dimension-preserving readouts $m_l=d$.
\end{theorem}

If $\rho_l\le1$ and $\beta_l+\|W^{(l)}\|_2\sqrt{m_l/S_l}\le\eta$ uniformly, the expected output error is at most $L\eta$, so local errors accumulate at most linearly; bounded VQC readouts alone do not guarantee $\rho_l\le1$.

\paragraph{Pointer to details.}
The assumptions, recursion, and proof are provided in Appendix~\ref{app:layerwise_error_propagation}.

\subsection{Block Information Exchange: Receptive-Field Expansion}
\label{subsec:block_information_exchange}

Let $\mathcal R^{(L)}(b)$ denote the input blocks connected to output block $b$ by paths in the block-routing graph; actual functional dependence may be smaller. The evaluated architectures use an unmixed first VQC stage.

\begin{theorem}[Receptive-field growth under sliding-window mixing]
\label{thm:rf_sliding_window_main}
Suppose $g^{(0)}=I$ and every later routing step $g^{(1)},\ldots,g^{(L-1)}$ is sliding-window with block radius $r$. Then, for every output block $b$,
$\mathcal R^{(L)}(b)\subseteq
\{b-(L-1)r,\ldots,b+(L-1)r\}\pmod B$, and consequently
\begin{equation}
|\mathcal R^{(L)}(b)|\le\min\{B,2(L-1)r+1\}.
\label{eq:rf_bound_sw_main}
\end{equation}
\end{theorem}

\begin{theorem}[Global receptive field under fully connected mixing]
\label{thm:rf_fully_connected_main}
Suppose $g^{(0)}=I$ and the first nontrivial routing step $g^{(1)}=g_{\mathrm{fc}}$ supplies an edge from every source block to every destination block. If every later block has at least one incoming routing edge, then $\mathcal R^{(1)}(b)=\{b\}$ and, for every $L\ge2$,
\begin{equation}
\mathcal R^{(L)}(b)=\{1,\ldots,B\}.
\end{equation}
Thus global block-level access begins after the first fully connected routing step.
\end{theorem}

\paragraph{Pointer to details.}
Formal routing definitions and proofs are given in Appendix~\ref{sec:theory_exchange}.

\subsection{Support-Induced Approximation Lower Bound}
\label{subsec:support_lower_bound}

Let $\mathcal S_b$ be the input blocks accessible to output block $b$, and let $T_b\in L_2$ be its target. For a PDE/BSDE time step, $T_b$ may denote the corresponding block of the exact $Z$ vector. Let $\mathcal H_b(\mathcal S_b)$ be a class of square-integrable predictors depending only on $X_{\mathcal S_b}$.

\begin{theorem}[Support-induced approximation lower bound]
\label{thm:support_lower_bound}
For squared loss,
\begin{equation}
\inf_{F_b\in\mathcal H_b(\mathcal S_b)}
\mathbb E\|F_b(X_{\mathcal S_b})-T_b\|_2^2
\ge
\mathbb E\|T_b-\mathbb E[T_b\mid X_{\mathcal S_b}]\|_2^2.
\label{eq:support_lower_bound}
\end{equation}
At depth $L$, write $\mathcal S_b^{(L)}:=\mathcal R^{(L)}(b)$ and let $\mathcal E_L^{\mathrm{supp}}$ denote the right-hand side of \eqref{eq:support_lower_bound} with $\mathcal S_b=\mathcal S_b^{(L)}$. If $\mathcal S_b^{(L)}\subseteq\mathcal S_b^{(L+1)}$, then $\mathcal E_{L+1}^{\mathrm{supp}}\le\mathcal E_L^{\mathrm{supp}}$.
\end{theorem}

Thus broader nested access cannot increase the support-induced approximation floor. This does not imply lower error for a finite trained FC-VQC.

\paragraph{Pointer to details.}
The proof, nested-information identity, and connection to FC-VQC receptive fields are provided in Appendix~\ref{sec:theory_support_mismatch}.

\section{Discussion, Limitations, and Conclusion}
\label{sec:conclusion}

\paragraph{Conclusion.}
FC-VQC supports circuit-based learning through local VQC blocks
connected by deterministic mixing, without trainable classical
neural components. SW and PL accommodate larger inputs through
additional fixed-width blocks, whereas FC uses
$q=\lceil\sqrt{d}\rceil$ qubits per block to enable global
block information exchange.

FC-VQC achieves higher mean test performance than the evaluated
monolithic VQC and standard DNN baselines on both tabular tasks.
For the PDE benchmarks, an FC-VQC variant achieves the lowest
five-seed mean trajectory RelMAE among the evaluated models
in all six settings at $d\in\{81,144\}$, with FC-VQC-FC leading
in five. These results support modular VQCs as a promising
approach to high-dimensional function approximation.
The theoretical analysis characterises conditional error
propagation, receptive fields, and support-induced approximation lower bounds, while gradient and depolarizing-noise studies
provide complementary empirical diagnostics.

\paragraph{Limitations.}
At $d=9$, FC-VQC underperforms the DNN baselines on the PDE
benchmarks. Limited trainable capacity in architectures with
few local blocks may contribute to this gap.
Table~\ref{tab:architecture_evolution_concrete} shows that
test performance on Concrete Strength improves alongside
increasing parameter counts across the selected Types~1--4,
although architecture and feature expansion also change.
In contrast, an FC-VQC variant achieves the lowest mean
trajectory RelMAE among the evaluated models for each PDE
at $d\in\{81,144\}$. These observations motivate further
investigation of capacity allocation for low-dimensional inputs.

A second limitation is potential information loss during
dimensionality reduction. Retaining fewer observables and
reducing intermediate feature dimensions may discard
task-relevant information. More informative readouts and
carefully chosen intermediate widths therefore warrant
further investigation.

\paragraph{Future work.}
All experiments reported here use classical simulation.
We will implement and evaluate FC-VQC on NISQ hardware,
examining finite-shot training and the effects of device noise
and connectivity constraints. We will also investigate capacity
allocation and readouts that better preserve task-relevant
information, and extend the evaluation to additional
high-dimensional learning tasks.

\bibliographystyle{conference}
\clearpage
\bibliography{bib/references}

\clearpage
\appendix
\etocdepthtag.toc{appendix}

\section*{Appendix Contents}
\label{sec:appendix_contents}

\begingroup
\small
\etocsettagdepth{main}{none}
\etocsettagdepth{appendix}{section}
\etocsettocstyle{}{}
\tableofcontents
\endgroup

\clearpage
\section{Experimental Setup}
\label{sec:appendix_experimental_setup}

This appendix provides the experimental details for the tabular benchmarks and the spatio-temporal PDE function-approximation tasks used in the main paper.

\subsection{Tabular Benchmarks}
\label{app:tabular_setup}

We evaluate FC-VQC on two low-dimensional diagnostic tasks: Concrete Compressive Strength regression and Red Wine Quality classification. These tasks are used to compare standard monolithic VQC, modular FC-VQC, and standard DNN baselines in settings where monolithic VQC simulation remains feasible.

\paragraph{Concrete Strength.}
The Concrete Compressive Strength dataset~\cite{yeh1998modeling} contains $1{,}030$ samples with $d=8$ numerical input features. The task is scalar regression, and we train models using mean squared error (MSE). Performance is reported using the test coefficient of determination,
\begin{equation}
    R^2
    =
    1 -
    \frac{\sum_i (y_i-\hat y_i)^2}{\sum_i (y_i-\bar y)^2}.
\end{equation}

Here, the sums are over the test samples, and $\bar y$ is the mean of their observed target values.

\paragraph{Wine Quality.}
The Red Wine Quality dataset~\cite{cortez2009modeling} contains $1{,}599$ samples with $d=11$ physicochemical input features. The quality score is treated as a six-class classification target. Models are trained using cross-entropy loss, and performance is reported using test accuracy,
\begin{equation}
    \mathrm{Acc.}
    =
    \frac{1}{N_{\rm test}}
    \sum_{i=1}^{N_{\rm test}}
    \mathbf{1}\{\hat y_i = y_i\}.
\end{equation}

For both tabular datasets, we use a 70\%/15\%/15\% train/validation/test
split. Input features are min--max scaled using training-set statistics
to the angle-encoding range and clipped to that range. For Concrete,
the target is scaled analogously during training and inverse-transformed
before evaluation. Wine labels are retained as categorical class labels.
All reported tabular results are computed over five random seeds.

\subsection{General BSDE/PDE Formulation}
\label{app:bsde_formulation}

Following the nonlinear Feynman--Kac framework~\cite{pardoux1992backward,pardoux1999forward} and Deep BSDE solvers~\cite{e2017deep,han2018solving,su2025quantum}, we consider semilinear parabolic PDEs of the form
\begin{align}
    \frac{\partial u}{\partial t}(t,x)
    &+
    \frac{1}{2}
    \mathrm{Tr}
    \left[
    \sigma(t,x)\sigma(t,x)^{\top}
    \mathrm{Hess}_x u(t,x)
    \right]
    +
    \nabla_x u(t,x)\cdot \mu(t,x) \nonumber\\
    &+
    f\left(t,x,u(t,x),\sigma(t,x)^{\top}\nabla_x u(t,x)\right)
    =
    0,
    \qquad
    u(T,x)=g(x),
    \label{eq:app_general_pde}
\end{align}
where $x\in\mathbb{R}^d$, $t\in[0,T]$, $\mu(t,x)$ is the drift, $\sigma(t,x)$ is the diffusion matrix, and $g$ is the terminal condition.

By the nonlinear Feynman--Kac correspondence, the PDE can be represented by the forward-backward SDE system
\begin{equation}
    dX_t = \mu(t,X_t)\,dt + \sigma(t,X_t)\,dW_t,
    \qquad X_0=\xi,
    \label{eq:app_forward_sde}
\end{equation}
and
\begin{equation}
    dY_t
    =
    -f(t,X_t,Y_t,Z_t)\,dt
    +
    Z_t^{\top}dW_t,
    \qquad
    Y_T=g(X_T),
    \label{eq:app_backward_sde}
\end{equation}
with
\begin{equation}
    Y_t = u(t,X_t),
    \qquad
    Z_t = \sigma(t,X_t)^{\top}\nabla_x u(t,X_t).
    \label{eq:app_yz_relation}
\end{equation}

Here, $W_t$ is a standard $d$-dimensional Brownian motion,
and $\xi\in\mathbb R^d$ is the prescribed initial state.

We discretize $[0,T]$ into $N$ equal time steps $0=t_0<t_1<\cdots<t_N=T$ with $\Delta t=T/N$. The forward process is simulated by Euler--Maruyama:
\begin{equation}
    X_{t_{n+1}}
    =
    X_{t_n}
    +
    \mu(t_n,X_{t_n})\Delta t
    +
    \sigma(t_n,X_{t_n})\Delta W_n,
    \qquad
    \Delta W_n\sim \mathcal{N}(0,\Delta t\,I_d).
    \label{eq:app_forward_euler}
\end{equation}

Starting from an initial-value prediction $\widehat Y_0$,
the differentiable solver uses
\begin{equation}
    \widehat Z_n
    =
    G_{\theta_n}(X_{t_n}),
    \qquad
    G_{\theta_n}:\mathbb{R}^{d}\to\mathbb{R}^{d},
\end{equation}
and propagates
\begin{equation}
    \widehat Y_{n+1}
    =
    \widehat Y_n
    -
    f(t_n,X_{t_n},\widehat Y_n,\widehat Z_n)\Delta t
    +
    \widehat Z_n^{\top}\Delta W_n,
    \qquad n=0,\ldots,N-1.
    \label{eq:app_backward_euler}
\end{equation}
Here, $\widehat Y_n$ and $\widehat Z_n$ denote numerical
predictions at time $t_n$. For differentiable models,
$G_{\theta_n}$ is a neural-network, explicit-feature,
tensor-network, or VQC-based module. Its parameters are
optimized through the terminal loss
\begin{equation}
    \mathcal{L}(\theta)
    =
    \mathbb{E}
    \left[
        \left|\widehat Y_N-g(X_{t_N})\right|^2
    \right].
    \label{eq:app_terminal_loss}
\end{equation}
The tree-based baselines instead use the backward-regression
procedure described in Appendix~\ref{app:training_protocol}.

\subsection{PDE Benchmark Definitions}
\label{app:pde_definitions}

Each benchmark has an analytical solution, enabling
trajectory-level evaluation against a known reference.

\subsubsection{Black--Scholes PDE}
\label{app:black_scholes_pde}

The multidimensional Black--Scholes PDE is based on the classical Black--Scholes option-pricing model~\cite{black1973pricing}:
\begin{equation}
    \frac{\partial u}{\partial t}(t,x)
    +
    r\sum_{i=1}^{d}x_i\frac{\partial u}{\partial x_i}(t,x)
    +
    \frac{1}{2}\sum_{i=1}^{d}\sigma_i^2 x_i^2
    \frac{\partial^2 u}{\partial x_i^2}(t,x)
    -
    r u(t,x)
    =
    0,
    \label{eq:app_bs_pde}
\end{equation}
with terminal payoff
\begin{equation}
    u(T,x)=g(x)
    =
    \sum_{i=1}^{d}(x_i-E_i)^{+},
    \qquad
    a^{+}:=\max\{a,0\}.
\end{equation}
The corresponding forward SDE is the component-wise geometric Brownian motion
\begin{equation}
    d(X_t)_i
    =
    r(X_t)_i\,dt
    +
    \sigma_i (X_t)_i\,d(W_t)_i.
    \label{eq:app_bs_forward}
\end{equation}
The BSDE generator is
\begin{equation}
    f(t,X_t,Y_t,Z_t)=-rY_t,
\end{equation}
so that
\begin{equation}
    dY_t = rY_t\,dt + Z_t^{\top}dW_t,
    \qquad
    Y_T=g(X_T).
    \label{eq:app_bs_bsde}
\end{equation}

In our experiments, we use initial stock price $(X_0)_i=1$, strike price $E_i=1$, risk-free rate $r=0.1$, volatility $\sigma_i=0.2$, and terminal time $T=1$ for all dimensions $i=1,\ldots,d$.

For a portfolio of independent European call options, the exact solution is the sum of the Black--Scholes formula across dimensions:
\begin{equation}
    u(t,x)
    =
    \sum_{i=1}^{d}
    \left[
    x_i\Phi(d_{1,i})
    -
    E_i e^{-r(T-t)}\Phi(d_{2,i})
    \right],
    \label{eq:app_bs_exact}
\end{equation}
where
\begin{equation}
    d_{1,i}
    =
    \frac{
    \log(x_i/E_i) + \left(r+\frac{1}{2}\sigma_i^2\right)(T-t)
    }{
    \sigma_i\sqrt{T-t}
    },
    \qquad
    d_{2,i}
    =
    d_{1,i}-\sigma_i\sqrt{T-t}.
\end{equation}

Here, $\Phi$ denotes the standard normal cumulative
distribution function. The pricing formula is evaluated
for $t<T$ and $x_i>0$; at $t=T$, we use the terminal payoff
$g(x)$.

\subsubsection{Burgers PDE}
\label{app:burgers_pde}

The Burgers-type PDE with explicit solution follows the benchmark formulation used in BSDE numerical analysis~\cite{chassagneux2014linear}:
\begin{equation}
    \frac{\partial u}{\partial t}(t,x)
    +
    \frac{d^2}{2}\Delta_x u(t,x)
    +
    \left(
    u(t,x)-\frac{2+d}{2d}
    \right)
    d\sum_{i=1}^{d}
    \frac{\partial u}{\partial x_i}(t,x)
    =
    0,
    \label{eq:app_burgers_pde}
\end{equation}
with terminal condition
\begin{equation}
    u(T,x)
    =
    \frac{
    \exp\left(T+\frac{1}{d}\sum_{i=1}^{d}x_i\right)
    }{
    1+\exp\left(T+\frac{1}{d}\sum_{i=1}^{d}x_i\right)
    }.
\end{equation}
The forward SDE is
\begin{equation}
    dX_t = d \, dW_t,
    \qquad X_0=\xi.
    \label{eq:app_burgers_forward}
\end{equation}
The backward process is
\begin{equation}
    \mathrm{d}Y_t
    =
    -
    \left(
    Y_t-\frac{d+2}{2d}
    \right)
    \left(
    \sum_{i=1}^{d}(Z_t)_i
    \right)\mathrm{d}t
    +
    Z_t^{\top}\mathrm{d}W_t,
    \qquad
    Y_T=u(T,X_T).
    \label{eq:app_burgers_bsde}
\end{equation}
The exact solution is
\begin{equation}
    u(t,x)
    =
    \frac{
    \exp\left(t+\frac{1}{d}\sum_{i=1}^{d}x_i\right)
    }{
    1+\exp\left(t+\frac{1}{d}\sum_{i=1}^{d}x_i\right)
    }.
    \label{eq:app_burgers_exact}
\end{equation}

\subsubsection{Oscillatory PDE}
\label{app:oscillatory_pde}

The time-dependent reaction--diffusion-type PDE with oscillating explicit solution follows the benchmark~\cite{gobet2017adaptive}. It is defined with $\kappa=0.6$ and $\lambda=1/\sqrt{d}$:
\begin{align}
    \frac{\partial u}{\partial t}(t,x)
    +
    \frac{1}{2}\Delta_x u(t,x)
    +
    \min\Bigg\{
    1,
    \Bigg[
    u(t,x)-\kappa-1
    -
    \sin\left(\lambda\sum_{i=1}^{d}x_i\right)
    \exp\left(\frac{\lambda^2 d(t-T)}{2}\right)
    \Bigg]^2
    \Bigg\}
    =
    0,
    \label{eq:app_osc_pde}
\end{align}
with terminal condition
\begin{equation}
    u(T,x)
    =
    1+\kappa+\sin\left(\lambda\sum_{i=1}^{d}x_i\right).
\end{equation}
The forward process is standard Brownian motion,
\begin{equation}
    dX_t=dW_t,
    \qquad X_0=\xi.
    \label{eq:app_osc_forward}
\end{equation}
The backward process is
\begin{align}
    dY_t
    =
    -
    \min\Bigg\{
    1,
    \Bigg[
    Y_t-\kappa-1
    -
    \sin\left(\lambda\sum_{i=1}^{d}(X_t)_i\right)
    \exp\left(\frac{\lambda^2 d(t-T)}{2}\right)
    \Bigg]^2
    \Bigg\}dt
    +
    Z_t^{\top}dW_t,
    \label{eq:app_osc_bsde}
\end{align}
with $Y_T=u(T,X_T)$. The exact solution is
\begin{equation}
    u(t,x)
    =
    1+\kappa+
    \sin\left(\lambda\sum_{i=1}^{d}x_i\right)
    \exp\left(\frac{\lambda^2 d(t-T)}{2}\right).
    \label{eq:app_osc_exact}
\end{equation}

\clearpage

\subsection{Training Protocol and Hyperparameters}
\label{app:training_protocol}

\paragraph{Data and PDE discretization.}
Tabular datasets, splits, and preprocessing are described in
Appendix~\ref{app:tabular_setup}. The PDE experiments use the
Black--Scholes, Burgers, and time-dependent oscillatory problems
in Appendix~\ref{app:pde_definitions}, with
$d\in\{9,36,81,144\}$, terminal time $T=1$, $N=10$ time steps,
and $\Delta t=0.1$. For each seed, a fixed set of $1{,}000$
Monte Carlo trajectories is used for both training and
trajectory-error evaluation; no independent validation or
test trajectories are used. Trajectory RelMAE is evaluated
against the analytical solutions as defined in
Appendix~\ref{app:evaluation_metrics}.

\paragraph{Differentiable training.}
Differentiable models use Adam. Tabular regression and
classification use mean squared error and cross-entropy,
respectively. For the PDE tasks, DNN, parameter-matched DNN,
FC-DNN, FC-Angle-Linear, MPS, and VQC-based models supply
state-dependent predictions to the BSDE recursion and are
optimized through the terminal loss in
\eqref{eq:app_terminal_loss}. The main VQC experiments use
noiseless state-vector simulation; the separate depolarizing-noise
study is described in Section~\ref{subsec:nisq_robustness}.

\paragraph{Tree-based training.}
XGBoost and CatBoost instead use backward regression.
Starting from the terminal payoff, regressors estimate the
conditional next-step value and martingale term from state
coordinates and simulated increments. The PDE-specific driver
then determines the preceding value estimate. Targets are
constructed from terminal payoffs and backward estimates,
without analytical solution or gradient labels. These models
use tree boosting rather than terminal-loss Adam optimization.

\paragraph{Hyperparameters.}
Table~\ref{tab:app_hyperparameters} summarizes common settings,
and Table~\ref{tab:app_pde_search_grid} gives the PDE training
budgets and dimension-dependent search grids. The table note
defines the model-specific meanings of $L$ and $K$ and the
exceptions for monolithic, stacked, and non-VQC baselines.

\begin{table}[htbp]
\centering
\caption{Common experimental settings. Adam settings apply to
 differentiable models; tree-based models use the separate backward-regression
 procedure described above. Dimension-dependent PDE settings are given
 in Table~\ref{tab:app_pde_search_grid}.}
\label{tab:app_hyperparameters}
\footnotesize
\setlength{\tabcolsep}{3pt}
\renewcommand{\arraystretch}{1.08}
\begin{tabularx}{\linewidth}{@{}lccX@{}}
\toprule
Setting & Concrete & Wine & PDE tasks \\
\midrule
Input dimension & $8$ & $11$ & $\{9,36,81,144\}$ \\
Output dimension & $1$ & $6$ & $d$ for the BSDE integrand $Z$ \\
Standard DNN hidden width & $64$ & $64$ & $64$ \\
Samples / paths & $1{,}030$ & $1{,}599$ & $1{,}000$ Monte Carlo simulation paths \\
Train/val/test split & $70/15/15$ & $70/15/15$ & No split (simulated paths) \\
Loss & MSE & Cross-entropy & Terminal squared error; regression losses for trees \\
Metric & Test $R^2$ & Test accuracy & Trajectory RelMAE \\
Optimizer & Adam & Adam & Adam for differentiable models \\
Adam learning rate & $0.005$ & $0.005$ & $0.005$ \\
Batch size & Full batch & Full batch & Table~\ref{tab:app_pde_search_grid} \\
Epochs & $5{,}000$ & $5{,}000$ & Table~\ref{tab:app_pde_search_grid} \\
Random seeds & \multicolumn{3}{c}{$\{42,123,456,789,1024\}$} \\
\bottomrule
\end{tabularx}
\end{table}

\begin{table}[htbp]
\centering
\caption{Dimension-dependent PDE training settings and hyperparameter grids.
Batch size and epochs apply to differentiable models.}
\label{tab:app_pde_search_grid}
\small
\setlength{\tabcolsep}{5pt}
\renewcommand{\arraystretch}{1.08}
\begin{tabular*}{\linewidth}{@{\extracolsep{\fill}}lcccc@{}}
\toprule
Setting & $d=9$ & $d=36$ & $d=81$ & $d=144$ \\
\midrule
Epochs & $10{,}000$ & $10{,}000$ & $10{,}000$ & $3{,}000$ \\
Batch size & $256$ & $256$ & $128$ & $16$ \\
Layer grid $L$ & $\{3,5,7,9\}$ & $\{3,5,7,9\}$ & $\{3,5\}$ & $\{3,5\}$ \\
Grid for $K$ & $\{3,5,7,9\}$ & $\{3,5,7,9\}$ & $\{3,5\}$ & $\{3,5\}$ \\
\bottomrule
\end{tabular*}
\vspace{1mm}
\begin{minipage}{\linewidth}
\footnotesize
\textit{Note.} $L$ denotes the number of stacked stages;
Type~1 uses $L=1$, and PDE Type~2 is evaluated only at $d=9$
with $L\in\{3,5\}$. $K$ denotes VQC circuit depth, FC-DNN
local hidden-layer count, MPS bond dimension, or maximum
XGBoost/CatBoost tree depth. $L$ does not apply to MPS or
the tree baselines. FC-Angle-Linear uses the displayed $L$
grid and the same training budget and batch size, but has
no independent $K$ or hidden-width hyperparameter; its results
are displayed once per $L$ across the $K$ columns. Only evaluated
model--dimension combinations are reported.
\end{minipage}
\end{table}

\paragraph{Random seeds and configuration selection.}
All reported results use the five seeds
$\{42,123,456,789,1024\}$. For each model, PDE, and dimension,
the main tables report the tested configuration with the lowest
mean trajectory RelMAE across these seeds, together with its
corresponding standard deviation. FC-Angle-Linear is selected
over $L$ only: $L\in\{3,5,7,9\}$ at $d=9,36$ and
$L\in\{3,5\}$ at $d=81,144$. At $d=9$, SW and FC use the same local width and number of
blocks and both attain global block-level receptive-field
coverage after routing, although their component-routing maps
are different.

\subsection{Model Architectures and Implementation Details}
\label{app:model_architectures}

We describe the classical baselines, VQC architectures, and
software implementation below. Training procedures and search
grids are specified in Appendix~\ref{app:training_protocol}.

\subsubsection{Classical baselines}

\paragraph{Dense DNN baselines.}
The standard DNN is a fully connected feedforward network with
ReLU hidden activations and hidden width $64$. It produces a
scalar regression output or six classification logits for the
tabular tasks, and a $d$-dimensional state-dependent output
at each PDE time step. The parameter-matched DNN uses the
same model family with an adjusted hidden width to obtain a
parameter count comparable to the reference FC-VQC configuration.

\paragraph{FC-DNN.}
FC-DNN retains the three-coordinate partition and sliding-window
routing of FC-VQC-SW, replacing each local VQC with a small
classical network. Each block maps three inputs to three outputs
through hidden layers of width $h$. The outer stage count and
local hidden-layer count are varied, providing a classical
comparison with the same block structure.

\paragraph{FC-Angle-Linear.}
FC-Angle-Linear retains the same three-coordinate partition and
circular sliding-window routing, but replaces each local VQC
with a trainable linear map over $27$ product-angle features:
\begin{equation}
    \Phi_{27}(h_b)
    =
    \bigotimes_{j=1}^{3}
    \frac{[1,\sin(h_{b,j}),\cos(h_{b,j})]}{\sqrt{2}}
    \in\mathbb R^{27},
    \qquad h_b\in\mathbb R^3.
    \label{eq:app_angle_linear_features}
\end{equation}
At time step $t_n$ and stage $\ell$, the local update is
\begin{equation}
    h_b^{(\ell+1)}
    =
    \Phi_{27}\!\left(\widetilde h_b^{(\ell)}\right)
    W_{n,\ell,b},
    \qquad
    W_{n,\ell,b}\in\mathbb R^{27\times3},
    \label{eq:app_angle_linear_update}
\end{equation}
where $\widetilde h_b^{(\ell)}$ is the routed input to block $b$.
Features are recomputed at every stage after routing. The
linear maps have no hidden layers or output activation, and
the coefficient matrices are independent across time steps,
stages, and blocks. They are trained jointly through the
terminal BSDE loss, not by separate local regressions.
This baseline controls the explicit angle features and SW
graph; it is not a width-matched replacement for FC-VQC-FC
at $d>9$.

\paragraph{Matrix product state (MPS).}
MPS represents the mapping through a chain of contracted
tensor cores~\citep{stoudenmire2016supervised}. Each input
coordinate is mapped to
$\varphi(x_i)=(\cos(x_i/2),\sin(x_i/2))$ before contracting
trainable cores to produce a $d$-dimensional output.
Parameters are independent across time steps. The bond
dimension $\chi$ controls the internal tensor dimensions and
is reported under $K$ in Table~\ref{tab:app_pde_search_grid}.

\paragraph{XGBoost and CatBoost.}
XGBoost~\citep{chen2016xgboost} and
CatBoost~\citep{prokhorenkova2018catboost} provide gradient-boosted
regressors for the backward procedure in
Appendix~\ref{app:training_protocol}. Model capacity is controlled
by tree depth, boosting rounds, and regularization; maximum
tree depth is reported under $K$.

\subsubsection{VQC architectures and readouts}

Local VQC blocks use angle encoding followed by trainable
rotation and entangling layers of the StronglyEntanglingLayers
form. Circuit depth is denoted by $K$ and the number of
stacked VQC stages by $L$.

\paragraph{Full-width baselines (Types~1 and~2).}
Type~1 uses a single $d$-qubit circuit, whereas Type~2 stacks
full-width VQCs through measurement and re-encoding.
Both are evaluated on the tabular tasks and on the PDEs at
$d=9$. Full-width PDE models at $d\ge36$ exceed the practical
resources of our direct state-vector simulation and training
setup and are not reported.

\paragraph{Modular PDE models (Type~3).}
All three variants preserve the $d$-dimensional output.
PL propagates blocks independently, SW exchanges components
between neighbouring blocks, and FC supplies each block with
components from every preceding-layer block. SW and PL use
$q=3$ and $B=d/3$; FC uses $q=B=\sqrt d$ for the four tested
perfect-square dimensions. Table~\ref{tab:app_pde_block_sizes}
lists the corresponding widths and block counts. At $d=9$, SW and FC have the same local width and number of
blocks, but use different component-routing maps.

\begin{table}[htbp]
\centering
\caption{Circuit widths and block counts per Type~3 VQC stage in
 the PDE experiments. All variants preserve the $d$-dimensional output.}
\label{tab:app_pde_block_sizes}
\small
\setlength{\tabcolsep}{9pt}
\begin{tabular}{@{}lcccc@{}}
\toprule
Setting & $d=9$ & $d=36$ & $d=81$ & $d=144$ \\
\midrule
SW/PL qubits per block $q$ & $3$ & $3$ & $3$ & $3$ \\
SW/PL blocks per stage $B$ & $3$ & $12$ & $27$ & $48$ \\
FC qubits per block $q$ & $3$ & $6$ & $9$ & $12$ \\
FC blocks per stage $B$ & $3$ & $6$ & $9$ & $12$ \\
\bottomrule
\end{tabular}
\end{table}

\paragraph{Tabular models and dimensionality reduction.}
Type~3 partitions the original features into local blocks,
with zero-padding where required. Type~4 first applies
deterministic feature expansion. Both use staged readouts
to produce a scalar regression output or six classification
logits. Table~\ref{tab:app_fcvqc_architectures} summarizes the
evaluated architectures.

\begin{table}[h]
\centering
\caption{Detailed FC-VQC architecture specifications.}
\label{tab:app_fcvqc_architectures}
\scriptsize
\setlength{\tabcolsep}{2.5pt}
\renewcommand{\arraystretch}{1.12}
\begin{tabularx}{\textwidth}{@{}p{0.08\textwidth}X X X@{}}
\toprule
Type & Concrete Strength & Wine Quality & PDE tasks \\
\midrule

Type~1
& $8t1$: one $8$-qubit monolithic VQC.
& $11t6$: one $11$-qubit monolithic VQC.
& One $9$-qubit circuit at $d=9$; not reported at $d\ge36$. \\
\midrule

Type~2
& $8t1$: stacked monolithic VQC with measure-and-re-encode interfaces.
& $11t6$: stacked monolithic VQC with measure-and-re-encode interfaces.
& Stacked $9$-qubit circuits at $d=9$; not reported at $d\ge36$. \\
\midrule

Type~3
& $8t3t1$: $8\!\to\!9$ padding; $3$ local $Q_3$ blocks; output $3\!\to\!1$.
& $12t8t6$: $11\!\to\!12$ padding; $4$ local $Q_3$ blocks; output $8\!\to\!6$.
& Dimension-preserving PL, SW, and FC variants at $d=9,36,81,144$; block sizes are listed in Table~\ref{tab:app_pde_block_sizes}. \\
\midrule

Type~4
& $16t4t1$: $8\!\to\!16$, $4$ local $Q_4$ blocks.\par
$24t8t3t1$: $8\!\to\!24$, $8$ local $Q_3$ blocks.\par
$32t11t4t1$: $8\!\to\!32$, $11$ local $Q_3$ blocks.\par
$40t14t5t1$: $8\!\to\!40$, $14$ local $Q_3$ blocks.
& $22t8t6$: $11\!\to\!22$, $8$ local $Q_3$ blocks.\par
$33t12t8t6$: $11\!\to\!33$, $11$ local $Q_3$ blocks.\par
$44t15t10t8t6$: $11\!\to\!44$, $15$ local $Q_3$ blocks.
& --- \\
\bottomrule
\end{tabularx}
\vspace{1mm}

\begin{minipage}{0.96\textwidth}
\footnotesize
\emph{Note.} $Q_n$ denotes an $n$-qubit local VQC block. A topology $d_{\rm in}t d_1t\cdots t d_{\rm out}$ denotes staged dimensionality reduction from $d_{\rm in}$ to $d_{\rm out}$. When the feature dimension is not divisible by the block size, zero-padding is applied before block partitioning.\\
\end{minipage}
\vspace{-2mm}
\end{table}

\begin{figure}[h]
    \centering
    \includegraphics[width=0.88\textwidth]{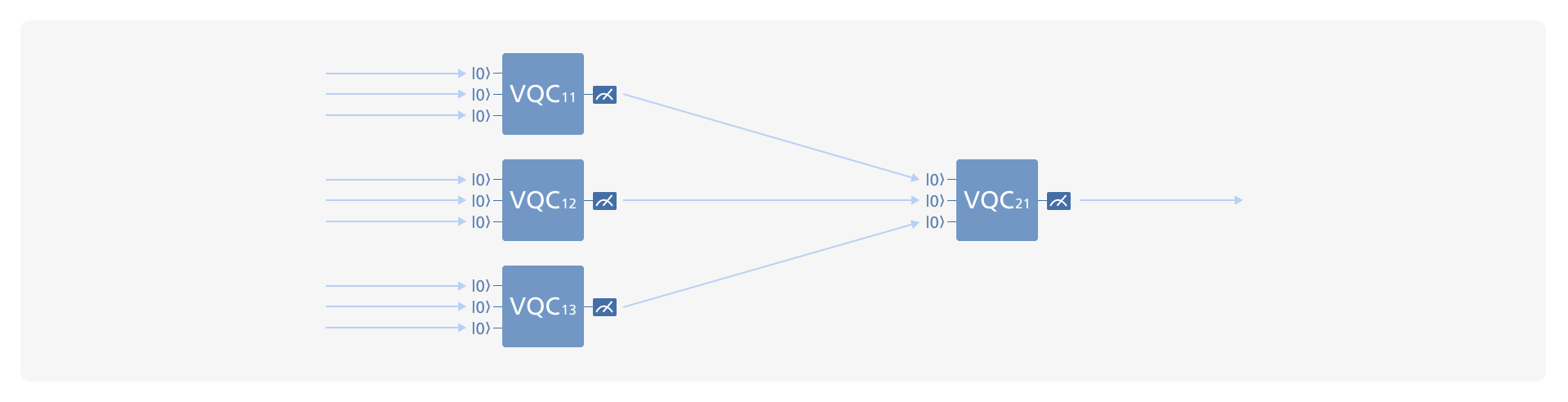}
    \caption{\textbf{Example of staged dimensionality reduction in FC-VQC.}
    The figure illustrates a $9$t$3$t$1$ reduction. Three local $Q_3$ blocks first process the $9$-dimensional input and produce $3$ intermediate outputs; a final $Q_3$ block then maps these intermediate features to a scalar output. The Concrete $8$t$3$t$1$ architecture uses the same reduction after zero-padding $8$ input features to $9$.}
    \label{fig:appendix_dim_reduction}
\end{figure}

Figure~\ref{fig:appendix_dim_reduction} illustrates reduction
from $9$ features to a scalar: three $Q_3$ blocks produce
three intermediate outputs, and a final $Q_3$ block maps
these to one output. Concrete Strength uses this structure
after padding its eight features to nine. For Wine Quality,
topologies such as $12$t$8$t$6$ instead end with an eight-qubit
VQC that returns six class logits.

\paragraph{Software implementation.}
The tabular experiments use PennyLane and PyTorch. PDE
experiments use JAX for differentiable models and VQC
state-vector simulation, together with the native XGBoost
and CatBoost libraries for tree-based models.

\subsection{Evaluation Metrics}
\label{app:evaluation_metrics}

For tabular regression and classification, we report test
$R^2$ and test accuracy, respectively. For PDE tasks, we
evaluate predicted solution trajectories using trajectory
relative mean absolute error (RelMAE). For one random seed
and $M$ simulated paths, we compute
\begin{equation}
\mathrm{RelMAE}
=
\frac{
\displaystyle
\sum_{n=0}^{N}\sum_{m=1}^{M}
\left|
\widehat{Y}^{(m)}_n-u(t_n,X^{(m)}_{t_n})
\right|
}{
\displaystyle
\sum_{n=0}^{N}\sum_{m=1}^{M}
\left|
u(t_n,X^{(m)}_{t_n})
\right|
}.
\end{equation}

where $\widehat{Y}^{(m)}_n$ is the predicted solution at time $t_n$
on path $m$, and $u(t_n,X^{(m)}_{t_n})$ is the corresponding
analytical value. The denominator is assumed nonzero.

The trajectory RelMAE aggregates the absolute error and analytical
magnitude over all simulated paths and all time steps before taking
their ratio. Lower values indicate better approximation. We compute
the metric separately for each seed and report its mean and standard
deviation across the five seeds. The trajectory plots in
Appendix~\ref{app:trajectory_plots} instead show the corresponding
relative MAE separately at each time step.
\clearpage
\section{Full Experimental Results}
\label{sec:appendix_full_results}

This appendix reports the full predictive results and
representative computational-resource measurements.
Predictive results are reported as mean $\pm$ standard
deviation across five random seeds: test $R^2$ for Concrete
Strength, test accuracy for Wine Quality, and trajectory
RelMAE for the PDE benchmarks. Runtime and memory measurements
are reported separately for representative single runs.
Dashes indicate inapplicable hyperparameters or unevaluated
configurations, as appropriate.

\subsection{Tabular Diagnostic Benchmarks}
\label{app:results_tabular}

Tables~\ref{tab:app_concrete_results} and~\ref{tab:app_wine_results} report the full tabular benchmark results across evaluated VQC depths, FC-VQC layers, and architecture types. These results support the low-dimensional diagnostic comparisons in the main paper.

\begin{table}[h]
\centering
\caption{Full test $R^2$ results for Concrete Strength.}
\label{tab:app_concrete_results}
\scriptsize
\setlength{\tabcolsep}{2.5pt}
\renewcommand{\arraystretch}{1.06}
\begin{tabular*}{\textwidth}{@{\extracolsep{\fill}}llccccc@{}}
\toprule
Type & Architecture & Layer & Depth $=3$ & Depth $=5$ & Depth $=7$ & Depth $=9$ \\
\midrule
DNN & DNN & -- 
& $0.8333 \pm 0.0555$ 
& $0.8397 \pm 0.0435$ 
& $0.8486 \pm 0.0291$ 
& $0.8292 \pm 0.0432$ \\
\midrule

Type~1 & \texttt{8t1} & $1$ 
& $0.4636 \pm 0.0264$ 
& $0.5880 \pm 0.0744$ 
& $0.6743 \pm 0.0211$ 
& $0.6768 \pm 0.0218$ \\
\midrule

\multirow{4}{*}{Type~2}
& \multirow{4}{*}{\texttt{8t1}} & $3$ 
& $0.5655 \pm 0.0715$ 
& $0.7051 \pm 0.0308$ 
& $0.7323 \pm 0.0605$ 
& $0.7360 \pm 0.0439$ \\
&  & $5$ 
& $0.5632 \pm 0.1354$ 
& $0.7287 \pm 0.0279$ 
& $0.6818 \pm 0.0923$ 
& $0.6960 \pm 0.1383$ \\
&  & $7$ 
& $0.5884 \pm 0.1560$ 
& $0.6570 \pm 0.0857$ 
& $0.6422 \pm 0.1385$ 
& $0.6811 \pm 0.0513$ \\
&  & $9$ 
& $0.6413 \pm 0.0425$ 
& $0.5543 \pm 0.2305$ 
& $0.5860 \pm 0.0837$ 
& $0.5782 \pm 0.3292$ \\
\midrule

\multirow{4}{*}{Type~3}
& \multirow{4}{*}{\texttt{8t3t1}} & $3$ 
& $0.8010 \pm 0.0460$ 
& $0.8141 \pm 0.0453$ 
& $0.7914 \pm 0.0634$ 
& $0.7901 \pm 0.0783$ \\
&  & $5$ 
& $0.7757 \pm 0.0633$ 
& $0.8005 \pm 0.0525$ 
& $0.8021 \pm 0.0767$ 
& $0.7876 \pm 0.0679$ \\
&  & $7$ 
& $0.7897 \pm 0.0553$ 
& $0.7685 \pm 0.0564$ 
& $0.8007 \pm 0.0463$ 
& $0.7857 \pm 0.0798$ \\
&  & $9$ 
& $0.7478 \pm 0.0813$ 
& $0.8096 \pm 0.0520$ 
& $0.7802 \pm 0.0565$ 
& $0.7953 \pm 0.0621$ \\
\midrule

\multirow{16}{*}{Type~4}
& \multirow{4}{*}{\texttt{16t4t1}} & $3$ 
& $0.8538 \pm 0.0272$ 
& $0.8517 \pm 0.0343$ 
& $0.8728 \pm 0.0112$ 
& $0.8491 \pm 0.0434$ \\
&  & $5$ 
& $0.8409 \pm 0.0249$ 
& $0.8124 \pm 0.0554$ 
& $0.8487 \pm 0.0328$ 
& $0.8709 \pm 0.0279$ \\
&  & $7$ 
& $0.8303 \pm 0.0341$ 
& $0.8560 \pm 0.0194$ 
& $0.8240 \pm 0.0511$ 
& $0.8259 \pm 0.0402$ \\
&  & $9$ 
& $0.8432 \pm 0.0393$ 
& $0.8293 \pm 0.0489$ 
& $0.8329 \pm 0.0162$ 
& $0.8183 \pm 0.0512$ \\
\cmidrule(lr){2-7}
& \multirow{4}{*}{\texttt{24t8t3t1}} & $3$ 
& $0.8594 \pm 0.0282$ 
& $0.8704 \pm 0.0359$ 
& $0.8708 \pm 0.0345$ 
& $0.8652 \pm 0.0384$ \\
&  & $5$ 
& $0.8637 \pm 0.0231$ 
& $0.8435 \pm 0.0505$ 
& $0.8551 \pm 0.0405$ 
& $0.8627 \pm 0.0359$ \\
&  & $7$ 
& $0.8355 \pm 0.0518$ 
& $0.8532 \pm 0.0223$ 
& $0.8269 \pm 0.0454$ 
& $0.8438 \pm 0.0282$ \\
&  & $9$ 
& $0.8348 \pm 0.0500$ 
& $0.8519 \pm 0.0394$ 
& $0.8498 \pm 0.0285$ 
& $0.8523 \pm 0.0500$ \\
\cmidrule(lr){2-7}
& \multirow{4}{*}{\texttt{32t11t4t1}} & $3$ 
& $0.8868 \pm 0.0092$ 
& $0.8773 \pm 0.0348$ 
& $0.8791 \pm 0.0261$ 
& $\mathbf{0.8928 \pm 0.0189}$ \\
&  & $5$ 
& $0.8680 \pm 0.0392$ 
& $0.8587 \pm 0.0541$ 
& $0.8591 \pm 0.0388$ 
& $0.8577 \pm 0.0301$ \\
&  & $7$ 
& $0.8768 \pm 0.0457$ 
& $0.8428 \pm 0.0358$ 
& $0.8625 \pm 0.0511$ 
& $0.8483 \pm 0.0475$ \\
&  & $9$ 
& $0.8624 \pm 0.0320$ 
& $0.8480 \pm 0.0355$ 
& $0.8505 \pm 0.0372$ 
& $0.8509 \pm 0.0281$ \\
\cmidrule(lr){2-7}
& \multirow{4}{*}{\texttt{40t14t5t1}} & $3$ 
& $0.8833 \pm 0.0182$ 
& $0.8794 \pm 0.0287$ 
& $0.8763 \pm 0.0533$ 
& $0.8705 \pm 0.0312$ \\
&  & $5$ 
& $0.8666 \pm 0.0364$ 
& $0.8612 \pm 0.0470$ 
& $0.8728 \pm 0.0366$ 
& $0.8573 \pm 0.0427$ \\
&  & $7$ 
& $0.8319 \pm 0.0304$ 
& $0.8534 \pm 0.0483$ 
& $0.8717 \pm 0.0473$ 
& $0.8297 \pm 0.0540$ \\
&  & $9$ 
& $0.8268 \pm 0.0578$ 
& $0.8486 \pm 0.0413$ 
& $0.8477 \pm 0.0465$ 
& $0.8196 \pm 0.0435$ \\
\bottomrule
\end{tabular*}
\vspace{1mm}

\begin{minipage}{0.96\textwidth}
\footnotesize
\emph{Note.} For quantum models, Depth denotes VQC circuit depth. For the DNN row, the four depth columns correspond to hidden-layer counts $L_{\rm DNN}\in\{3,5,7,9\}$, transposed into the same columns for compact presentation. Bold indicates the best mean $R^2$ in the table.
\end{minipage}
\vspace{-2mm}
\end{table}

\begin{table}[h]
\centering
\caption{Full test accuracy results for Wine Quality.}
\label{tab:app_wine_results}
\scriptsize
\setlength{\tabcolsep}{2.5pt}
\renewcommand{\arraystretch}{1.06}
\begin{tabular*}{\textwidth}{@{\extracolsep{\fill}}llccccc@{}}
\toprule
Type & Architecture & Layer & Depth $=3$ & Depth $=5$ & Depth $=7$ & Depth $=9$ \\
\midrule
DNN & DNN & -- 
& $58.4\% \pm 3.1\%$ 
& $57.8\% \pm 2.6\%$ 
& $56.6\% \pm 3.8\%$ 
& $56.7\% \pm 1.3\%$ \\
\midrule

Type~1 & \texttt{11t6} & $1$ 
& $48.2\% \pm 0.6\%$ 
& $55.4\% \pm 1.1\%$ 
& $57.2\% \pm 1.6\%$ 
& $56.1\% \pm 1.3\%$ \\
\midrule

\multirow{4}{*}{Type~2}
&\multirow{4}{*}{\texttt{11t6}} & $3$ 
& $58.3\% \pm 0.7\%$ 
& $59.1\% \pm 2.1\%$ 
& $59.7\% \pm 2.1\%$ 
& $59.8\% \pm 1.4\%$ \\
&  & $5$ 
& $57.7\% \pm 2.4\%$ 
& $60.7\% \pm 2.8\%$ 
& $60.0\% \pm 2.8\%$ 
& $60.2\% \pm 0.4\%$ \\
&  & $7$ 
& $57.4\% \pm 2.2\%$ 
& $59.8\% \pm 1.0\%$ 
& $62.2\% \pm 1.4\%$ 
& $59.8\% \pm 1.4\%$ \\
&  & $9$ 
& $57.1\% \pm 3.5\%$ 
& $60.7\% \pm 1.3\%$ 
& $60.0\% \pm 2.2\%$ 
& $59.3\% \pm 2.3\%$ \\
\midrule

\multirow{4}{*}{Type~3}
& \multirow{4}{*}{\texttt{12t8t6}} & $3$ 
& $60.5\% \pm 1.0\%$ 
& $60.2\% \pm 1.2\%$ 
& $61.4\% \pm 2.2\%$ 
& $60.4\% \pm 2.1\%$ \\
&  & $5$ 
& $60.9\% \pm 2.0\%$ 
& $58.5\% \pm 1.1\%$ 
& $60.8\% \pm 2.5\%$ 
& $59.8\% \pm 2.6\%$ \\
&  & $7$ 
& $60.0\% \pm 2.1\%$ 
& $61.5\% \pm 0.8\%$ 
& $60.2\% \pm 1.2\%$ 
& $58.1\% \pm 2.7\%$ \\
&  & $9$ 
& $58.8\% \pm 1.5\%$ 
& $60.9\% \pm 2.0\%$ 
& $59.0\% \pm 2.9\%$ 
& $57.2\% \pm 1.5\%$ \\
\midrule

\multirow{12}{*}{Type~4}
& \multirow{4}{*}{\texttt{22t8t6}} & $3$ 
& $59.3\% \pm 1.2\%$ 
& $59.9\% \pm 2.6\%$ 
& $58.7\% \pm 1.4\%$ 
& $59.7\% \pm 1.4\%$ \\
&  & $5$ 
& $57.9\% \pm 1.9\%$ 
& $60.2\% \pm 1.5\%$ 
& $59.4\% \pm 2.0\%$ 
& $59.1\% \pm 2.0\%$ \\
&  & $7$ 
& $58.0\% \pm 1.7\%$ 
& $56.5\% \pm 1.5\%$ 
& $59.2\% \pm 1.4\%$ 
& $59.0\% \pm 3.5\%$ \\
&  & $9$ 
& $57.8\% \pm 1.8\%$ 
& $55.1\% \pm 0.6\%$ 
& $57.7\% \pm 2.1\%$ 
& $60.5\% \pm 2.6\%$ \\
\cmidrule(lr){2-7}
& \multirow{4}{*}{\texttt{33t12t8t6}} & $3$ 
& $\mathbf{63.6\% \pm 1.1\%}$ 
& $62.8\% \pm 0.5\%$ 
& $61.4\% \pm 1.4\%$ 
& $62.0\% \pm 0.7\%$ \\
& & $5$ 
& $59.1\% \pm 2.2\%$ 
& $59.5\% \pm 1.2\%$ 
& $60.1\% \pm 2.1\%$ 
& $62.6\% \pm 1.5\%$ \\
& & $7$ 
& $59.4\% \pm 2.1\%$ 
& $60.7\% \pm 2.2\%$ 
& $59.3\% \pm 2.9\%$ 
& $62.7\% \pm 1.0\%$ \\
&  & $9$ 
& $59.2\% \pm 1.9\%$ 
& $58.7\% \pm 4.1\%$ 
& $59.7\% \pm 3.4\%$ 
& $59.9\% \pm 2.3\%$ \\
\cmidrule(lr){2-7}
& \multirow{4}{*}{\texttt{44t15t10t8t6}} & $3$ 
& $62.2\% \pm 1.4\%$ 
& $60.0\% \pm 0.9\%$ 
& $58.2\% \pm 2.5\%$ 
& $60.7\% \pm 2.2\%$ \\
&  & $5$ 
& $61.0\% \pm 1.3\%$ 
& $58.2\% \pm 1.8\%$ 
& $60.2\% \pm 3.0\%$ 
& $60.4\% \pm 2.2\%$ \\
&  & $7$ 
& $60.6\% \pm 1.2\%$ 
& $58.0\% \pm 1.2\%$ 
& $58.7\% \pm 1.9\%$ 
& $57.8\% \pm 2.8\%$ \\
&  & $9$ 
& $58.8\% \pm 1.8\%$ 
& $57.6\% \pm 1.7\%$ 
& $57.2\% \pm 3.0\%$ 
& $61.5\% \pm 1.4\%$ \\
\bottomrule
\end{tabular*}
\vspace{1mm}

\begin{minipage}{0.96\textwidth}
\footnotesize
\emph{Note.} For quantum models, Depth denotes VQC circuit depth. For the DNN row, the four depth columns correspond to hidden-layer counts $L_{\rm DNN}\in\{3,5,7,9\}$, transposed into the same columns for compact presentation. Bold indicates the best mean accuracy in the table.
\end{minipage}
\vspace{-2mm}
\end{table}

\clearpage

\subsection{BSDE/PDE Benchmarks}
\label{app:bsde_pde_benchmarks}

Tables~\ref{tab:app_pde_d9_black_scholes}--\ref{tab:app_pde_d144_oscillatory}
report trajectory RelMAE for the Black--Scholes, Burgers, and
time-dependent oscillatory PDEs at $d\in\{9,36,81,144\}$.
Each entry gives the mean $\pm$ standard deviation across seeds
$\{42,123,456,789,1024\}$ for one configuration. All available
configurations are reported, with values rounded to four decimal places.
Lower values are better. Boldface identifies the lowest mean within each
model, and underlining identifies the overall lowest mean in each
PDE--dimension table. Both use unrounded values and do not imply statistical
significance. Horizontal rules separate neural-network and explicit-feature baselines,
tensor-network/tree baselines, and VQC-based models.

For VQC models, $L$ denotes the number of stages and $K$ the circuit
depth. For FC-DNN, $L$ and $K$ denote outer stages and local hidden-layer
count. Parameter-matched DNN configurations are indexed by the reference
$(L,K)$ settings used for parameter matching. For MPS, $K$ denotes bond
dimension; for XGBoost and CatBoost, it denotes maximum tree depth.
The layer parameter $L$ does not apply to these three models, as indicated
by a dash. Standard DNN and FC-Angle-Linear results are indexed by
layer count and displayed once across the $K$ columns, since $K$ is
not a separate hyperparameter for either model. FC-Angle-Linear uses
$L\in\{3,5,7,9\}$ at $d=9,36$ and $L\in\{3,5\}$ at $d=81,144$.

The monolithic VQC (Type~1, $L=1$) and stacked full-width VQC
(Type~2, $L>1$) are reported at $d=9$. FC-VQC-PL, FC-VQC-SW,
and FC-VQC-FC are Type~3 variants. At $d=9$, FC-VQC-SW is not reported separately;
the FC-VQC-FC results are shown for that dimension. The standard DNN
uses hidden width $64$.

\begin{table}[!htbp]
\centering
\caption{Black--Scholes PDE at $d=9$: trajectory RelMAE (mean $\pm$ standard deviation). Bold: best per model; underline: best overall. Lower is better.}
\label{tab:app_pde_d9_black_scholes}
\footnotesize
\setlength{\tabcolsep}{1.5pt}
\renewcommand{\arraystretch}{1.08}
\begin{tabular*}{\linewidth}{@{\extracolsep{\fill}}lccccc@{}}
\toprule
Model & $L$ & $K=3$ & $K=5$ & $K=7$ & $K=9$ \\
\midrule
DNN & $3$ & \multicolumn{4}{c}{$\underline{\mathbf{0.0374\pm0.0010}}$} \\ 
 & $5$ & \multicolumn{4}{c}{$0.0410\pm0.0009$} \\ 
 & $7$ & \multicolumn{4}{c}{$0.0467\pm0.0021$} \\ 
 & $9$ & \multicolumn{4}{c}{$0.0475\pm0.0019$} \\ 
\addlinespace[2pt]
Parameter-matched DNN & $3$ & $0.0504\pm0.0017$ & $0.0444\pm0.0013$ & $\mathbf{0.0421\pm0.0011}$ & $0.0422\pm0.0011$ \\ 
 & $5$ & $0.0552\pm0.0019$ & $0.0498\pm0.0012$ & $0.0462\pm0.0011$ & $0.0461\pm0.0014$ \\ 
 & $7$ & $0.0568\pm0.0023$ & $0.0532\pm0.0025$ & $0.0506\pm0.0016$ & $0.0501\pm0.0026$ \\ 
 & $9$ & $0.0567\pm0.0027$ & $0.0551\pm0.0032$ & $0.0513\pm0.0035$ & $0.0510\pm0.0018$ \\ 
\addlinespace[2pt]
FC-DNN & $3$ & $\mathbf{0.1612\pm0.0093}$ & $0.1760\pm0.0075$ & $0.1924\pm0.0079$ & $0.2024\pm0.0030$ \\ 
 & $5$ & $0.1723\pm0.0122$ & $0.1969\pm0.0025$ & $0.2024\pm0.0031$ & $0.2054\pm0.0024$ \\ 
 & $7$ & $0.1887\pm0.0126$ & $0.2043\pm0.0029$ & $0.2057\pm0.0026$ & $0.2057\pm0.0026$ \\ 
 & $9$ & $0.1927\pm0.0063$ & $0.2057\pm0.0026$ & $0.2053\pm0.0023$ & $0.2057\pm0.0026$ \\ 
\addlinespace[2pt]
FC-Angle-Linear & $3$ & \multicolumn{4}{c}{$0.0416\pm0.0006$} \\ 
 & $5$ & \multicolumn{4}{c}{$0.0419\pm0.0032$} \\ 
 & $7$ & \multicolumn{4}{c}{$\mathbf{0.0408\pm0.0015}$} \\ 
 & $9$ & \multicolumn{4}{c}{$0.0438\pm0.0036$} \\ 
\midrule
MPS & --- & $0.0691\pm0.0042$ & $0.0475\pm0.0018$ & $0.0454\pm0.0012$ & $\mathbf{0.0439\pm0.0017}$ \\ 
\addlinespace[2pt]
XGBoost & --- & $\mathbf{0.0553\pm0.0011}$ & $0.0962\pm0.0017$ & $0.1541\pm0.0028$ & $0.1796\pm0.0036$ \\ 
\addlinespace[2pt]
CatBoost & --- & $0.0543\pm0.0104$ & $\mathbf{0.0471\pm0.0018}$ & $0.0686\pm0.0015$ & $0.1057\pm0.0019$ \\ 
\midrule
Monolithic VQC (Type~1) & $1$ & $0.0537\pm0.0010$ & $0.0537\pm0.0008$ & $0.0484\pm0.0017$ & $\mathbf{0.0471\pm0.0013}$ \\ 
\addlinespace[2pt]
FC-VQC-Type2 & $3$ & $0.0495\pm0.0010$ & $0.0501\pm0.0006$ & $0.0485\pm0.0011$ & $\mathbf{0.0478\pm0.0015}$ \\ 
 & $5$ & $0.0505\pm0.0016$ & $0.0495\pm0.0016$ & $0.0519\pm0.0027$ & $0.0516\pm0.0039$ \\ 
\addlinespace[2pt]
FC-VQC-PL & $3$ & $0.0404\pm0.0010$ & $0.0394\pm0.0014$ & $\mathbf{0.0387\pm0.0011}$ & $0.0399\pm0.0021$ \\ 
 & $5$ & $0.0448\pm0.0022$ & $0.0435\pm0.0016$ & $0.0444\pm0.0006$ & $0.0434\pm0.0013$ \\ 
 & $7$ & $0.0447\pm0.0018$ & $0.0442\pm0.0010$ & $0.0442\pm0.0013$ & $0.0439\pm0.0017$ \\ 
 & $9$ & $0.0452\pm0.0009$ & $0.0437\pm0.0014$ & $0.0436\pm0.0006$ & $0.0435\pm0.0015$ \\ 
\addlinespace[2pt]
FC-VQC-FC & $3$ & $0.0410\pm0.0014$ & $0.0392\pm0.0013$ & $0.0395\pm0.0004$ & $\mathbf{0.0391\pm0.0018}$ \\ 
 & $5$ & $0.0434\pm0.0015$ & $0.0412\pm0.0013$ & $0.0416\pm0.0013$ & $0.0418\pm0.0014$ \\ 
 & $7$ & $0.0456\pm0.0016$ & $0.0468\pm0.0022$ & $0.0467\pm0.0009$ & $0.0464\pm0.0025$ \\ 
 & $9$ & $0.0510\pm0.0024$ & $0.0501\pm0.0023$ & $0.0502\pm0.0015$ & $0.0498\pm0.0030$ \\ 
\bottomrule
\end{tabular*}
\end{table}

\begin{table}[!htbp]
\centering
\caption{Burgers PDE at $d=9$: trajectory RelMAE (mean $\pm$ standard deviation). Bold: best per model; underline: best overall. Lower is better.}
\label{tab:app_pde_d9_burgers}
\footnotesize
\setlength{\tabcolsep}{1.5pt}
\renewcommand{\arraystretch}{1.08}
\begin{tabular*}{\linewidth}{@{\extracolsep{\fill}}lccccc@{}}
\toprule
Model & $L$ & $K=3$ & $K=5$ & $K=7$ & $K=9$ \\
\midrule
DNN & $3$ & \multicolumn{4}{c}{$0.1074\pm0.0060$} \\ 
 & $5$ & \multicolumn{4}{c}{$0.1055\pm0.0074$} \\ 
 & $7$ & \multicolumn{4}{c}{$\mathbf{0.1040\pm0.0050}$} \\ 
 & $9$ & \multicolumn{4}{c}{$0.1122\pm0.0142$} \\ 
\addlinespace[2pt]
Parameter-matched DNN & $3$ & $0.1433\pm0.0121$ & $0.1133\pm0.0080$ & $0.1175\pm0.0065$ & $0.1150\pm0.0082$ \\ 
 & $5$ & $0.1275\pm0.0158$ & $0.1092\pm0.0041$ & $0.1096\pm0.0076$ & $0.1065\pm0.0080$ \\ 
 & $7$ & $0.1409\pm0.0094$ & $0.1149\pm0.0162$ & $0.1179\pm0.0045$ & $\underline{\mathbf{0.1032\pm0.0067}}$ \\ 
 & $9$ & $0.1390\pm0.0093$ & $0.1199\pm0.0172$ & $0.1249\pm0.0099$ & $0.1218\pm0.0174$ \\ 
\addlinespace[2pt]
FC-DNN & $3$ & $\mathbf{0.3492\pm0.0230}$ & $0.3935\pm0.0111$ & $0.4408\pm0.0111$ & $0.4476\pm0.0138$ \\ 
 & $5$ & $0.3820\pm0.0250$ & $0.4386\pm0.0062$ & $0.4516\pm0.0106$ & $0.4550\pm0.0095$ \\ 
 & $7$ & $0.4326\pm0.0277$ & $0.4538\pm0.0083$ & $0.4559\pm0.0096$ & $0.4561\pm0.0096$ \\ 
 & $9$ & $0.4285\pm0.0283$ & $0.4553\pm0.0095$ & $0.4555\pm0.0095$ & $0.4561\pm0.0096$ \\ 
\addlinespace[2pt]
FC-Angle-Linear & $3$ & \multicolumn{4}{c}{$0.1362\pm0.0036$} \\ 
 & $5$ & \multicolumn{4}{c}{$\mathbf{0.1156\pm0.0070}$} \\ 
 & $7$ & \multicolumn{4}{c}{$0.1180\pm0.0077$} \\ 
 & $9$ & \multicolumn{4}{c}{$0.1832\pm0.0165$} \\ 
\midrule
MPS & --- & $0.2280\pm0.0093$ & $0.1646\pm0.0055$ & $0.1433\pm0.0038$ & $\mathbf{0.1360\pm0.0032}$ \\ 
\addlinespace[2pt]
XGBoost & --- & $\mathbf{0.1407\pm0.0178}$ & $0.2378\pm0.0052$ & $0.3482\pm0.0028$ & $0.3610\pm0.0053$ \\ 
\addlinespace[2pt]
CatBoost & --- & $0.1735\pm0.0316$ & $\mathbf{0.1311\pm0.0100}$ & $0.1762\pm0.0082$ & $0.2796\pm0.0038$ \\ 
\midrule
Monolithic VQC (Type~1) & $1$ & $0.1375\pm0.0066$ & $0.1341\pm0.0070$ & $0.1325\pm0.0075$ & $\mathbf{0.1313\pm0.0080}$ \\ 
\addlinespace[2pt]
FC-VQC-Type2 & $3$ & $0.1349\pm0.0074$ & $0.1314\pm0.0072$ & $0.1277\pm0.0066$ & $0.1298\pm0.0069$ \\ 
 & $5$ & $0.1349\pm0.0086$ & $0.1298\pm0.0093$ & $0.1259\pm0.0094$ & $\mathbf{0.1244\pm0.0091}$ \\ 
\addlinespace[2pt]
FC-VQC-PL & $3$ & $0.1469\pm0.0058$ & $0.1470\pm0.0048$ & $\mathbf{0.1462\pm0.0039}$ & $0.1464\pm0.0042$ \\ 
 & $5$ & $0.1496\pm0.0065$ & $0.1555\pm0.0047$ & $0.1538\pm0.0045$ & $0.1522\pm0.0055$ \\ 
 & $7$ & $0.1576\pm0.0070$ & $0.1594\pm0.0062$ & $0.1624\pm0.0067$ & $0.1566\pm0.0016$ \\ 
 & $9$ & $0.1562\pm0.0044$ & $0.1629\pm0.0053$ & $0.1547\pm0.0084$ & $0.1550\pm0.0073$ \\ 
\addlinespace[2pt]
FC-VQC-FC & $3$ & $0.1439\pm0.0060$ & $0.1501\pm0.0118$ & $0.1489\pm0.0087$ & $0.1478\pm0.0105$ \\ 
 & $5$ & $0.1400\pm0.0096$ & $0.1410\pm0.0110$ & $0.1396\pm0.0071$ & $0.1387\pm0.0083$ \\ 
 & $7$ & $0.1224\pm0.0105$ & $0.1211\pm0.0070$ & $0.1299\pm0.0074$ & $0.1278\pm0.0099$ \\ 
 & $9$ & $\mathbf{0.1155\pm0.0084}$ & $0.1226\pm0.0191$ & $0.1271\pm0.0221$ & $0.1266\pm0.0208$ \\ 
\bottomrule
\end{tabular*}
\end{table}

\begin{table}[!htbp]
\centering
\caption{Time-dependent oscillatory PDE at $d=9$: trajectory RelMAE (mean $\pm$ standard deviation). Bold: best per model; underline: best overall. Lower is better.}
\label{tab:app_pde_d9_oscillatory}
\footnotesize
\setlength{\tabcolsep}{1.5pt}
\renewcommand{\arraystretch}{1.08}
\begin{tabular*}{\linewidth}{@{\extracolsep{\fill}}lccccc@{}}
\toprule
Model & $L$ & $K=3$ & $K=5$ & $K=7$ & $K=9$ \\
\midrule
DNN & $3$ & \multicolumn{4}{c}{$0.0268\pm0.0028$} \\ 
 & $5$ & \multicolumn{4}{c}{$0.0239\pm0.0026$} \\ 
 & $7$ & \multicolumn{4}{c}{$0.0239\pm0.0018$} \\ 
 & $9$ & \multicolumn{4}{c}{$\mathbf{0.0234\pm0.0012}$} \\ 
\addlinespace[2pt]
Parameter-matched DNN & $3$ & $0.0297\pm0.0030$ & $0.0240\pm0.0020$ & $0.0251\pm0.0027$ & $0.0243\pm0.0016$ \\ 
 & $5$ & $0.0255\pm0.0025$ & $0.0236\pm0.0036$ & $0.0247\pm0.0034$ & $0.0224\pm0.0008$ \\ 
 & $7$ & $0.0299\pm0.0064$ & $0.0231\pm0.0023$ & $0.0229\pm0.0011$ & $\underline{\mathbf{0.0219\pm0.0007}}$ \\ 
 & $9$ & $0.0287\pm0.0049$ & $0.0259\pm0.0038$ & $0.0238\pm0.0020$ & $0.0230\pm0.0010$ \\ 
\addlinespace[2pt]
FC-DNN & $3$ & $\mathbf{0.2051\pm0.0075}$ & $0.2410\pm0.0136$ & $0.2750\pm0.0085$ & $0.2790\pm0.0070$ \\ 
 & $5$ & $0.2380\pm0.0133$ & $0.2747\pm0.0061$ & $0.2818\pm0.0073$ & $0.2846\pm0.0059$ \\ 
 & $7$ & $0.2697\pm0.0165$ & $0.2822\pm0.0053$ & $0.2852\pm0.0062$ & $0.2853\pm0.0063$ \\ 
 & $9$ & $0.2669\pm0.0132$ & $0.2848\pm0.0056$ & $0.2850\pm0.0061$ & $0.2853\pm0.0063$ \\ 
\addlinespace[2pt]
FC-Angle-Linear & $3$ & \multicolumn{4}{c}{$0.0323\pm0.0028$} \\ 
 & $5$ & \multicolumn{4}{c}{$0.0305\pm0.0030$} \\ 
 & $7$ & \multicolumn{4}{c}{$\mathbf{0.0299\pm0.0042}$} \\ 
 & $9$ & \multicolumn{4}{c}{$0.0351\pm0.0047$} \\ 
\midrule
MPS & --- & $0.0479\pm0.0043$ & $0.0313\pm0.0010$ & $0.0296\pm0.0017$ & $\mathbf{0.0296\pm0.0020}$ \\ 
\addlinespace[2pt]
XGBoost & --- & $\mathbf{0.0643\pm0.0091}$ & $0.0987\pm0.0031$ & $0.1783\pm0.0026$ & $0.2051\pm0.0010$ \\ 
\addlinespace[2pt]
CatBoost & --- & $0.0796\pm0.0165$ & $\mathbf{0.0569\pm0.0037}$ & $0.0715\pm0.0030$ & $0.1154\pm0.0042$ \\ 
\midrule
Monolithic VQC (Type~1) & $1$ & $0.0474\pm0.0042$ & $0.0459\pm0.0045$ & $0.0453\pm0.0038$ & $\mathbf{0.0446\pm0.0038}$ \\ 
\addlinespace[2pt]
FC-VQC-Type2 & $3$ & $0.0458\pm0.0028$ & $0.0437\pm0.0020$ & $0.0365\pm0.0009$ & $0.0382\pm0.0016$ \\ 
 & $5$ & $0.0405\pm0.0020$ & $0.0348\pm0.0019$ & $\mathbf{0.0291\pm0.0012}$ & $0.0293\pm0.0013$ \\ 
\addlinespace[2pt]
FC-VQC-PL & $3$ & $0.0512\pm0.0030$ & $0.0512\pm0.0024$ & $0.0517\pm0.0028$ & $\mathbf{0.0510\pm0.0019}$ \\ 
 & $5$ & $0.0529\pm0.0029$ & $0.0530\pm0.0029$ & $0.0520\pm0.0022$ & $0.0530\pm0.0029$ \\ 
 & $7$ & $0.0531\pm0.0026$ & $0.0524\pm0.0016$ & $0.0538\pm0.0014$ & $0.0525\pm0.0037$ \\ 
 & $9$ & $0.0519\pm0.0028$ & $0.0522\pm0.0028$ & $0.0530\pm0.0035$ & $0.0525\pm0.0034$ \\ 
\addlinespace[2pt]
FC-VQC-FC & $3$ & $0.0510\pm0.0025$ & $0.0505\pm0.0039$ & $0.0504\pm0.0032$ & $0.0472\pm0.0041$ \\ 
 & $5$ & $0.0348\pm0.0045$ & $0.0346\pm0.0036$ & $0.0296\pm0.0030$ & $0.0290\pm0.0026$ \\ 
 & $7$ & $0.0295\pm0.0019$ & $0.0283\pm0.0039$ & $0.0321\pm0.0044$ & $0.0306\pm0.0043$ \\ 
 & $9$ & $0.0255\pm0.0020$ & $0.0286\pm0.0048$ & $0.0292\pm0.0062$ & $\mathbf{0.0252\pm0.0015}$ \\ 
\bottomrule
\end{tabular*}
\end{table}

\begin{table}[!htbp]
\centering
\caption{Black--Scholes PDE at $d=36$: trajectory RelMAE (mean $\pm$ standard deviation). Bold: best per model; underline: best overall. Lower is better.}
\label{tab:app_pde_d36_black_scholes}
\footnotesize
\setlength{\tabcolsep}{1.5pt}
\renewcommand{\arraystretch}{1.08}
\begin{tabular*}{\linewidth}{@{\extracolsep{\fill}}lccccc@{}}
\toprule
Model & $L$ & $K=3$ & $K=5$ & $K=7$ & $K=9$ \\
\midrule
DNN & $3$ & \multicolumn{4}{c}{$\mathbf{0.0306\pm0.0024}$} \\ 
 & $5$ & \multicolumn{4}{c}{$0.0385\pm0.0022$} \\ 
 & $7$ & \multicolumn{4}{c}{$0.0409\pm0.0038$} \\ 
 & $9$ & \multicolumn{4}{c}{$0.0341\pm0.0016$} \\ 
\addlinespace[2pt]
Parameter-matched DNN & $3$ & $0.0306\pm0.0027$ & $0.0307\pm0.0027$ & $0.0305\pm0.0030$ & $\mathbf{0.0303\pm0.0028}$ \\ 
 & $5$ & $0.0374\pm0.0027$ & $0.0378\pm0.0026$ & $0.0378\pm0.0025$ & $0.0373\pm0.0019$ \\ 
 & $7$ & $0.0398\pm0.0039$ & $0.0414\pm0.0034$ & $0.0394\pm0.0036$ & $0.0394\pm0.0043$ \\ 
 & $9$ & $0.0338\pm0.0018$ & $0.0341\pm0.0023$ & $0.0337\pm0.0017$ & $0.0335\pm0.0024$ \\ 
\addlinespace[2pt]
FC-DNN & $3$ & $\mathbf{0.0799\pm0.0038}$ & $0.0908\pm0.0016$ & $0.0972\pm0.0033$ & $0.1008\pm0.0019$ \\ 
 & $5$ & $0.0886\pm0.0032$ & $0.0977\pm0.0019$ & $0.1018\pm0.0011$ & $0.1023\pm0.0017$ \\ 
 & $7$ & $0.0893\pm0.0034$ & $0.1007\pm0.0023$ & $0.1025\pm0.0009$ & $0.1026\pm0.0013$ \\ 
 & $9$ & $0.0959\pm0.0033$ & $0.1022\pm0.0011$ & $0.1027\pm0.0013$ & $0.1027\pm0.0013$ \\ 
\addlinespace[2pt]
FC-Angle-Linear & $3$ & \multicolumn{4}{c}{$0.0250\pm0.0008$} \\ 
 & $5$ & \multicolumn{4}{c}{$\mathbf{0.0249\pm0.0012}$} \\ 
 & $7$ & \multicolumn{4}{c}{$0.0250\pm0.0011$} \\ 
 & $9$ & \multicolumn{4}{c}{$0.0254\pm0.0010$} \\ 
\midrule
MPS & --- & $0.0247\pm0.0007$ & $\mathbf{0.0235\pm0.0006}$ & $0.0235\pm0.0004$ & $0.0235\pm0.0005$ \\ 
\addlinespace[2pt]
XGBoost & --- & $\mathbf{0.0466\pm0.0049}$ & $0.0640\pm0.0016$ & $0.0876\pm0.0015$ & $0.0919\pm0.0024$ \\ 
\addlinespace[2pt]
CatBoost & --- & $0.0630\pm0.0044$ & $\mathbf{0.0458\pm0.0011}$ & $0.0483\pm0.0017$ & $0.0690\pm0.0085$ \\ 
\midrule
FC-VQC-PL & $3$ & $\mathbf{0.0225\pm0.0005}$ & $0.0227\pm0.0005$ & $0.0227\pm0.0006$ & $0.0226\pm0.0008$ \\ 
 & $5$ & $0.0244\pm0.0012$ & $0.0244\pm0.0005$ & $0.0238\pm0.0004$ & $0.0235\pm0.0007$ \\ 
 & $7$ & $0.0262\pm0.0024$ & $0.0259\pm0.0025$ & $0.0305\pm0.0043$ & $0.0604\pm0.0765$ \\ 
 & $9$ & $0.0271\pm0.0031$ & $0.0279\pm0.0050$ & $0.0265\pm0.0014$ & $0.0278\pm0.0030$ \\ 
\addlinespace[2pt]
FC-VQC-SW & $3$ & $0.0226\pm0.0002$ & $\underline{\mathbf{0.0224\pm0.0004}}$ & $0.0228\pm0.0004$ & $0.0226\pm0.0004$ \\ 
 & $5$ & $0.0239\pm0.0006$ & $0.0231\pm0.0008$ & $0.0232\pm0.0011$ & $0.0231\pm0.0004$ \\ 
 & $7$ & $0.0315\pm0.0060$ & $0.0249\pm0.0018$ & $0.0261\pm0.0019$ & $0.0312\pm0.0119$ \\ 
 & $9$ & $0.0309\pm0.0071$ & $0.0252\pm0.0022$ & $0.0267\pm0.0028$ & $0.0277\pm0.0019$ \\ 
\addlinespace[2pt]
FC-VQC-FC & $3$ & $0.0244\pm0.0004$ & $0.0238\pm0.0003$ & $\mathbf{0.0232\pm0.0003}$ & $0.0234\pm0.0005$ \\ 
 & $5$ & $0.0242\pm0.0007$ & $0.0246\pm0.0007$ & $0.0249\pm0.0010$ & $0.0262\pm0.0007$ \\ 
 & $7$ & $0.0269\pm0.0024$ & $0.0284\pm0.0020$ & $0.0274\pm0.0012$ & $0.0294\pm0.0028$ \\ 
 & $9$ & $0.0275\pm0.0038$ & $0.0272\pm0.0013$ & $0.0272\pm0.0018$ & $0.0267\pm0.0008$ \\ 
\bottomrule
\end{tabular*}
\end{table}

\begin{table}[!htbp]
\centering
\caption{Burgers PDE at $d=36$: trajectory RelMAE (mean $\pm$ standard deviation). Bold: best per model; underline: best overall. Lower is better.}
\label{tab:app_pde_d36_burgers}
\footnotesize
\setlength{\tabcolsep}{1.5pt}
\renewcommand{\arraystretch}{1.08}
\begin{tabular*}{\linewidth}{@{\extracolsep{\fill}}lccccc@{}}
\toprule
Model & $L$ & $K=3$ & $K=5$ & $K=7$ & $K=9$ \\
\midrule
DNN & $3$ & \multicolumn{4}{c}{$\mathbf{0.3609\pm0.0246}$} \\ 
 & $5$ & \multicolumn{4}{c}{$0.4035\pm0.0078$} \\ 
 & $7$ & \multicolumn{4}{c}{$0.3912\pm0.0175$} \\ 
 & $9$ & \multicolumn{4}{c}{$0.3912\pm0.0108$} \\ 
\addlinespace[2pt]
Parameter-matched DNN & $3$ & $0.3609\pm0.0246$ & $0.3609\pm0.0246$ & $0.3609\pm0.0246$ & $\mathbf{0.3601\pm0.0283}$ \\ 
 & $5$ & $0.4014\pm0.0075$ & $0.4046\pm0.0143$ & $0.4045\pm0.0129$ & $0.4001\pm0.0128$ \\ 
 & $7$ & $0.4011\pm0.0165$ & $0.4041\pm0.0185$ & $0.3992\pm0.0188$ & $0.3992\pm0.0188$ \\ 
 & $9$ & $0.3898\pm0.0136$ & $0.3821\pm0.0142$ & $0.3991\pm0.0099$ & $0.3888\pm0.0137$ \\ 
\addlinespace[2pt]
FC-DNN & $3$ & $\mathbf{0.4518\pm0.0174}$ & $0.5649\pm0.0112$ & $0.5986\pm0.0108$ & $0.6146\pm0.0087$ \\ 
 & $5$ & $0.5421\pm0.0113$ & $0.6002\pm0.0210$ & $0.6230\pm0.0098$ & $0.6264\pm0.0092$ \\ 
 & $7$ & $0.5597\pm0.0089$ & $0.6113\pm0.0068$ & $0.6277\pm0.0097$ & $0.6280\pm0.0067$ \\ 
 & $9$ & $0.5760\pm0.0078$ & $0.6228\pm0.0114$ & $0.6288\pm0.0084$ & $0.6288\pm0.0084$ \\ 
\addlinespace[2pt]
FC-Angle-Linear & $3$ & \multicolumn{4}{c}{$0.4431\pm0.0108$} \\ 
 & $5$ & \multicolumn{4}{c}{$0.4336\pm0.0112$} \\ 
 & $7$ & \multicolumn{4}{c}{$0.4356\pm0.0144$} \\ 
 & $9$ & \multicolumn{4}{c}{$\mathbf{0.4122\pm0.0161}$} \\ 
\midrule
MPS & --- & $0.2962\pm0.0086$ & $0.2818\pm0.0061$ & $0.2808\pm0.0091$ & $\underline{\mathbf{0.2776\pm0.0098}}$ \\ 
\addlinespace[2pt]
XGBoost & --- & $\mathbf{0.2964\pm0.0289}$ & $0.3855\pm0.0145$ & $0.4543\pm0.0156$ & $0.4869\pm0.0358$ \\ 
\addlinespace[2pt]
CatBoost & --- & $0.4202\pm0.0277$ & $\mathbf{0.3121\pm0.0085}$ & $0.3878\pm0.0138$ & $0.4783\pm0.0165$ \\ 
\midrule
FC-VQC-PL & $3$ & $0.3931\pm0.0118$ & $0.3775\pm0.0112$ & $0.3732\pm0.0136$ & $0.3705\pm0.0120$ \\ 
 & $5$ & $0.3899\pm0.0119$ & $0.3804\pm0.0150$ & $0.3761\pm0.0104$ & $0.3714\pm0.0093$ \\ 
 & $7$ & $0.3906\pm0.0095$ & $0.3732\pm0.0135$ & $0.3668\pm0.0117$ & $0.3551\pm0.0131$ \\ 
 & $9$ & $0.3762\pm0.0121$ & $0.3648\pm0.0091$ & $0.3559\pm0.0125$ & $\mathbf{0.3528\pm0.0141}$ \\ 
\addlinespace[2pt]
FC-VQC-SW & $3$ & $0.3582\pm0.0077$ & $0.3368\pm0.0080$ & $0.3363\pm0.0081$ & $0.3324\pm0.0057$ \\ 
 & $5$ & $0.3593\pm0.0094$ & $0.3404\pm0.0130$ & $0.3283\pm0.0074$ & $0.3322\pm0.0060$ \\ 
 & $7$ & $0.3641\pm0.0034$ & $0.3461\pm0.0055$ & $0.3357\pm0.0039$ & $0.3321\pm0.0086$ \\ 
 & $9$ & $0.3593\pm0.0070$ & $0.3425\pm0.0015$ & $0.3287\pm0.0076$ & $\mathbf{0.3260\pm0.0103}$ \\ 
\addlinespace[2pt]
FC-VQC-FC & $3$ & $0.3629\pm0.0124$ & $0.3452\pm0.0087$ & $0.3348\pm0.0071$ & $\mathbf{0.3245\pm0.0050}$ \\ 
 & $5$ & $0.3680\pm0.0054$ & $0.3498\pm0.0078$ & $0.3457\pm0.0112$ & $0.3339\pm0.0041$ \\ 
 & $7$ & $0.3663\pm0.0081$ & $0.3525\pm0.0062$ & $0.3398\pm0.0066$ & $0.3481\pm0.0074$ \\ 
 & $9$ & $0.3791\pm0.0099$ & $0.3469\pm0.0080$ & $0.3450\pm0.0114$ & $0.3416\pm0.0106$ \\ 
\bottomrule
\end{tabular*}
\end{table}

\begin{table}[!htbp]
\centering
\caption{Time-dependent oscillatory PDE at $d=36$: trajectory RelMAE (mean $\pm$ standard deviation). Bold: best per model; underline: best overall. Lower is better.}
\label{tab:app_pde_d36_oscillatory}
\footnotesize
\setlength{\tabcolsep}{1.5pt}
\renewcommand{\arraystretch}{1.08}
\begin{tabular*}{\linewidth}{@{\extracolsep{\fill}}lccccc@{}}
\toprule
Model & $L$ & $K=3$ & $K=5$ & $K=7$ & $K=9$ \\
\midrule
DNN & $3$ & \multicolumn{4}{c}{$0.0770\pm0.0061$} \\ 
 & $5$ & \multicolumn{4}{c}{$0.0835\pm0.0111$} \\ 
 & $7$ & \multicolumn{4}{c}{$0.0778\pm0.0053$} \\ 
 & $9$ & \multicolumn{4}{c}{$\mathbf{0.0749\pm0.0107}$} \\ 
\addlinespace[2pt]
Parameter-matched DNN & $3$ & $0.0747\pm0.0059$ & $0.0759\pm0.0057$ & $0.0747\pm0.0059$ & $0.0770\pm0.0061$ \\ 
 & $5$ & $0.0831\pm0.0126$ & $0.0819\pm0.0098$ & $0.0835\pm0.0111$ & $0.0851\pm0.0112$ \\ 
 & $7$ & $0.0664\pm0.0045$ & $0.0654\pm0.0055$ & $0.0639\pm0.0028$ & $0.0653\pm0.0021$ \\ 
 & $9$ & $0.0664\pm0.0031$ & $\mathbf{0.0636\pm0.0018}$ & $0.0662\pm0.0039$ & $0.0753\pm0.0092$ \\ 
\addlinespace[2pt]
FC-DNN & $3$ & $\mathbf{0.1850\pm0.0112}$ & $0.2479\pm0.0162$ & $0.2689\pm0.0106$ & $0.2741\pm0.0099$ \\ 
 & $5$ & $0.2366\pm0.0118$ & $0.2654\pm0.0164$ & $0.2799\pm0.0079$ & $0.2805\pm0.0082$ \\ 
 & $7$ & $0.2485\pm0.0097$ & $0.2732\pm0.0069$ & $0.2807\pm0.0069$ & $0.2808\pm0.0059$ \\ 
 & $9$ & $0.2591\pm0.0051$ & $0.2790\pm0.0084$ & $0.2815\pm0.0071$ & $0.2815\pm0.0071$ \\ 
\addlinespace[2pt]
FC-Angle-Linear & $3$ & \multicolumn{4}{c}{$0.3462\pm0.0062$} \\ 
 & $5$ & \multicolumn{4}{c}{$0.3532\pm0.0034$} \\ 
 & $7$ & \multicolumn{4}{c}{$0.3367\pm0.0049$} \\ 
 & $9$ & \multicolumn{4}{c}{$\mathbf{0.1641\pm0.0271}$} \\ 
\midrule
MPS & --- & $0.0487\pm0.0016$ & $0.0461\pm0.0028$ & $0.0469\pm0.0019$ & $\underline{\mathbf{0.0459\pm0.0008}}$ \\ 
\addlinespace[2pt]
XGBoost & --- & $\mathbf{0.1008\pm0.0130}$ & $0.1396\pm0.0036$ & $0.1996\pm0.0044$ & $0.2058\pm0.0035$ \\ 
\addlinespace[2pt]
CatBoost & --- & $0.1481\pm0.0093$ & $0.1024\pm0.0017$ & $\mathbf{0.1013\pm0.0023}$ & $0.1578\pm0.0205$ \\ 
\midrule
FC-VQC-PL & $3$ & $0.0737\pm0.0043$ & $0.0702\pm0.0038$ & $0.0699\pm0.0035$ & $0.0684\pm0.0037$ \\ 
 & $5$ & $0.0724\pm0.0036$ & $0.0696\pm0.0023$ & $0.0679\pm0.0038$ & $0.0669\pm0.0052$ \\ 
 & $7$ & $0.0703\pm0.0045$ & $0.0675\pm0.0041$ & $0.0631\pm0.0027$ & $\mathbf{0.0629\pm0.0056}$ \\ 
 & $9$ & $0.0665\pm0.0036$ & $0.0632\pm0.0030$ & $0.0644\pm0.0019$ & $0.0662\pm0.0026$ \\ 
\addlinespace[2pt]
FC-VQC-SW & $3$ & $0.0663\pm0.0031$ & $0.0633\pm0.0027$ & $0.0609\pm0.0026$ & $0.0612\pm0.0020$ \\ 
 & $5$ & $0.0621\pm0.0018$ & $0.0574\pm0.0023$ & $0.0591\pm0.0056$ & $0.0571\pm0.0032$ \\ 
 & $7$ & $0.0613\pm0.0031$ & $0.0567\pm0.0029$ & $0.0557\pm0.0035$ & $0.0558\pm0.0011$ \\ 
 & $9$ & $0.0596\pm0.0034$ & $\mathbf{0.0543\pm0.0022}$ & $0.0547\pm0.0021$ & $0.0560\pm0.0015$ \\ 
\addlinespace[2pt]
FC-VQC-FC & $3$ & $0.0575\pm0.0019$ & $0.0576\pm0.0024$ & $0.0565\pm0.0033$ & $0.0577\pm0.0023$ \\ 
 & $5$ & $0.0572\pm0.0032$ & $0.0574\pm0.0025$ & $\mathbf{0.0553\pm0.0022}$ & $0.0567\pm0.0013$ \\ 
 & $7$ & $0.0597\pm0.0033$ & $0.0596\pm0.0027$ & $0.0589\pm0.0019$ & $0.0616\pm0.0032$ \\ 
 & $9$ & $0.0637\pm0.0035$ & $0.0618\pm0.0023$ & $0.0568\pm0.0079$ & $0.0574\pm0.0047$ \\ 
\bottomrule
\end{tabular*}
\end{table}

\begin{table}[!htbp]
\centering
\caption{Black--Scholes PDE at $d=81$: trajectory RelMAE (mean $\pm$ standard deviation). Bold: best per model; underline: best overall. Lower is better.}
\label{tab:app_pde_d81_black_scholes}
\footnotesize
\setlength{\tabcolsep}{1.5pt}
\renewcommand{\arraystretch}{1.08}
\begin{tabular*}{\linewidth}{@{\extracolsep{\fill}}lccc@{}}
\toprule
Model & $L$ & $K=3$ & $K=5$ \\
\midrule
DNN & $3$ & \multicolumn{2}{c}{$\mathbf{0.0239\pm0.0003}$} \\ 
 & $5$ & \multicolumn{2}{c}{$0.0255\pm0.0008$} \\ 
\addlinespace[2pt]
Parameter-matched DNN & $3$ & $\mathbf{0.0237\pm0.0006}$ & $\mathbf{0.0237\pm0.0006}$ \\ 
 & $5$ & $0.0256\pm0.0013$ & $0.0256\pm0.0013$ \\ 
\addlinespace[2pt]
FC-DNN & $3$ & $\mathbf{0.0534\pm0.0021}$ & $0.0611\pm0.0041$ \\ 
 & $5$ & $0.0586\pm0.0018$ & $0.0690\pm0.0083$ \\ 
\addlinespace[2pt]
FC-Angle-Linear & $3$ & \multicolumn{2}{c}{$\mathbf{0.0252\pm0.0007}$} \\ 
 & $5$ & \multicolumn{2}{c}{$0.0258\pm0.0009$} \\ 
\midrule
MPS & --- & $\mathbf{0.0275\pm0.0012}$ & $0.0292\pm0.0021$ \\ 
\addlinespace[2pt]
XGBoost & --- & $\mathbf{0.0366\pm0.0026}$ & $0.0466\pm0.0011$ \\ 
\addlinespace[2pt]
CatBoost & --- & $0.0479\pm0.0019$ & $\mathbf{0.0377\pm0.0006}$ \\ 
\midrule
FC-VQC-PL & $3$ & $0.0210\pm0.0055$ & $\mathbf{0.0179\pm0.0003}$ \\ 
 & $5$ & $0.0187\pm0.0024$ & $0.0189\pm0.0017$ \\ 
\addlinespace[2pt]
FC-VQC-SW & $3$ & $0.0193\pm0.0019$ & $\mathbf{0.0178\pm0.0004}$ \\ 
 & $5$ & $0.0186\pm0.0020$ & $0.0183\pm0.0006$ \\ 
\addlinespace[2pt]
FC-VQC-FC & $3$ & $\underline{\mathbf{0.0165\pm0.0007}}$ & $0.0189\pm0.0022$ \\ 
 & $5$ & $0.0168\pm0.0010$ & $0.0171\pm0.0006$ \\ 
\bottomrule
\end{tabular*}
\end{table}

\begin{table}[!htbp]
\centering
\caption{Burgers PDE at $d=81$: trajectory RelMAE (mean $\pm$ standard deviation). Bold: best per model; underline: best overall. Lower is better.}
\label{tab:app_pde_d81_burgers}
\footnotesize
\setlength{\tabcolsep}{1.5pt}
\renewcommand{\arraystretch}{1.08}
\begin{tabular*}{\linewidth}{@{\extracolsep{\fill}}lccc@{}}
\toprule
Model & $L$ & $K=3$ & $K=5$ \\
\midrule
DNN & $3$ & \multicolumn{2}{c}{$0.4467\pm0.0180$} \\ 
 & $5$ & \multicolumn{2}{c}{$\mathbf{0.4410\pm0.0185}$} \\ 
\addlinespace[2pt]
Parameter-matched DNN & $3$ & $0.4475\pm0.0155$ & $0.4475\pm0.0155$ \\ 
 & $5$ & $\mathbf{0.4364\pm0.0144}$ & $\mathbf{0.4364\pm0.0144}$ \\ 
\addlinespace[2pt]
FC-DNN & $3$ & $\mathbf{0.5017\pm0.0155}$ & $0.6227\pm0.0181$ \\ 
 & $5$ & $0.6127\pm0.0103$ & $0.6749\pm0.0079$ \\ 
\addlinespace[2pt]
FC-Angle-Linear & $3$ & \multicolumn{2}{c}{$\mathbf{0.5733\pm0.0050}$} \\ 
 & $5$ & \multicolumn{2}{c}{$0.5741\pm0.0078$} \\ 
\midrule
MPS & --- & $\mathbf{0.4236\pm0.0159}$ & $0.4404\pm0.0431$ \\ 
\addlinespace[2pt]
XGBoost & --- & $\mathbf{0.4184\pm0.0159}$ & $0.5090\pm0.0150$ \\ 
\addlinespace[2pt]
CatBoost & --- & $0.5080\pm0.0225$ & $\mathbf{0.4348\pm0.0173}$ \\ 
\midrule
FC-VQC-PL & $3$ & $0.4296\pm0.0123$ & $\mathbf{0.4233\pm0.0076}$ \\ 
 & $5$ & $0.4361\pm0.0038$ & $0.4323\pm0.0082$ \\ 
\addlinespace[2pt]
FC-VQC-SW & $3$ & $\underline{\mathbf{0.4136\pm0.0090}}$ & $0.4193\pm0.0106$ \\ 
 & $5$ & $0.4243\pm0.0102$ & $0.4213\pm0.0095$ \\ 
\addlinespace[2pt]
FC-VQC-FC & $3$ & $0.4327\pm0.0122$ & $\mathbf{0.4275\pm0.0123}$ \\ 
 & $5$ & $0.4434\pm0.0141$ & $0.4462\pm0.0148$ \\ 
\bottomrule
\end{tabular*}
\end{table}

\begin{table}[!htbp]
\centering
\caption{Time-dependent oscillatory PDE at $d=81$: trajectory RelMAE (mean $\pm$ standard deviation). Bold: best per model; underline: best overall. Lower is better.}
\label{tab:app_pde_d81_oscillatory}
\footnotesize
\setlength{\tabcolsep}{1.5pt}
\renewcommand{\arraystretch}{1.08}
\begin{tabular*}{\linewidth}{@{\extracolsep{\fill}}lccc@{}}
\toprule
Model & $L$ & $K=3$ & $K=5$ \\
\midrule
DNN & $3$ & \multicolumn{2}{c}{$0.2079\pm0.0196$} \\ 
 & $5$ & \multicolumn{2}{c}{$\mathbf{0.1445\pm0.0227}$} \\ 
\addlinespace[2pt]
Parameter-matched DNN & $3$ & $0.2001\pm0.0169$ & $0.2001\pm0.0169$ \\ 
 & $5$ & $\mathbf{0.1339\pm0.0226}$ & $\mathbf{0.1339\pm0.0226}$ \\ 
\addlinespace[2pt]
FC-DNN & $3$ & $\mathbf{0.1809\pm0.0091}$ & $0.2461\pm0.0013$ \\ 
 & $5$ & $0.2451\pm0.0025$ & $0.2687\pm0.0070$ \\ 
\addlinespace[2pt]
FC-Angle-Linear & $3$ & \multicolumn{2}{c}{$\mathbf{0.3143\pm0.0155}$} \\ 
 & $5$ & \multicolumn{2}{c}{$0.3399\pm0.0217$} \\ 
\midrule
MPS & --- & $\mathbf{0.1239\pm0.0143}$ & $0.1351\pm0.0131$ \\ 
\addlinespace[2pt]
XGBoost & --- & $\mathbf{0.1141\pm0.0008}$ & $0.1490\pm0.0201$ \\ 
\addlinespace[2pt]
CatBoost & --- & $0.1629\pm0.0026$ & $\mathbf{0.1304\pm0.0132}$ \\ 
\midrule
FC-VQC-PL & $3$ & $\mathbf{0.0944\pm0.0055}$ & $0.0990\pm0.0055$ \\ 
 & $5$ & $0.0944\pm0.0047$ & $0.0983\pm0.0048$ \\ 
\addlinespace[2pt]
FC-VQC-SW & $3$ & $0.0938\pm0.0043$ & $0.1002\pm0.0038$ \\ 
 & $5$ & $\mathbf{0.0909\pm0.0043}$ & $0.0955\pm0.0038$ \\ 
\addlinespace[2pt]
FC-VQC-FC & $3$ & $\underline{\mathbf{0.0816\pm0.0066}}$ & $0.0818\pm0.0059$ \\ 
 & $5$ & $0.0830\pm0.0059$ & $0.0826\pm0.0063$ \\ 
\bottomrule
\end{tabular*}
\end{table}

\begin{table}[!htbp]
\centering
\caption{Black--Scholes PDE at $d=144$: trajectory RelMAE (mean $\pm$ standard deviation). Bold: best per model; underline: best overall. Lower is better.}
\label{tab:app_pde_d144_black_scholes}
\footnotesize
\setlength{\tabcolsep}{1.5pt}
\renewcommand{\arraystretch}{1.08}
\begin{tabular*}{\linewidth}{@{\extracolsep{\fill}}lccc@{}}
\toprule
Model & $L$ & $K=3$ & $K=5$ \\
\midrule
DNN & $3$ & \multicolumn{2}{c}{$\mathbf{0.0230\pm0.0004}$} \\ 
 & $5$ & \multicolumn{2}{c}{$0.0235\pm0.0009$} \\ 
\addlinespace[2pt]
Parameter-matched DNN & $3$ & $\mathbf{0.0229\pm0.0007}$ & $0.0230\pm0.0006$ \\ 
 & $5$ & $0.0229\pm0.0007$ & $0.0230\pm0.0007$ \\ 
\addlinespace[2pt]
FC-DNN & $3$ & $\mathbf{0.0457\pm0.0049}$ & $0.0492\pm0.0022$ \\ 
 & $5$ & $0.0486\pm0.0014$ & $0.0517\pm0.0011$ \\ 
\addlinespace[2pt]
FC-Angle-Linear & $3$ & \multicolumn{2}{c}{$\mathbf{0.0252\pm0.0021}$} \\ 
 & $5$ & \multicolumn{2}{c}{$0.0258\pm0.0009$} \\ 
\midrule
MPS & --- & $0.0235\pm0.0012$ & $\mathbf{0.0231\pm0.0008}$ \\ 
\addlinespace[2pt]
XGBoost & --- & $\mathbf{0.0293\pm0.0017}$ & $0.0367\pm0.0005$ \\ 
\addlinespace[2pt]
CatBoost & --- & $0.0373\pm0.0012$ & $\mathbf{0.0303\pm0.0005}$ \\ 
\midrule
FC-VQC-PL & $3$ & $\mathbf{0.0174\pm0.0005}$ & $0.0188\pm0.0011$ \\ 
 & $5$ & $0.0178\pm0.0006$ & $0.0203\pm0.0016$ \\ 
\addlinespace[2pt]
FC-VQC-SW & $3$ & $0.0182\pm0.0013$ & $\mathbf{0.0175\pm0.0003}$ \\ 
 & $5$ & $0.0188\pm0.0021$ & $0.0193\pm0.0009$ \\ 
\addlinespace[2pt]
FC-VQC-FC & $3$ & $\underline{\mathbf{0.0133\pm0.0003}}$ & $0.0148\pm0.0012$ \\ 
 & $5$ & $0.0136\pm0.0009$ & $0.0148\pm0.0015$ \\ 
\bottomrule
\end{tabular*}
\end{table}

\begin{table}[!htbp]
\centering
\caption{Burgers PDE at $d=144$: trajectory RelMAE (mean $\pm$ standard deviation). Bold: best per model; underline: best overall. Lower is better.}
\label{tab:app_pde_d144_burgers}
\footnotesize
\setlength{\tabcolsep}{1.5pt}
\renewcommand{\arraystretch}{1.08}
\begin{tabular*}{\linewidth}{@{\extracolsep{\fill}}lccc@{}}
\toprule
Model & $L$ & $K=3$ & $K=5$ \\
\midrule
DNN & $3$ & \multicolumn{2}{c}{$\mathbf{0.5752\pm0.0246}$} \\ 
 & $5$ & \multicolumn{2}{c}{$0.5780\pm0.0359$} \\ 
\addlinespace[2pt]
Parameter-matched DNN & $3$ & $\mathbf{0.5705\pm0.0311}$ & $\mathbf{0.5705\pm0.0311}$ \\ 
 & $5$ & $0.5897\pm0.0391$ & $0.5897\pm0.0391$ \\ 
\addlinespace[2pt]
FC-DNN & $3$ & $\mathbf{0.6500\pm0.0115}$ & $0.7064\pm0.0187$ \\ 
 & $5$ & $0.6828\pm0.0082$ & $0.7333\pm0.0083$ \\ 
\addlinespace[2pt]
FC-Angle-Linear & $3$ & \multicolumn{2}{c}{$0.6943\pm0.0154$} \\ 
 & $5$ & \multicolumn{2}{c}{$\mathbf{0.6913\pm0.0362}$} \\ 
\midrule
MPS & --- & $\mathbf{0.6347\pm0.0394}$ & $0.7090\pm0.1473$ \\ 
\addlinespace[2pt]
XGBoost & --- & $\mathbf{0.4926\pm0.0259}$ & $0.5940\pm0.0140$ \\ 
\addlinespace[2pt]
CatBoost & --- & $0.5663\pm0.0254$ & $\mathbf{0.5394\pm0.0166}$ \\ 
\midrule
FC-VQC-PL & $3$ & $0.5843\pm0.1222$ & $0.5502\pm0.0641$ \\ 
 & $5$ & $\mathbf{0.5074\pm0.0118}$ & $0.5535\pm0.0432$ \\ 
\addlinespace[2pt]
FC-VQC-SW & $3$ & $0.5352\pm0.0910$ & $0.5574\pm0.0460$ \\ 
 & $5$ & $\mathbf{0.4957\pm0.0076}$ & $0.5221\pm0.0165$ \\ 
\addlinespace[2pt]
FC-VQC-FC & $3$ & $0.4895\pm0.0141$ & $\underline{\mathbf{0.4850\pm0.0121}}$ \\ 
 & $5$ & $0.4881\pm0.0146$ & $0.4875\pm0.0107$ \\ 
\bottomrule
\end{tabular*}
\end{table}

\begin{table}[!htbp]
\centering
\caption{Time-dependent oscillatory PDE at $d=144$: trajectory RelMAE (mean $\pm$ standard deviation). Bold: best per model; underline: best overall. Lower is better.}
\label{tab:app_pde_d144_oscillatory}
\footnotesize
\setlength{\tabcolsep}{1.5pt}
\renewcommand{\arraystretch}{1.08}
\begin{tabular*}{\linewidth}{@{\extracolsep{\fill}}lccc@{}}
\toprule
Model & $L$ & $K=3$ & $K=5$ \\
\midrule
DNN & $3$ & \multicolumn{2}{c}{$0.2812\pm0.0315$} \\ 
 & $5$ & \multicolumn{2}{c}{$\mathbf{0.2487\pm0.0060}$} \\ 
\addlinespace[2pt]
Parameter-matched DNN & $3$ & $0.2784\pm0.0321$ & $0.2784\pm0.0321$ \\ 
 & $5$ & $\mathbf{0.2523\pm0.0140}$ & $0.2648\pm0.0218$ \\ 
\addlinespace[2pt]
FC-DNN & $3$ & $\mathbf{0.2506\pm0.0084}$ & $0.2720\pm0.0071$ \\ 
 & $5$ & $0.2662\pm0.0062$ & $0.2850\pm0.0064$ \\ 
\addlinespace[2pt]
FC-Angle-Linear & $3$ & \multicolumn{2}{c}{$\mathbf{0.3982\pm0.1492}$} \\ 
 & $5$ & \multicolumn{2}{c}{$0.4393\pm0.0138$} \\ 
\midrule
MPS & --- & $\mathbf{0.2505\pm0.0064}$ & $0.2598\pm0.0118$ \\ 
\addlinespace[2pt]
XGBoost & --- & $\mathbf{0.1303\pm0.0114}$ & $0.1680\pm0.0038$ \\ 
\addlinespace[2pt]
CatBoost & --- & $0.1774\pm0.0082$ & $\mathbf{0.1369\pm0.0027}$ \\ 
\midrule
FC-VQC-PL & $3$ & $0.1574\pm0.0062$ & $0.1565\pm0.0060$ \\ 
 & $5$ & $\mathbf{0.1461\pm0.0061}$ & $0.1561\pm0.0075$ \\ 
\addlinespace[2pt]
FC-VQC-SW & $3$ & $0.1938\pm0.0496$ & $0.1581\pm0.0071$ \\ 
 & $5$ & $\mathbf{0.1477\pm0.0035}$ & $0.1852\pm0.0597$ \\ 
\addlinespace[2pt]
FC-VQC-FC & $3$ & $\underline{\mathbf{0.1004\pm0.0049}}$ & $0.1148\pm0.0062$ \\ 
 & $5$ & $0.1059\pm0.0078$ & $0.1163\pm0.0053$ \\ 
\bottomrule
\end{tabular*}
\end{table}

\clearpage
\subsection{Trajectory-Level Error Plots}
\label{app:trajectory_plots}

\begin{figure}[!htbp]
    \centering
    \includegraphics[width=\linewidth]{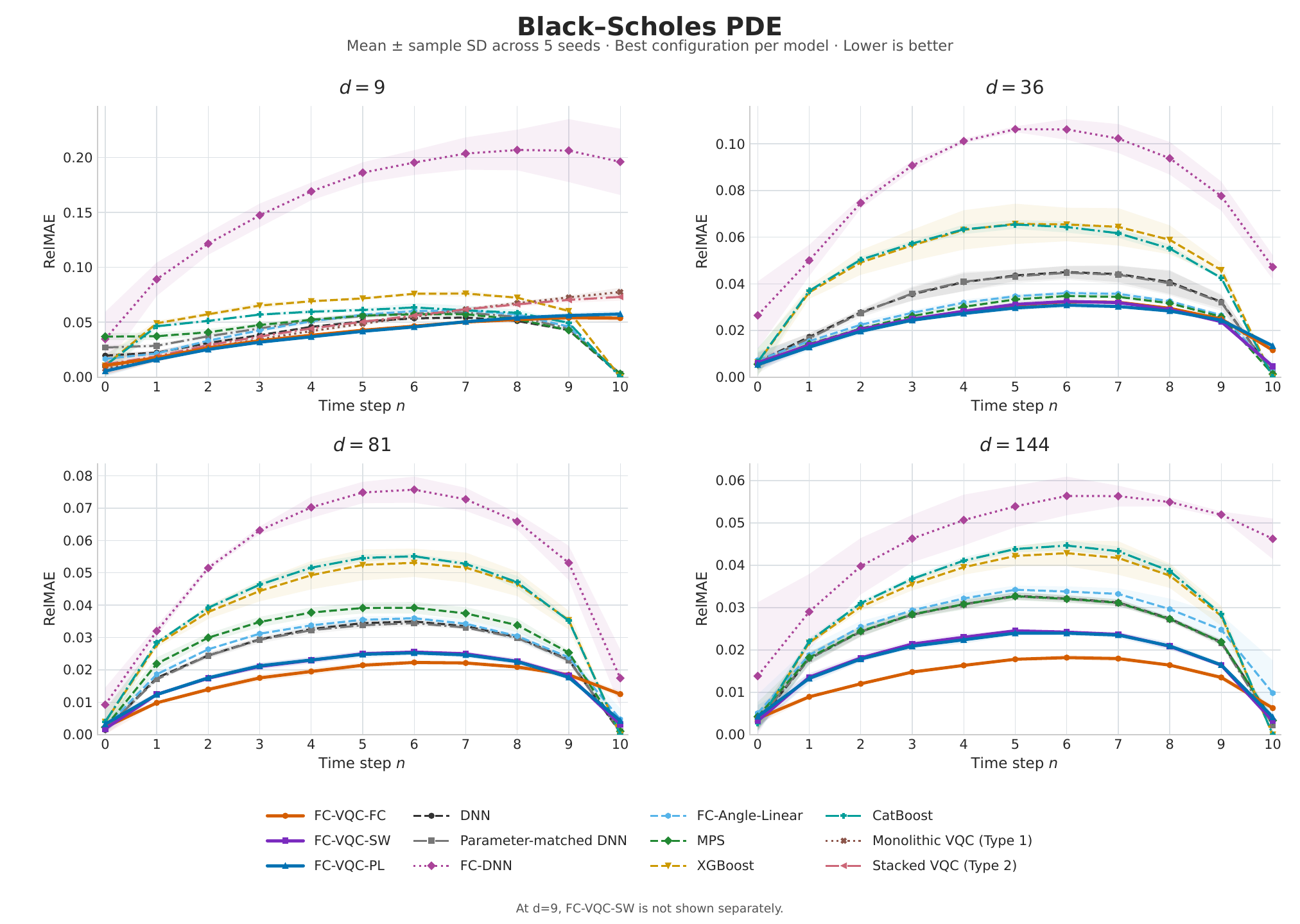}
    \caption{\textbf{Time-resolved RelMAE for the Black--Scholes PDE.}
    The panels show $d=9,36,81,144$ in reading order.
    Each model uses its selected configuration; lines and shading
    show the mean and $\pm$ one sample standard deviation across
    five seeds. Lower is better.}
    \label{fig:trajectory_black_scholes}
\end{figure}

\begin{figure}[!htbp]
    \centering
    \includegraphics[width=\linewidth]{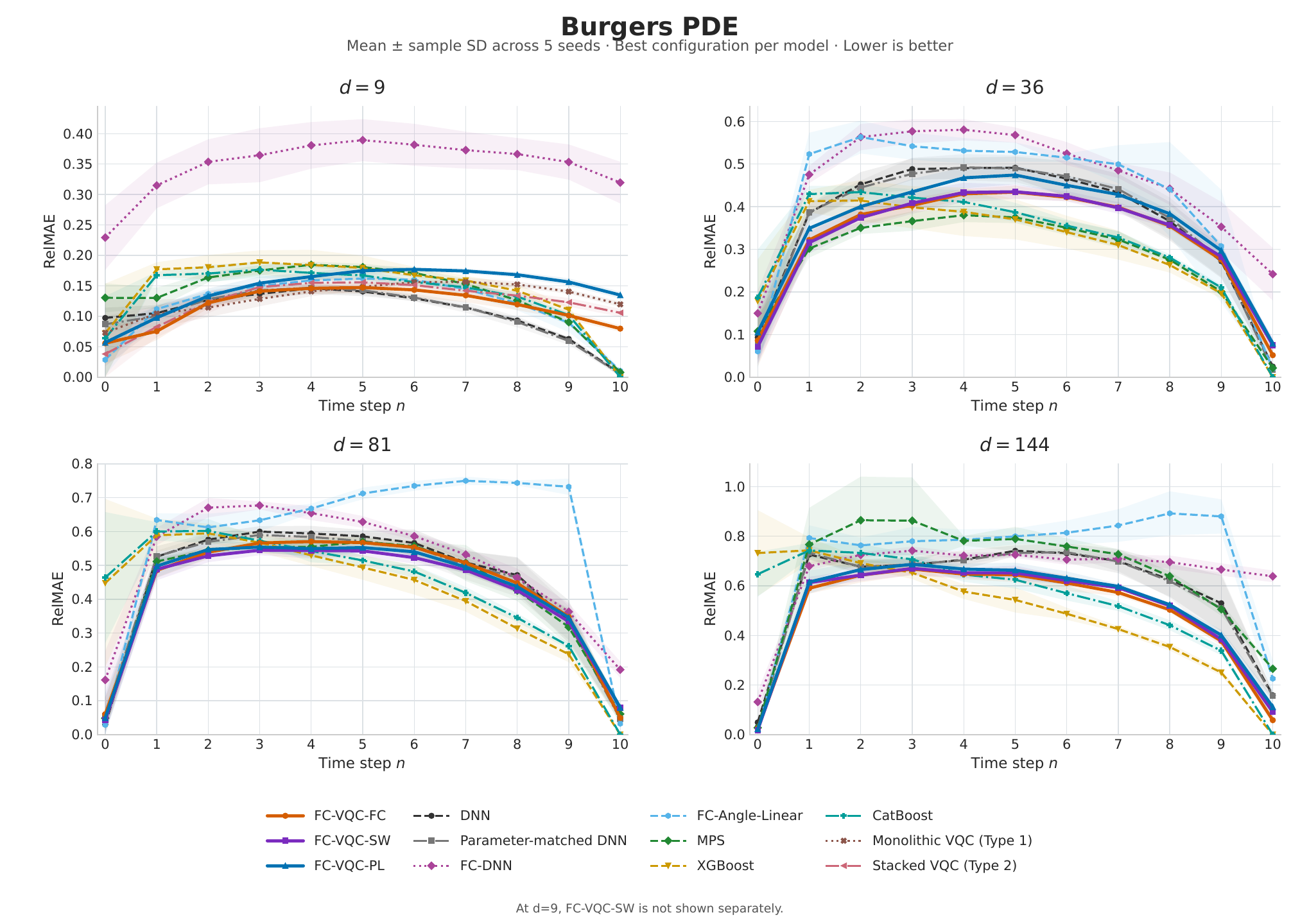}
    \caption{\textbf{Time-resolved RelMAE for the Burgers PDE.}
    The panels show $d=9,36,81,144$ in reading order.
    Each model uses its selected configuration; lines and shading
    show the mean and $\pm$ one sample standard deviation across
    five seeds. Lower is better.}
    \label{fig:trajectory_burgers}
\end{figure}

\begin{figure}[!htbp]
    \centering
    \includegraphics[width=\linewidth]{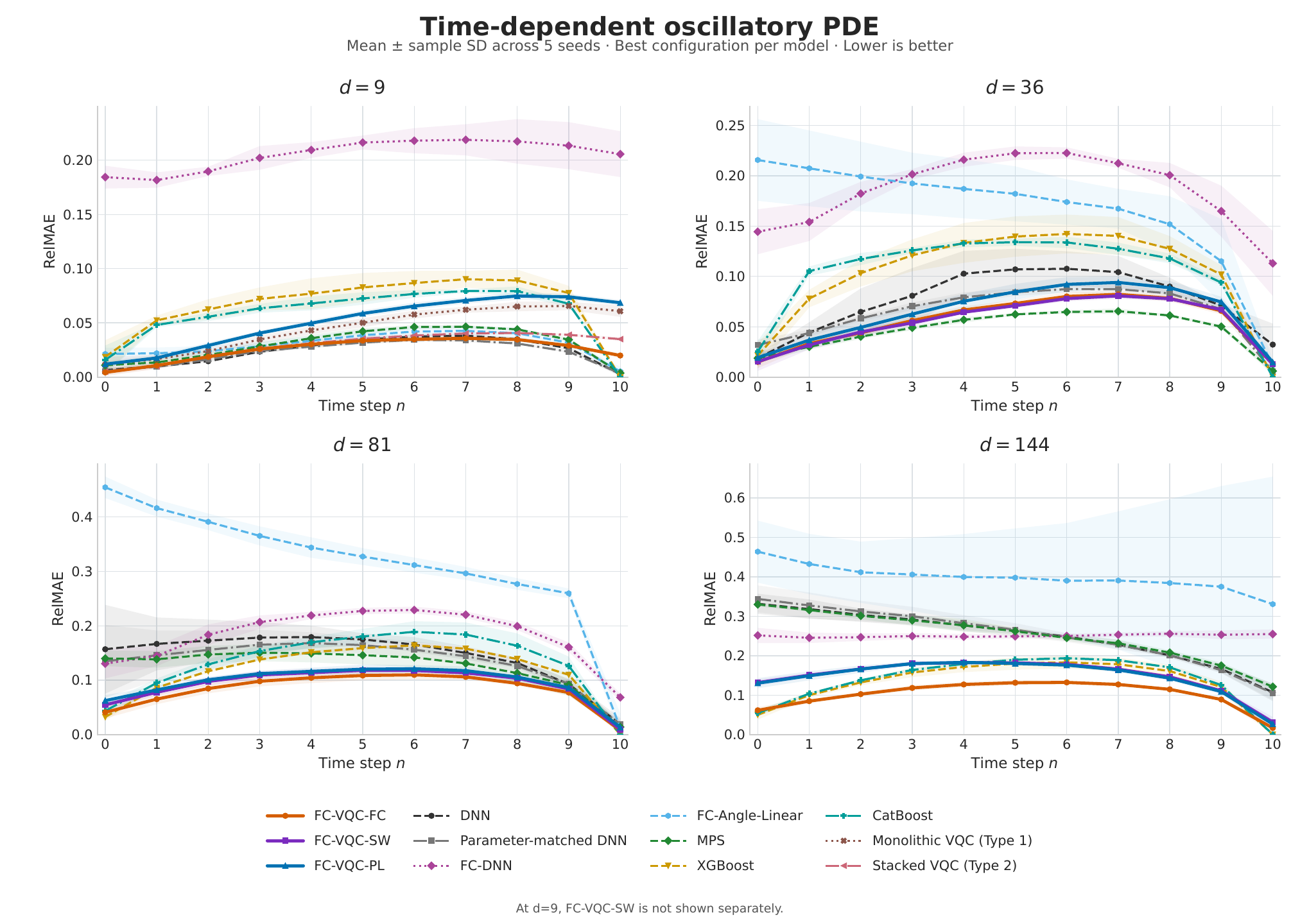}
    \caption{\textbf{Time-resolved RelMAE for the time-dependent oscillatory PDE.}
    The panels show $d=9,36,81,144$ in reading order.
    Each model uses its selected configuration; lines and shading
    show the mean and $\pm$ one sample standard deviation across
    five seeds. Lower is better.}
    \label{fig:trajectory_oscillatory}
\end{figure}

Figures~\ref{fig:trajectory_black_scholes}--\ref{fig:trajectory_oscillatory}
show per-time-step relative mean absolute error (RelMAE) for the
Black--Scholes, Burgers, and time-dependent oscillatory PDEs at
$d\in\{9,36,81,144\}$, over time steps $n=0,\ldots,10$.
For each PDE, dimension, and model, we select the configuration with
the lowest mean overall trajectory RelMAE across the five random seeds
$\{42,123,456,789,1024\}$. This configuration is held fixed across
all time steps and seeds.

Lines show the mean per-time-step RelMAE across seeds, and shaded
regions indicate $\pm$ one sample standard deviation. Lower values
indicate better approximation to the analytical solution.
At $d=9$, FC-VQC-SW is not plotted separately;
the FC-VQC-FC curve is shown for that dimension. Monolithic VQC (Type~1) and Stacked VQC (Type~2)
are included only at $d=9$, where their results are available.

\clearpage

\clearpage
\clearpage

\subsection{Representative Runtime and GPU Memory}
\label{app:runtime_memory}

We report representative single-run computational-resource
measurements from a separate benchmark on the oscillatory PDE
at $d=36$. The experiments use JAX/JIT on an AMD Ryzen 9 9950X
CPU and an NVIDIA RTX 5090 GPU, with $10{,}000$ training epochs,
batch size $256$, and $L=3$, $K=3$ where applicable.
Runtime includes the initial JIT compilation and the complete
training loop. GPU memory preallocation is disabled.

\begin{table}[htbp]
\centering
\caption{Representative single-run runtime and peak GPU memory
on the oscillatory PDE at $d=36$. These fixed-configuration
resource measurements are separate from the best-configuration
predictive comparisons.}
\label{tab:runtime_memory}
\small
\setlength{\tabcolsep}{6pt}
\renewcommand{\arraystretch}{1.10}
\begin{tabular*}{\linewidth}
{@{\extracolsep{\fill}}lrr@{}}
\toprule
Model & Runtime (s) & Peak GPU memory (MB) \\
\midrule
DNN             & $24.40$  & $1{,}114$ \\
FC-DNN          & $28.21$  & $860$     \\
FC-Angle-Linear & $25.72$  & $881$     \\
MPS             & $46.45$  & $858$     \\
\midrule
FC-VQC-PL       & $130.50$ & $2{,}652$ \\
FC-VQC-SW       & $168.79$ & $2{,}656$ \\
FC-VQC-FC       & $739.29$ & $2{,}658$ \\
\bottomrule
\end{tabular*}
\end{table}

Under this benchmark, the FC-VQC variants require more runtime
than the DNN, FC-DNN, FC-Angle-Linear, and MPS baselines. FC-VQC-FC uses six-qubit
local circuits, whereas FC-VQC-SW and FC-VQC-PL use three-qubit
local circuits. The comparison therefore includes differences
in local circuit width as well as routing.

These measurements characterize the execution costs of the
specified implementations and configurations. They are not
comparisons at matched predictive accuracy, and a single
dimension does not establish an empirical scaling law.
Peak GPU memory is an end-to-end implementation measurement,
rather than the size of an individual circuit state vector.

\clearpage
\section{Parameter Counting}
\label{sec:appendix_parameter_counting}

We first state the general counting rules, then apply them separately
to Concrete Strength, Wine Quality, and the PDE/BSDE models.
Tabular counts describe a complete predictor. PDE counts describe
one time-step module unless explicitly labelled as full-solver totals.

\subsection{General Counting Rules and Notation}
\label{app:parameter_counting_general}

\paragraph{Counting convention.}
We count scalar rotation parameters for VQCs and weights and biases
for DNNs. For modular tabular VQCs, the reported totals count all
rotation entries in blocks executed during the forward pass and
exclude unused tensors allocated by the implementation. These are
gate-parameter counts, not counts of functionally independent degrees
of freedom. Fixed encoding, padding, routing, readout selection,
and fixed output rescaling add no trainable parameters.
For PDE models, stored and loss-participating counts are distinguished
in Section~\ref{app:pde_solver_parameter_counts}.

\paragraph{VQC blocks.}
A block with $q$ qubits and circuit depth $K$ contains three trainable
rotation angles per qubit and circuit layer:
\begin{equation}
    P_{\mathrm{block}}(q,K)=3qK.
    \label{eq:app_local_vqc_parameter_count}
\end{equation}
Counts add across independently parameterized blocks. A full-width
$q$-qubit model with $L$ stages and no separate input or reduction
circuits therefore has
\begin{equation}
    P_{\mathrm{mono}}(q,L,K)=3qKL,
    \label{eq:app_monolithic_parameter_count}
\end{equation}
with $L=1$ for Type~1 and $L>1$ for Type~2.

\paragraph{Dense neural networks.}
For input dimension $d_{\mathrm{in}}$, output dimension
$d_{\mathrm{out}}$, hidden width $h$, and $H\ge1$ total hidden
layers, including all biases gives
\begin{equation}
    P_{\mathrm{DNN}}(d_{\mathrm{in}},d_{\mathrm{out}},h,H)
    =d_{\mathrm{in}}h+h+(H-1)(h^2+h)
      +hd_{\mathrm{out}}+d_{\mathrm{out}}.
    \label{eq:app_tabular_dnn_count}
\end{equation}
The tabular implementations use an initial input-to-hidden layer
followed by $\ell$ additional hidden-to-hidden layers, where
$\ell$ is the code argument \texttt{layers}; hence $H=\ell+1$.
The PDE DNN instead uses exactly $L=\texttt{num\_layers}$ hidden
layers, so $H=L$. The meanings of $L$ for the VQC models are
specified separately below.

\subsection{Concrete Strength: Complete-Predictor Counts}
\label{app:concrete_parameter_counts}

\paragraph{DNN.}
The Concrete DNN maps eight inputs to one output with hidden width
$64$. Its initial $8\to64$ layer, $\ell$ additional $64\to64$
layers, and final $64\to1$ layer give
\begin{equation}
    P_{\mathrm{Concrete,DNN}}(\ell)
    =(8\cdot64+64)+\ell(64^2+64)+(64+1)
    =641+4160\ell.
    \label{eq:app_concrete_dnn_count}
\end{equation}

\paragraph{VQCs.}
For the full-width $8t1$ model, $L$ counts all eight-qubit stages
and $P_{8t1}=24LK$. For modular Types~3--4, $L$ counts only the
width-preserving hidden stages; the separate input and reduction
stages must also be counted. Summing the parameters of the executed
blocks gives
\begin{align}
    P_{8t3t1}       &=9K(3L+7),\notag\\
    P_{16t4t1}      &=12K(4L+9),\notag\\
    P_{24t8t3t1}    &=9K(8L+20),\notag\\
    P_{32t11t4t1}   &=9K(11L+26)+12K,\notag\\
    P_{40t14t5t1}   &=9K(14L+33)+15K.
    \label{eq:app_concrete_vqc_counts}
\end{align}
For example, $8t3t1$ uses three input blocks, $3L$ hidden blocks,
and $3+1$ reduction blocks, all with three qubits. The $12K$ and
$15K$ terms in the last two formulas count the final four- and
five-qubit reduction circuits. Table~\ref{tab:app_param_concrete}
reports the complete-predictor counts.

\begin{table}[!htbp]
\centering
\caption{Parameter counts for Concrete Strength. VQC totals count rotation parameters in executed blocks; DNN totals include all weights and biases.}
\label{tab:app_param_concrete}
\scriptsize
\setlength{\tabcolsep}{3pt}
\renewcommand{\arraystretch}{1.06}
\begin{tabular*}{\textwidth}{@{\extracolsep{\fill}}llccccc@{}}
\toprule
\multirow{2}{*}{Type} & \multirow{2}{*}{Architecture}
& \multirow{2}{*}{$L$}
& \multicolumn{4}{c}{Depth: $K$ for VQCs; $\ell$ for DNN} \\
\cmidrule(lr){4-7}
& & & $3$ & $5$ & $7$ & $9$ \\
\midrule
DNN & DNN & -- 
& $13{,}121$ & $21{,}441$ & $29{,}761$ & $38{,}081$ \\
\midrule
Type~1 & $8t1$ & $1$
& $72$ & $120$ & $168$ & $216$ \\
\midrule
\multirow{4}{*}{Type~2}
& \multirow{4}{*}{8t1} & $3$ & $216$ & $360$ & $504$ & $648$ \\
&  & $5$ & $360$ & $600$ & $840$ & $1{,}080$ \\
&  & $7$ & $504$ & $840$ & $1{,}176$ & $1{,}512$ \\
&  & $9$ & $648$ & $1{,}080$ & $1{,}512$ & $1{,}944$ \\
\midrule
\multirow{4}{*}{Type~3}
& \multirow{4}{*}{8t3t1} & $3$ & $432$ & $720$ & $1{,}008$ & $1{,}296$ \\
&  & $5$ & $594$ & $990$ & $1{,}386$ & $1{,}782$ \\
&  & $7$ & $756$ & $1{,}260$ & $1{,}764$ & $2{,}268$ \\
&  & $9$ & $918$ & $1{,}530$ & $2{,}142$ & $2{,}754$ \\
\midrule
\multirow{16}{*}{Type~4}
& \multirow{4}{*}{16t4t1} & $3$ & $756$ & $1{,}260$ & $1{,}764$ & $2{,}268$ \\
&  & $5$ & $1{,}044$ & $1{,}740$ & $2{,}436$ & $3{,}132$ \\
&  & $7$ & $1{,}332$ & $2{,}220$ & $3{,}108$ & $3{,}996$ \\
&  & $9$ & $1{,}620$ & $2{,}700$ & $3{,}780$ & $4{,}860$ \\
\cmidrule(lr){2-7}
& \multirow{4}{*}{24t8t3t1} & $3$ & $1{,}188$ & $1{,}980$ & $2{,}772$ & $3{,}564$ \\
&  & $5$ & $1{,}620$ & $2{,}700$ & $3{,}780$ & $4{,}860$ \\
&  & $7$ & $2{,}052$ & $3{,}420$ & $4{,}788$ & $6{,}156$ \\
&  & $9$ & $2{,}484$ & $4{,}140$ & $5{,}796$ & $7{,}452$ \\
\cmidrule(lr){2-7}
& \multirow{4}{*}{32t11t4t1} & $3$ & $1{,}629$ & $2{,}715$ & $3{,}801$ & $4{,}887$ \\
& & $5$ & $2{,}223$ & $3{,}705$ & $5{,}187$ & $6{,}669$ \\
&  & $7$ & $2{,}817$ & $4{,}695$ & $6{,}573$ & $8{,}451$ \\
&  & $9$ & $3{,}411$ & $5{,}685$ & $7{,}959$ & $10{,}233$ \\
\cmidrule(lr){2-7}
& \multirow{4}{*}{40t14t5t1} & $3$ & $2{,}070$ & $3{,}450$ & $4{,}830$ & $6{,}210$ \\
& & $5$ & $2{,}826$ & $4{,}710$ & $6{,}594$ & $8{,}478$ \\
&  & $7$ & $3{,}582$ & $5{,}970$ & $8{,}358$ & $10{,}746$ \\
&  & $9$ & $4{,}338$ & $7{,}230$ & $10{,}122$ & $13{,}014$ \\
\bottomrule
\end{tabular*}
\vspace{1mm}

\begin{minipage}{0.96\textwidth}
\footnotesize
\emph{Note.} For VQCs, the four depth columns give circuit depth $K$.
For the DNN row, they give the code argument
$\ell\in\{3,5,7,9\}$, corresponding to
$H=\ell+1\in\{4,6,8,10\}$ total hidden layers.
For Types~1--2, $L$ counts all full-width VQC stages;
for Types~3--4, it counts width-preserving hidden stages only.
Separate input and reduction stages are included in the modular
VQC totals. Unused allocated tensors are excluded.
\end{minipage}
\vspace{-2mm}
\end{table}

\clearpage
\subsection{Wine Quality: Complete-Predictor Counts}
\label{app:wine_parameter_counts}

\paragraph{DNN.}
The Wine classifier maps eleven inputs to six class logits with
hidden width $64$. As in the Concrete implementation, the code
argument $\ell=\texttt{layers}$ counts the additional $64\to64$
layers after the initial hidden layer. Therefore,
\begin{equation}
    P_{\mathrm{Wine,DNN}}(\ell)
    =(11\cdot64+64)+\ell(64^2+64)+(64\cdot6+6)
    =1158+4160\ell.
    \label{eq:app_wine_dnn_count}
\end{equation}

\paragraph{VQCs.}
The full-width model is denoted $11t6$: it processes eleven
features with an eleven-qubit circuit and selects six final
readouts. Its parameter count is $P_{11t6}=33LK$, where $L$
counts the full-width stages. Selecting six readouts and multiplying
them by a fixed output scale introduce no additional parameters.

For modular Types~3--4, $L$ again counts only the width-preserving
hidden stages. Each topology below ends in an eight-qubit circuit
that supplies six class logits and contributes $24K$ parameters:
\begin{align}
    P_{12t8t6}
    &=9K(4L+8)+24K=K(36L+96),\notag\\
    P_{22t8t6}
    &=9K(8L+16)+24K=K(72L+168),\notag\\
    P_{33t12t8t6}
    &=9K(11L+26)+24K=K(99L+258),\notag\\
    P_{44t15t10t8t6}
    &=9K(15L+39)+24K=K(135L+375).
    \label{eq:app_wine_vqc_counts}
\end{align}
For example, $33t12t8t6$ uses eleven input blocks, $11L$ hidden
blocks, and $11+4$ reduction blocks with three qubits, followed by
the eight-qubit output circuit. As for Concrete, unused allocated
block tensors are excluded. Table~\ref{tab:app_param_wine} gives
the resulting complete-predictor counts.

\begin{table}[!htbp]
\centering
\caption{Parameter counts for Wine Quality. All models have six outputs. VQC totals count rotation parameters in executed blocks; DNN totals include all weights and biases.}
\label{tab:app_param_wine}
\scriptsize
\setlength{\tabcolsep}{3pt}
\renewcommand{\arraystretch}{1.06}
\begin{tabular*}{\textwidth}{@{\extracolsep{\fill}}llccccc@{}}
\toprule
\multirow{2}{*}{Type} & \multirow{2}{*}{Architecture}
& \multirow{2}{*}{$L$}
& \multicolumn{4}{c}{Depth: $K$ for VQCs; $\ell$ for DNN} \\
\cmidrule(lr){4-7}
& & & $3$ & $5$ & $7$ & $9$ \\
\midrule
DNN & DNN & -- 
& $13{,}638$ & $21{,}958$ & $30{,}278$ & $38{,}598$ \\
\midrule
Type~1 & \texttt{11t6} & $1$
& $99$ & $165$ & $231$ & $297$ \\
\midrule
\multirow{4}{*}{Type~2}
& \multirow{4}{*}{\texttt{11t6}} & $3$ & $297$ & $495$ & $693$ & $891$ \\
&  & $5$ & $495$ & $825$ & $1{,}155$ & $1{,}485$ \\
& & $7$ & $693$ & $1{,}155$ & $1{,}617$ & $2{,}079$ \\
& & $9$ & $891$ & $1{,}485$ & $2{,}079$ & $2{,}673$ \\
\midrule
\multirow{4}{*}{Type~3}
& \multirow{4}{*}{12t8t6}& $3$ & $612$ & $1{,}020$ & $1{,}428$ & $1{,}836$ \\
&  & $5$ & $828$ & $1{,}380$ & $1{,}932$ & $2{,}484$ \\
&  & $7$ & $1{,}044$ & $1{,}740$ & $2{,}436$ & $3{,}132$ \\
& & $9$ & $1{,}260$ & $2{,}100$ & $2{,}940$ & $3{,}780$ \\
\midrule
\multirow{12}{*}{Type~4}
& \multirow{4}{*}{22t8t6}& $3$ & $1{,}152$ & $1{,}920$ & $2{,}688$ & $3{,}456$ \\
&  & $5$ & $1{,}584$ & $2{,}640$ & $3{,}696$ & $4{,}752$ \\
& & $7$ & $2{,}016$ & $3{,}360$ & $4{,}704$ & $6{,}048$ \\
&  & $9$ & $2{,}448$ & $4{,}080$ & $5{,}712$ & $7{,}344$ \\
\cmidrule(lr){2-7}
& \multirow{4}{*}{33t12t8t6} & $3$ & $1{,}665$ & $2{,}775$ & $3{,}885$ & $4{,}995$ \\
&  & $5$ & $2{,}259$ & $3{,}765$ & $5{,}271$ & $6{,}777$ \\
&  & $7$ & $2{,}853$ & $4{,}755$ & $6{,}657$ & $8{,}559$ \\
&  & $9$ & $3{,}447$ & $5{,}745$ & $8{,}043$ & $10{,}341$ \\
\cmidrule(lr){2-7}
& \multirow{4}{*}{44t15t10t8t6}& $3$ & $2{,}340$ & $3{,}900$ & $5{,}460$ & $7{,}020$ \\
&  & $5$ & $3{,}150$ & $5{,}250$ & $7{,}350$ & $9{,}450$ \\
& & $7$ & $3{,}960$ & $6{,}600$ & $9{,}240$ & $11{,}880$ \\
& & $9$ & $4{,}770$ & $7{,}950$ & $11{,}130$ & $14{,}310$ \\
\bottomrule
\end{tabular*}
\vspace{1mm}

\begin{minipage}{0.96\textwidth}
\footnotesize
\emph{Note.} For VQCs, the four depth columns give circuit depth $K$.
For the DNN row, they give the code argument
$\ell\in\{3,5,7,9\}$, corresponding to
$H=\ell+1\in\{4,6,8,10\}$ total hidden layers.
For Types~1--2, $L$ counts all full-width VQC stages;
for Types~3--4, it counts width-preserving hidden stages only.
Separate input and reduction stages are included in the modular
VQC totals. Unused allocated tensors are excluded.
\end{minipage}
\vspace{-2mm}
\end{table}

\clearpage
\subsection{PDE/BSDE Models: Per-Time-Step and Full-Solver Counts}
\label{app:pde_parameter_counts}

The Black--Scholes, Burgers, and oscillatory tasks use the same
parameter-count formulas when the architecture settings are fixed.
In this subsection, $P$ denotes the count for one state-dependent
$d\mapsto d$ time-step module, excluding the initial-value model
and duplication across time. Full-solver totals are given in
Section~\ref{app:pde_solver_parameter_counts}.
Unlike the modular tabular architectures, the dimension-preserving
PDE VQCs execute exactly $L$ VQC stages; there are no extra input
or dimensionality-reduction stages to add to $L$.

\subsubsection{Dense DNNs}
\label{app:pde_dnn_parameter_counts}

Here $L$ is the total number of hidden layers. For hidden width $h$,
the general DNN rule gives
\begin{align}
    P_{\mathrm{DNN}}(d,L,h)
    &=2dh+(L-1)h^2+Lh+d,\notag\\
    P_{\mathrm{DNN}}(d,L,64)
    &=129d+4160L-4096.
    \label{eq:app_dnn_parameter_count}
\end{align}
Table~\ref{tab:app_dnn_parameter_grid} uses $h=64$.
The same general expression applies to the parameter-matched DNN
with its run-specific hidden width substituted for $h$; the table
does not give counts for those reduced-width models.

\begin{table}[htbp]
\centering
\caption{Parameters in one PDE time-step DNN with hidden width $64$, including biases. Here $L$ is the total number of hidden layers. Counts exclude time-step duplication and the initial-value model. Dashes indicate settings outside the reported grid.}
\label{tab:app_dnn_parameter_grid}
\small
\setlength{\tabcolsep}{5pt}
\begin{tabular*}{\linewidth}{@{\extracolsep{\fill}}lrrrr@{}}
\toprule
Hidden layers $L$ & $d=9$ & $d=36$ & $d=81$ & $d=144$ \\
\midrule
$3$ & $9{,}545$ & $13{,}028$ & $18{,}833$ & $26{,}960$ \\
$5$ & $17{,}865$ & $21{,}348$ & $27{,}153$ & $35{,}280$ \\
$7$ & $26{,}185$ & $29{,}668$ & --- & --- \\
$9$ & $34{,}505$ & $37{,}988$ & --- & --- \\
\bottomrule
\end{tabular*}
\end{table}

\subsubsection{Modular FC-VQCs}
\label{app:pde_vqc_parameter_counts}

Each of the $L$ stages contains $B$ independent $q$-qubit blocks
with circuit depth $K$. In the evaluated configurations, $Bq=d$:
SW and PL use $q=3$ and $B=d/3$, while FC uses
$q=B=\sqrt d$ for the perfect-square dimensions
$d\in\{9,36,81,144\}$. Thus,
\begin{equation}
    P_{\mathrm{FC}}=P_{\mathrm{SW}}=P_{\mathrm{PL}}
    =3BLKq=3dLK.
    \label{eq:app_fcvqc_parameter_count}
\end{equation}
The three variants have the same gate-parameter count at fixed
$(d,L,K)$ despite their different circuit widths and routing.
Table~\ref{tab:app_fcvqc_parameter_grid} reports these per-time-step
counts, not full BSDE solver totals.

\begin{table}[htbp]
\centering
\caption{Parameters in one PDE time-step FC-VQC model. Here $L$ counts all dimension-preserving VQC stages. Each entry applies equally to FC-VQC-FC, FC-VQC-SW, and FC-VQC-PL and excludes time-step duplication and the initial-value model. Dashes indicate settings outside the reported grid.}
\label{tab:app_fcvqc_parameter_grid}
\small
\setlength{\tabcolsep}{5pt}
\begin{tabular*}{\linewidth}{@{\extracolsep{\fill}}lrrrrr@{}}
\toprule
$d$ & $L$ & $K=3$ & $K=5$ & $K=7$ & $K=9$ \\
\midrule
$9$ & $3$ & $243$ & $405$ & $567$ & $729$ \\
 & $5$ & $405$ & $675$ & $945$ & $1{,}215$ \\
 & $7$ & $567$ & $945$ & $1{,}323$ & $1{,}701$ \\
 & $9$ & $729$ & $1{,}215$ & $1{,}701$ & $2{,}187$ \\
\midrule
$36$ & $3$ & $972$ & $1{,}620$ & $2{,}268$ & $2{,}916$ \\
 & $5$ & $1{,}620$ & $2{,}700$ & $3{,}780$ & $4{,}860$ \\
 & $7$ & $2{,}268$ & $3{,}780$ & $5{,}292$ & $6{,}804$ \\
 & $9$ & $2{,}916$ & $4{,}860$ & $6{,}804$ & $8{,}748$ \\
\midrule
$81$ & $3$ & $2{,}187$ & $3{,}645$ & --- & --- \\
 & $5$ & $3{,}645$ & $6{,}075$ & --- & --- \\
\midrule
$144$ & $3$ & $3{,}888$ & $6{,}480$ & --- & --- \\
 & $5$ & $6{,}480$ & $10{,}800$ & --- & --- \\
\bottomrule
\end{tabular*}
\end{table}

\subsubsection{Other Differentiable Baselines}
\label{app:pde_other_parameter_counts}

\paragraph{FC-DNN.}
For this model, $L$ counts outer block stages, $K$ counts hidden
layers inside each local DNN, and $h$ is the local hidden width.
A three-input, three-output block with biases in every affine
layer has
\begin{equation}
    P_{\mathrm{FC\text{-}DNN,block}}
    =(3h+h)+(K-1)(h^2+h)+(3h+3)
    =(K-1)h^2+(K+6)h+3.
    \label{eq:app_fc_dnn_block_parameter_count}
\end{equation}
For $B=d/3$ blocks per stage,
\begin{equation}
    P_{\mathrm{FC\text{-}DNN}}
    =\frac{dL}{3}\left[(K-1)h^2+(K+6)h+3\right].
    \label{eq:app_fc_dnn_parameter_count}
\end{equation}
At $h=3$, this becomes $4dL(K+1)$. This architecture-matched
baseline is not parameter matched to a three-qubit VQC of depth
$K$, whose local count is $9K$.

\paragraph{FC-Angle-Linear.}
In the explicit-feature definition, each local map has one
$27\times3$ trainable coefficient matrix and no separate bias
vector. With $B=d/3$ blocks and $L$ stages, its coefficient count is
\begin{equation}
    P_{\mathrm{Angle}}=81BL=27dL.
    \label{eq:app_angle_linear_parameter_count}
\end{equation}
There is no independent $K$ or hidden-width parameter. This
per-time-step coefficient count equals the FC-VQC-SW count at
$K=9$ for the same $d$ and $L$; it does not imply an equal
full-solver count without also accounting for the initial-value
parameterization and time-step allocation.

\paragraph{MPS.}
For bond dimension $\chi$, the implemented time-step model has
one $(2,\chi)$ input core, $d-2$ cores of shape $(\chi,2,\chi)$,
one $(\chi,2,d)$ output core, and a $d$-component output bias.
Hence, for $d\ge2$,
\begin{equation}
    P_{\mathrm{MPS}}(d,\chi)
    =2\chi+2(d-2)\chi^2+2d\chi+d.
    \label{eq:app_mps_parameter_count}
\end{equation}
MPS has no outer-stage parameter $L$; its capacity is controlled
by $\chi$, which is indexed by $K$ in the experimental result tables.

\subsubsection{Initial Value and Full BSDE Solver}
\label{app:pde_solver_parameter_counts}

\paragraph{Stored versus loss-participating parameters.}
For the DNN, FC-DNN, MPS, and three FC-VQC implementations, the
state-dependent model allocates $N+1$ independent parameter slices.
The BSDE recursion uses only the first $N$ slices, at
$t_0,\ldots,t_{N-1}$; the terminal-time slice is stored but does not
contribute to the training loss. If $P$ is a per-time-step count
and $P_{Y_0}$ is the initial-value count, then
\begin{align}
    P^{\mathrm{stored}}&=(N+1)P+P_{Y_0},\notag\\
    P^{\mathrm{loss}}&=NP+P_{Y_0}.
    \label{eq:app_pde_stored_vs_loss_parameters}
\end{align}
Here $P^{\mathrm{loss}}$ counts the scalar entries in parameter
slices used by the loss, not functionally independent degrees of
freedom. The implementation's array-based parameter counter
returns $P^{\mathrm{stored}}$.

\paragraph{Initial-value models.}
The DNN, FC-DNN, and MPS use one trainable scalar for $Y_0$,
so $P_{Y_0}=1$. Each FC-VQC variant uses a one-qubit circuit
with circuit depth $K\ge1$, giving $P_{Y_0}=3K$.
Fixed initial-value multipliers add no trainable parameters.
For the evaluated $N=10$ time steps,
\begin{align}
    P^{\mathrm{stored}}_{\mathrm{DNN}}
    &=11P_{\mathrm{DNN}}+1,\notag\\
    P^{\mathrm{stored}}_{\mathrm{FC\text{-}DNN}}
    &=11P_{\mathrm{FC\text{-}DNN}}+1,\notag\\
    P^{\mathrm{stored}}_{\mathrm{MPS}}
    &=11P_{\mathrm{MPS}}+1,\notag\\
    P^{\mathrm{stored}}_{\mathrm{FC\text{-}VQC}}
    &=11P_{\mathrm{FC\text{-}VQC}}+3K.
    \label{eq:app_bsde_stored_parameter_count}
\end{align}
Replacing $11$ by $10$ gives the counts excluding the unused
terminal-time slice. These solver totals must not be compared
directly with per-time-step counts or with the complete-predictor
counts for the tabular tasks.

\clearpage
\section{Theoretical Details}
\label{sec:appendix_theory}

\subsection{Layerwise Error Propagation: Assumptions and Proof}
\label{app:layerwise_error_propagation}

This section proves Theorem~\ref{thm:layerwise_error_propagation}.
We consider forward evaluation with fixed circuit parameters and a
fixed input. Expectations are over measurement and implementation
randomness. The analysis applies to stacked full-width VQCs and to
modular FC-VQC, with each layer's readout collecting the outputs of
its constituent circuits. We count $L$ VQC evaluations, with fixed
routing $g^{(0)},\ldots,g^{(L-1)}$ before those evaluations.
The map $g^{(0)}$ may be the identity. For a direct final readout we set
$g^{(L)}=I$; any fixed linear output scaling can instead be included in
$W^{(L)}$. This adds no VQC evaluation or nontrivial routing step.

\subsubsection{Ideal and noisy layer maps}

Let $z^{(l)}:\mathcal D_{l-1}\to\mathbb R^{m_l}$ denote the ideal
quantum readout and let
$W^{(l)}\in\mathbb R^{d_l\times m_l}$ be the deterministic routing
matrix. The domain $\mathcal D_{l-1}\subseteq\mathbb R^{d_{l-1}}$
contains all ideal and noisy inputs reachable at layer $l$.
Each layer combines quantum readout and linear routing:
\begin{equation}
f_{\Theta^{(l)}}(h)
=g^{(l)}\bigl(z^{(l)}(h)\bigr),
\qquad
g^{(l)}(u)=W^{(l)}u.
\label{eq:app_layer_map}
\end{equation}
The ideal recursion is
\begin{equation}
H^{(l)}=f_{\Theta^{(l)}}(H^{(l-1)}),
\qquad H^{(0)}=g^{(0)}(x),\qquad l=1,\ldots,L.
\label{eq:app_layerwise_ideal}
\end{equation}
Thus, each stage groups a quantum evaluation with the routing that
prepares the next input. For $l<L$, $W^{(l)}$ is the matrix
of the outgoing linear map $g^{(l)}$; the initial map $g^{(0)}$
has already been applied to the common input.
For a direct final readout $W^{(L)}=I$; the same proof permits any
fixed linear final map $W^{(L)}$. Dimension-preserving readout and
routing correspond to $d_l=m_l=d$.

Let $\widetilde z^{(l)}$ be the implemented quantum readout,
including physical noise and finite-shot estimation. Define
\begin{equation}
\widetilde f_{\Theta^{(l)}}(h)=W^{(l)}\widetilde z^{(l)}(h),
\qquad
\widetilde H^{(l)}
=\widetilde f_{\Theta^{(l)}}(\widetilde H^{(l-1)}),
\qquad \widetilde H^{(0)}=g^{(0)}(x).
\label{eq:app_layerwise_noisy}
\end{equation}
Define the local implementation error
\begin{equation}
\varepsilon^{(l)}(h)
:=\widetilde f_{\Theta^{(l)}}(h)-f_{\Theta^{(l)}}(h).
\label{eq:app_implementation_error}
\end{equation}

Let $\mathcal F_{l-1}$ contain all randomness observed before layer
$l$. The incoming noisy state $\widetilde H^{(l-1)}$ is then
$\mathcal F_{l-1}$-measurable. For a fixed or
$\mathcal F_{l-1}$-measurable input $h$, define
\begin{align}
b^{(l)}(h)
&:=\mathbb E\!\left[\widetilde f_{\Theta^{(l)}}(h)
\mid\mathcal F_{l-1}\right]-f_{\Theta^{(l)}}(h),
\label{eq:app_layer_bias}\\
\xi^{(l)}(h)
&:=\widetilde f_{\Theta^{(l)}}(h)
-\mathbb E\!\left[\widetilde f_{\Theta^{(l)}}(h)
\mid\mathcal F_{l-1}\right].
\label{eq:app_layer_shot_noise}
\end{align}
Consequently,
\begin{equation}
\varepsilon^{(l)}(h)=b^{(l)}(h)+\xi^{(l)}(h),
\qquad
\mathbb E[\xi^{(l)}(h)\mid\mathcal F_{l-1}]=0.
\label{eq:app_readout_decomposition}
\end{equation}
Both $b^{(l)}$ and $\xi^{(l)}$ are errors \emph{after routing},
consistent with their definition through the complete layer map.
Their dependence on the preceding history is implicit.

\subsubsection{Assumptions}

\paragraph{A1 (Input sensitivity and routing).}
For every $h,h'\in\mathcal D_{l-1}$,
\begin{equation}
\|z^{(l)}(h)-z^{(l)}(h')\|_2
\le L_{z,l}\|h-h'\|_2.
\label{eq:app_readout_lipschitz}
\end{equation}
The routing map has operator norm $\|W^{(l)}\|_2$.
Hence the complete layer map has the valid Lipschitz bound
\begin{equation}
\rho_l=\|W^{(l)}\|_2L_{z,l}.
\label{eq:app_complete_sensitivity}
\end{equation}
All Lipschitz constants and routing matrices are deterministic for
the fixed model. In particular, $\|W^{(l)}\|_2$ alone does not control the sensitivity
of the complete quantum-and-routing layer. We do not assume
$\rho_l\le1$.

\paragraph{A2 (Bounded conditional layer bias).}
There is a deterministic constant $\beta_l\ge0$ such that
\begin{equation}
\|b^{(l)}(h)\|_2\le \beta_l
\quad\text{almost surely},
\label{eq:app_readout_bias}
\end{equation}
uniformly over reachable inputs and preceding histories. The bound
must be justified for the implementation and noise model considered.
Because $b^{(l)}$ is defined after routing, $\beta_l$ already includes
the effect of $W^{(l)}$ on the bias.

\paragraph{A3 (Conditional finite-shot variance).}
For each readout coordinate $i=1,\ldots,m_l$,
\begin{equation}
\operatorname{Var}\!\left(
\widetilde z^{(l)}_i(h)\mid\mathcal F_{l-1}
\right)\le\frac{1}{S_l}
\quad\text{almost surely},
\label{eq:app_readout_variance}
\end{equation}
uniformly over reachable inputs and histories. This holds when each
coordinate averages $S_l$ fresh, conditionally independent
measurement outcomes in $[-1,1]$.
Different coordinates need not be independent: commuting Pauli-$Z$
observables can share measurement shots. Correlations between
successive shots require a separate variance bound.

\subsubsection{Local implementation error}

Since $W^{(l)}$ is deterministic,
\begin{equation}
\xi^{(l)}(h)
=W^{(l)}\!\left(
\widetilde z^{(l)}(h)
-\mathbb E[\widetilde z^{(l)}(h)\mid\mathcal F_{l-1}]
\right).
\end{equation}
The operator-norm bound and conditional Jensen's inequality give
\begin{align}
\mathbb E[\|\xi^{(l)}(h)\|_2\mid\mathcal F_{l-1}]
&\le
\|W^{(l)}\|_2
\left(
\sum_{i=1}^{m_l}
\operatorname{Var}(\widetilde z^{(l)}_i(h)\mid\mathcal F_{l-1})
\right)^{1/2}
\nonumber\\
&\le \|W^{(l)}\|_2\sqrt{\frac{m_l}{S_l}}.
\label{eq:app_centered_readout_bound}
\end{align}
No independence between coordinates is used. Combining this with
A2 yields
\begin{equation}
\mathbb E[\|\varepsilon^{(l)}(h)\|_2\mid\mathcal F_{l-1}]
\le
\beta_l+\|W^{(l)}\|_2\sqrt{\frac{m_l}{S_l}}
=:\eta_l.
\label{eq:app_local_implementation_error}
\end{equation}
The routing norm multiplies the sampling term because A3 concerns
readouts before routing. It does not multiply $\beta_l$ again, since
$\beta_l$ already bounds the bias of the routed layer output.

\subsubsection{Proof of Theorem~\ref{thm:layerwise_error_propagation}}

For reference, under A1--A3 the propagated bound is
\begin{equation}
\mathbb E\|\widetilde H^{(L)}-H^{(L)}\|_2
\le
\sum_{l=1}^{L}
\left(\prod_{j=l+1}^{L}\rho_j\right)
\left(\beta_l+\|W^{(l)}\|_2\sqrt{\frac{m_l}{S_l}}\right),
\label{eq:app_layerwise_error_bound}
\end{equation}
where $\rho_j=\|W^{(j)}\|_2L_{z,j}$ and an empty product equals one.

\begin{proof}
Define the accumulated error
$\Delta^{(l)}:=\widetilde H^{(l)}-H^{(l)}$.
Adding and subtracting
$f_{\Theta^{(l)}}(\widetilde H^{(l-1)})$ gives
\begin{equation}
\Delta^{(l)}
=f_{\Theta^{(l)}}(\widetilde H^{(l-1)})
-f_{\Theta^{(l)}}(H^{(l-1)})
+\varepsilon^{(l)}(\widetilde H^{(l-1)}).
\label{eq:app_delta_decomposition}
\end{equation}
By A1,
\begin{equation}
\|\Delta^{(l)}\|_2
\le \rho_l\|\Delta^{(l-1)}\|_2
+\|\varepsilon^{(l)}(\widetilde H^{(l-1)})\|_2.
\end{equation}
The incoming error $\Delta^{(l-1)}$ is
$\mathcal F_{l-1}$-measurable. Taking conditional expectations and
using \eqref{eq:app_local_implementation_error} therefore yields
\begin{equation}
\mathbb E[\|\Delta^{(l)}\|_2\mid\mathcal F_{l-1}]
\le \rho_l\|\Delta^{(l-1)}\|_2+\eta_l.
\end{equation}
Taking expectations again gives
\begin{equation}
\mathbb E\|\Delta^{(l)}\|_2
\le
\rho_l\mathbb E\|\Delta^{(l-1)}\|_2+\eta_l.
\label{eq:app_error_recursion}
\end{equation}
The initial routing is deterministic and shared by both recursions,
so $\Delta^{(0)}=0$. Repeated substitution gives
\begin{equation}
\mathbb E\|\Delta^{(L)}\|_2
\le
\sum_{l=1}^{L}
\left(\prod_{j=l+1}^{L}\rho_j\right)\eta_l.
\end{equation}
Substituting the definition of $\eta_l$ proves
\eqref{eq:app_layerwise_error_bound}.
\end{proof}

\subsubsection{Input sensitivity of local VQC blocks}

For a single product angle-encoding layer, an explicit sufficient
Lipschitz bound can be derived. Consider a block with $q$ inputs and
$m$ readouts,
\begin{equation}
|\psi(h)\rangle
=\left(\bigotimes_{j=1}^{q}e^{-ih_jP_j/2}\right)|0\rangle^{\otimes q},
\qquad
z_a(h)=\langle\psi(h)|U^\dagger O_aU|\psi(h)\rangle,
\end{equation}
where each $P_j$ is a single-qubit Pauli operator,
$\|O_a\|_{\mathrm{op}}\le1$, and the trainable unitary $U$ is fixed
and independent of $h$. Set $A_a=U^\dagger O_aU$.
Differentiating gives
\begin{equation}
\left|\frac{\partial z_a}{\partial h_j}\right|
=\left|i\langle\psi(h)|[P_j/2,A_a]|\psi(h)\rangle\right|
\le\|[P_j/2,A_a]\|_{\mathrm{op}}
\le1.
\end{equation}
Here $P_j$ acts on the full register, and the last step follows from
$\|[A,B]\|_{\mathrm{op}}\le
2\|A\|_{\mathrm{op}}\|B\|_{\mathrm{op}}$.
Consequently, the readout Jacobian satisfies
\begin{equation}
\|J_z(h)\|_2\le\|J_z(h)\|_{\mathrm F}\le\sqrt{mq}.
\label{eq:app_block_lipschitz}
\end{equation}
The bound holds for every $h\in\mathbb R^q$. Integrating along the
line segment between two inputs gives the sufficient Lipschitz
constant $\sqrt{mq}$. This bound is conservative.
If the angles are $c_jh_j$, the corresponding bound is
$\sqrt{m\sum_j c_j^2}$. Repeated data encoding requires accounting
for every occurrence of each input angle.

For disjoint blocks evaluated in parallel, the pre-routing Jacobian
is block diagonal. A valid choice of Lipschitz upper bound under the unscaled single-encoding
assumptions is therefore
\begin{equation}
L_{z,l}=\max_b\sqrt{m_{lb}q_{lb}},
\end{equation}
where $q_{lb}$ and $m_{lb}$ are the input width and readout count
of block $b$ in layer $l$.
Identity and permutation routing have $\|W^{(l)}\|_2=1$;
feature duplication or rescaling must be included in the actual
routing norm.

Bounded readouts do not imply non-expansiveness.
For example, product $R_Y$ encoding on two qubits followed by a
CNOT from the first to the second gives Pauli-$Z$ readouts
\begin{equation}
z(h_1,h_2)=(\cos h_1,\cos h_1\cos h_2).
\end{equation}
Along $h=(t,0)$, the derivative norm is
$\sqrt{2}|\sin t|>1$ for $\pi/4<t<3\pi/4$, although both outputs lie in
$[-1,1]$. Thus $L_{z,l}\le1$ requires a separate justification.

\subsubsection{Depth dependence and finite-shot accuracy}

If $\rho_l\le\rho$ and $\eta_l\le\eta$ uniformly, then
\begin{equation}
\mathbb E\|\Delta^{(L)}\|_2
\le\eta\sum_{k=0}^{L-1}\rho^k
=
\begin{cases}
\displaystyle\eta\frac{1-\rho^L}{1-\rho}, & 0\le\rho<1,\\[4pt]
L\eta, & \rho=1,\\[2pt]
\displaystyle\eta\frac{\rho^L-1}{\rho-1}, & \rho>1.
\end{cases}
\label{eq:app_uniform_error_bound}
\end{equation}
Contractive layers give a uniform upper bound $\eta/(1-\rho)$;
non-expansive layers give an at-most-linear bound. When $\rho>1$,
the estimate permits amplification through depth but does not assert
exponential growth of the actual error. The conservative
input-sensitivity bounds above do not establish $\rho_l\le1$.

For example, suppose $\beta_l=0$, $\rho_l\le1$,
$\|W^{(l)}\|_2\le\overline w$, $m_l\le\overline m$, and $S_l=S$.
For a target expected error $\epsilon_{\mathrm{tol}}>0$,
\begin{equation}
\mathbb E\|\Delta^{(L)}\|_2
\le\overline w L\sqrt{\frac{\overline m}{S}},
\qquad
S\ge\frac{\overline w^{\,2}\overline m L^2}
{\epsilon_{\mathrm{tol}}^2}
\ \Longrightarrow\
\mathbb E\|\Delta^{(L)}\|_2\le\epsilon_{\mathrm{tol}}.
\label{eq:app_sufficient_shots}
\end{equation}
This is a sufficient condition for expected forward-output accuracy.
It is not a lower bound, a guarantee for gradient accuracy, or a
count of total training shots. With nonzero physical bias,
increasing $S$ reduces the sampling term but does not remove the
bias contribution from \eqref{eq:app_layerwise_error_bound}.

\subsubsection{Scope of the result}

Measurement confines coherent evolution to each circuit, while
errors in its readout affect later encodings. The theorem quantifies
this propagation through the complete layer sensitivities.
It establishes a conditional stability bound for a fixed model;
it does not establish greater noise robustness than coherent circuits.
Such a comparison requires comparable input--output maps,
specified noise channels, and matched resource assumptions.
No universal signal-contraction law for arbitrary local noise is
assumed or asserted.

\subsection{Quantum-Hardware Circuit and Shot Costs}
\label{app:hardware_cost}

The simulator measurements reported in this work do not measure
quantum-hardware execution time. Here we account for the circuit
settings and measurement shots required by a prospective
hardware implementation using two-term parameter-shift
derivatives and local-Jacobian propagation through re-encoding.

\paragraph{Forward evaluation.}
Consider $L$ stages, with $B_\ell$ local VQC blocks at stage
$\ell$. Block $(\ell,b)$ contains $q_{\ell b}$ qubits,
$K_{\ell b}$ circuit layers, and
\begin{equation}
    p_{\ell b}=3q_{\ell b}K_{\ell b}
\end{equation}
trainable rotation angles.

Let $n_b$ denote the batch size and $R$ the number of model
evaluations per sample: $R=1$ for a single tabular prediction,
and $R=N$ for the state-dependent modules in an $N$-step
BSDE rollout. One forward loss evaluation requires
\begin{equation}
    C_{\mathrm{fwd}}
    =
    n_b R
    \sum_{\ell=1}^{L} B_\ell
    \label{eq:hardware_forward_settings}
\end{equation}
local circuit settings. A circuit setting specifies the
encoded inputs and trainable gate angles for one local block.
All single-qubit Pauli-$Z$ readouts of that block can be
estimated from the same computational-basis measurement batch.

\paragraph{Gradient evaluation.}
Assume each differentiated angle occurs in a single
Pauli-rotation gate and admits a two-term parameter-shift
rule. With cached forward activations, evaluating the local
parameter Jacobian requires $2p_{\ell b}$ additional settings
per block. Propagating derivatives through a later stage also
requires its input-angle Jacobian, costing $2q_{\ell b}$
additional settings.

Define $I_\ell=0$ for the first stage, whose inputs are fixed
data, and $I_\ell=1$ for subsequent stages. The resulting
circuit-setting count per optimizer update is approximately
\begin{equation}
    C_{\mathrm{step}}
    \approx
    n_b R
    \sum_{\ell=1}^{L}
    \sum_{b=1}^{B_\ell}
    \left(
        1+2p_{\ell b}+2q_{\ell b}I_\ell
    \right).
    \label{eq:hardware_step_settings}
\end{equation}

This count refers to the specified local-Jacobian training
scheme, not to an optimal hardware differentiation strategy.
The counts for the BSDE state-dependent modules exclude any
additional circuits used to parameterize the initial value
$Y_0$, which must be counted separately.

\paragraph{Measurement shots and sequential stages.}
With $S$ shots per circuit setting and $T_{\mathrm{opt}}$
optimizer updates, the corresponding shot counts are
\begin{align}
    N_{\mathrm{shots,step}}
    &=
    S C_{\mathrm{step}},
    \label{eq:hardware_step_shots}\\
    N_{\mathrm{shots,train}}
    &=
    T_{\mathrm{opt}} S C_{\mathrm{step}}.
    \label{eq:hardware_training_shots}
\end{align}
Here, $T_{\mathrm{opt}}$ counts optimizer updates rather than
assuming that one epoch equals one update.

Each shot requires a new state preparation and circuit
execution. Blocks within a stage may be evaluated in parallel
when sufficient hardware is available, but the $L$
measurement--mixing--re-encoding stages remain sequential.
The counts do not include classical processing time,
device-control latency, or additional error-mitigation
overhead.

\paragraph{Accuracy dependence and scope.}
The shot budget must also account for error propagation
through successive stages. Under the zero-bias and
non-expansive assumptions stated in
Appendix~\ref{app:layerwise_error_propagation},
Eq.~\eqref{eq:app_sufficient_shots} provides a sufficient
shot budget for a specified expected forward-output error.
It does not establish a corresponding guarantee for gradient
accuracy or optimization convergence.

Consequently, small local circuit width does not imply
depth-independent shot efficiency or low total training cost.
This accounting describes prospective hardware resource
requirements; it is not a hardware experiment or evidence of
a hardware-training advantage.

\subsection{Locality, Gradient Propagation, and Classical Simulability}
\label{app:locality_trainability}

\paragraph{Relation to locality-based trainability.}
Local circuit structure is an established approach to
mitigating barren plateaus.
\citet{pesah2021absence} analyze QCNNs with hierarchical
convolution and pooling and establish polynomial, rather
than exponential, gradient-variance scaling under their
stated assumptions.
FC-VQC also restricts coherent evolution to local circuits,
but connects independently parameterized blocks through
measurement, deterministic mixing, and re-encoding.
Its architecture therefore differs from the QCNN construction,
and the QCNN result does not directly establish the absence
of barren plateaus in FC-VQC.

For FC-VQC-SW and FC-VQC-PL, increasing the input dimension
adds blocks without increasing the local circuit width.
Each block therefore continues to evolve a $2^q$-dimensional
state, rather than becoming a wider coherent circuit.
Classical communication between blocks does not preserve
inter-block quantum entanglement.
However, fixed local width alone does not guarantee
well-conditioned gradients throughout the composed model.
Local gradients can be small, and derivatives can be
attenuated through successive classical interfaces.
For FC-VQC-FC, the local width instead grows as
$q=\lceil\sqrt{d}\rceil$, so the fixed-width argument does
not apply as $d$ increases.

\paragraph{Gradient propagation through measured stages.}
Consider one FC-VQC module with $L$ stages and a fixed input.
Write its ideal expectation-value computation as
\begin{equation}
    H^{(\ell)}
    =
    F_\ell\!\left(H^{(\ell-1)};\theta_\ell\right),
    \qquad
    \ell=1,\ldots,L,
    \qquad H^{(0)}=x,
    \label{eq:app_trainability_recursion}
\end{equation}
where $F_\ell$ includes the deterministic routing, encoding,
local VQC transformations, and expectation-value readout.
Assume that these maps are differentiable and that parameters
are independent across stages.
For the parameter vector $\theta_{\ell,b}$ of block $b$
in stage $\ell$, define
\begin{equation}
    A_j
    :=
    \frac{\partial H^{(j)}}{\partial H^{(j-1)}},
    \qquad
    D_{\ell,b}
    :=
    \frac{\partial H^{(\ell)}}{\partial\theta_{\ell,b}}.
    \label{eq:app_trainability_jacobians}
\end{equation}

Let $\mathcal L$ be a differentiable scalar loss whose
dependence on these parameters is through $H^{(L)}$.
The chain rule gives
\begin{equation}
    \nabla_{\theta_{\ell,b}}\mathcal L
    =
    D_{\ell,b}^{\top}
    A_{\ell+1}^{\top}\cdots A_L^{\top}
    \nabla_{H^{(L)}}\mathcal L.
    \label{eq:app_gradient_chain}
\end{equation}
Using the Euclidean norm for vectors and the induced
spectral norm for matrices,
\begin{equation}
    \left\|\nabla_{\theta_{\ell,b}}\mathcal L\right\|_2
    \le
    \|D_{\ell,b}\|_2
    \left(
        \prod_{j=\ell+1}^{L}\|A_j\|_2
    \right)
    \left\|\nabla_{H^{(L)}}\mathcal L\right\|_2.
    \label{eq:app_gradient_propagation_bound}
\end{equation}
The product is interpreted as one when $\ell=L$.
For a time-step-specific BSDE module,
$\nabla_{H^{(L)}}\mathcal L$ includes the dependence of
the terminal loss on that module's output through the
BSDE recursion.

\paragraph{Conditional gradient attenuation.}
If all downstream Jacobians satisfy
$\|A_j\|_2\le\rho<1$, then
\begin{equation}
    \left\|\nabla_{\theta_{\ell,b}}\mathcal L\right\|_2
    \le
    \|D_{\ell,b}\|_2
    \rho^{L-\ell}
    \left\|\nabla_{H^{(L)}}\mathcal L\right\|_2.
    \label{eq:app_gradient_attenuation}
\end{equation}
Thus, when the remaining factors are uniformly bounded,
contractive downstream stages can suppress early-stage
gradients as the number of stages increases.
We do not assume or establish that this contraction condition
holds for the trained configurations.
The bound identifies a possible attenuation mechanism;
it is not a barren-plateau theorem and provides no lower
bound on gradient magnitude.
Finite-shot estimation and physical noise can further
complicate the resolution of small gradients.

\paragraph{Classical simulability.}
\citet{cerezo2025simulability} present case-by-case evidence
that structures associated with provable absence of barren
plateaus can also support classical simulation, potentially
after an initial quantum-data acquisition phase.
This is not a universal equivalence between trainability
and classical simulability.

For the present classical-input FC-VQC models, fixed-width
simulability follows directly from the explicit block
decomposition, without requiring a trainability theorem.
As derived in Section~\ref{subsec:simulator_complexity},
one forward evaluation of a dimension-preserving architecture
has direct state-vector cost
\begin{equation}
    C_{\mathrm{sim}}
    =
    O(LBKq2^q)
    =
    O(LKd2^q),
    \label{eq:app_locality_simulation_cost}
\end{equation}
where $Bq=d$ for the evaluated configurations.
For FC-VQC-SW and FC-VQC-PL, this is linear in $d$ at
fixed $q$, $L$, and $K$.
For FC-VQC-FC, substituting
$q=\lceil\sqrt{d}\rceil$ gives
\begin{equation}
    C_{\mathrm{sim,FC}}
    =
    O\!\left(LKd2^{\lceil\sqrt{d}\rceil}\right).
    \label{eq:app_locality_fc_simulation_cost}
\end{equation}
The fixed-width linear scaling therefore does not extend
to FC-VQC-FC.
Conversely, the increasing cost of this direct simulator
does not establish classical intractability of FC-VQC-FC.

\paragraph{Interpretation of the empirical diagnostics.}
Appendix~\ref{app:gradients_concrete} reports dispersion
across parameter-gradient components during training.
This statistic is neither the gradient norm nor gradient
variance over random circuit initializations.
Its evolution is therefore interpreted alongside predictive
performance, rather than as proof of barren-plateau
avoidance. The present analysis separates local circuit
width, gradient propagation across stages, and classical
simulation cost without asserting guaranteed trainability
or quantum computational advantage.

\subsection{Block Information Exchange: Definitions and Proofs}
\label{sec:theory_exchange}
\label{app:block_information_exchange}

We establish the structural receptive-field results in
Theorems~\ref{thm:rf_sliding_window_main}--\ref{thm:rf_fully_connected_main}.
The analysis concerns the block routing graph of dimension-preserving
architectures. It describes which input blocks have paths to each
output block, without assuming that the learned function uses every
available path.

\subsubsection{Blockwise model and stage counting}

Partition $x\in\mathbb R^d$ into $B$ blocks,
$x=(x_1,\ldots,x_B)$ with $x_b\in\mathbb R^q$ and $d=Bq$.
At stage $l$, apply
\begin{equation}
\Phi^{(l)}(H)
=\bigl(\phi_1^{(l)}(h_1),\ldots,\phi_B^{(l)}(h_B)\bigr),
\qquad
\phi_b^{(l)}:\mathbb R^q\to\mathbb R^q.
\end{equation}
Each $\phi_b^{(l)}$ is a local encode--evolve--measure VQC map.
The recursion is
\begin{equation}
H^{(l)}
=\Phi^{(l)}\!\left(g^{(l-1)}(H^{(l-1)})\right),
\qquad H^{(0)}=x,\qquad l=1,\ldots,L.
\label{eq:exchange_recursion}
\end{equation}
Here $L$ counts VQC evaluations and there are $L$ preceding routing
slots, indexed $0,\ldots,L-1$. A slot may apply the identity.
In particular, $g^{(0)}=I$ represents an unmixed first VQC layer.

In this subsection $H^{(l)}$ is the readout before outgoing routing.
For the error analysis in
Appendix~\ref{app:layerwise_error_propagation}, the intermediate
state is instead taken after outgoing routing. The two descriptions
evaluate the same sequence: initial routing $g^{(0)}$, first VQC,
routing $g^{(1)}$, second VQC, and so on, ending at VQC $L$.
The error analysis initializes its routed state at $g^{(0)}(x)$. For
the direct-readout case, $W^{(L)}=I$, so the final outputs agree; a
fixed linear final map can instead be included in $W^{(L)}$.

\subsubsection{Parallel blocks without exchange}

If every routing map is the identity, block $b$ only receives its own
preceding output. Therefore,
\begin{equation}
H^{(L)}(x)=\bigl(F_1(x_1),\ldots,F_B(x_B)\bigr),
\qquad
F_b=\phi_b^{(L)}\circ\cdots\circ\phi_b^{(1)}.
\label{eq:app_parallel_factorization}
\end{equation}
This follows directly by composing the independent block maps.
No output block has functional dependence on another input block.
The statement concerns this blockwise output; it does not include
an additional readout that combines distinct blocks.

\subsubsection{Structural block receptive fields}

Let $\mathcal N^{(l-1)}(b)\subseteq\{1,\ldots,B\}$ contain the source
blocks whose readout components are supplied by $g^{(l-1)}$ to
block $b$. These sets are specified by the routing architecture.
For linear routing, an edge is present when the corresponding
source-to-destination submatrix is nonzero.

Construct a graph with nodes $(l,b)$ and edges from $(l-1,a)$
to $(l,b)$ for $a\in\mathcal N^{(l-1)}(b)$.
The structural receptive field is
\begin{equation}
\mathcal R^{(0)}(b)=\{b\},
\qquad
\mathcal R^{(l)}(b)
=\bigcup_{a\in\mathcal N^{(l-1)}(b)}
\mathcal R^{(l-1)}(a).
\label{eq:app_structural_receptive_field}
\end{equation}
Equivalently, it is the set of input blocks with a directed path to
$(l,b)$. Purely parallel routing gives
$\mathcal R^{(l)}(b)=\{b\}$ at every stage.

If an input block is outside this set, no operation contributing
to the output receives information from that block. Induction
therefore shows that actual functional dependence is contained in
$\mathcal R^{(l)}(b)$. The inclusion need not be an equality:
constant outputs, discarded components, or cancellation can remove
functional dependencies despite the existence of graph paths.
The graph is defined at block level and does not assert that every
feature of an accessible block is retained.

\subsubsection{Sliding-Window Mixing: Locality and Receptive-Field Growth}

Let $r\ge1$ denote the sliding-window routing radius in blocks. The
sliding-window mixer satisfies
\begin{equation}
\bigl(g_{\mathrm{sw}}(H)\bigr)_b
=\mathcal M_b(h_{b-r},\ldots,h_{b+r}),
\qquad
\mathcal N_{\mathrm{sw}}(b)\subseteq\{b-r,\ldots,b+r\},
\label{eq:sliding_window_mixer}
\end{equation}
where indices are taken modulo $B$ and
$\mathcal M_b:(\mathbb R^q)^{2r+1}\to\mathbb R^q$ is a fixed
deterministic routing map. In the evaluated FC-VQC-SW architecture,
$r=1$: each destination block receives selected components from its
previous neighbour, itself, and its next neighbour on the ring. In
particular, the self-routing edge is retained.

\paragraph{Proof of Theorem~\ref{thm:rf_sliding_window_main}.}
Because $g^{(0)}=I$,
$\mathcal R^{(1)}(b)=\{b\}$. For some $l\ge2$, suppose that for every
block $a$,
\[
\mathcal R^{(l-1)}(a)
\subseteq
\{a-(l-2)r,\ldots,a+(l-2)r\}\pmod B.
\]
For any $c\in\mathcal R^{(l)}(b)$,
Eq.~\eqref{eq:app_structural_receptive_field} supplies a predecessor
$a\in\mathcal N_{\mathrm{sw}}(b)$ with
$c\in\mathcal R^{(l-1)}(a)$. Locality gives
$a\equiv b+u\pmod B$ for some $|u|\le r$, while the induction
hypothesis gives $c\equiv a+v\pmod B$ for some
$|v|\le(l-2)r$. Hence
$c\equiv b+(u+v)\pmod B$ with $|u+v|\le(l-1)r$.
Induction yields
\begin{equation}
\begin{aligned}
\mathcal R^{(L)}(b)
&\subseteq
\{b-(L-1)r,\ldots,b+(L-1)r\}\pmod B,\\
|\mathcal R^{(L)}(b)|
&\le\min\{B,2(L-1)r+1\}.
\end{aligned}
\label{eq:app_sw_receptive_field_bound}
\end{equation}
The bound does not guarantee that every permitted path is present or
used by the learned function. The condition $2(L-1)r+1\ge B$ allows
full-ring coverage but does not establish it under locality alone.

\subsubsection{Fully Connected Mixing: Global Structural Access}

A fully connected routing slot supplies components from every
preceding block to each next block:
\begin{equation}
\bigl(g_{\mathrm{fc}}(H)\bigr)_b
=\mathcal M_b(h_1,\ldots,h_B),
\qquad
\mathcal N_{\mathrm{fc}}(b)=\{1,\ldots,B\}.
\label{eq:fully_connected_mixer}
\end{equation}
Here $\mathcal M_b:(\mathbb R^q)^B\to\mathbb R^q$ is fixed. The
condition on $\mathcal N_{\mathrm{fc}}$ requires an incoming edge from
every source block; merely listing all blocks as arguments does not
impose this condition. Each local VQC may use or discard the supplied
components.

\paragraph{Proof of Theorem~\ref{thm:rf_fully_connected_main}.}
Since $g^{(0)}=I$, $\mathcal R^{(1)}(a)=\{a\}$. The first fully
connected routing step therefore gives
\begin{equation}
\mathcal R^{(2)}(b)
=\bigcup_{a=1}^{B}\mathcal R^{(1)}(a)
=\bigcup_{a=1}^{B}\{a\}
=\{1,\ldots,B\}.
\label{eq:app_fc_receptive_field}
\end{equation}
If every preceding block has this structural receptive field, a
nonempty union of their receptive fields is again $\{1,\ldots,B\}$.
The assumed incoming edge for each later block therefore proves the
claim for every $L\ge2$ by induction.

\subsubsection{Interpretation}

Parallel routing provides no cross-block paths; sliding-window
routing propagates structural access locally over successive routing
steps; and fully connected routing provides global block-level access
after its first fully connected routing step. These are structural
statements. They do not establish preservation of all functional
dependencies, universal approximation, nesting of finite VQC model
classes, or an ordering of trained prediction errors.

\subsection{Support-Induced Approximation Lower Bound}
\label{sec:theory_support_mismatch}

We relate the structural receptive fields of
Appendix~\ref{sec:theory_exchange} to the information available to
each output block. The result concerns an information-induced lower
bound and does not assume that finite FC-VQC hypothesis classes are
nested across depth or mixing strategies.

Let $X=(X_1,\ldots,X_B)$ be the block-partitioned input and let
$T_b\in L_2$ denote the target associated with output block $b$. For
a fixed PDE/BSDE time step, $T_b$ may be the corresponding block of
the exact $Z=\sigma^\top\nabla_x u(t,X)$ vector. If block $b$ can
access only $\mathcal S_b\subseteq\{1,\ldots,B\}$, define
\begin{equation}
\mathcal A_b=\sigma(X_j:j\in\mathcal S_b),
\qquad
\mu_b=\mathbb E[T_b\mid\mathcal A_b].
\end{equation}

\paragraph{Proof of Theorem~\ref{thm:support_lower_bound}.}
For any predictor $F_b\in L_2(\mathcal A_b)$,
\begin{align}
\mathbb E\|T_b-F_b\|_2^2
&=
\mathbb E\|T_b-\mu_b\|_2^2
+
\mathbb E\|\mu_b-F_b\|_2^2,
\label{eq:app_support_pythagorean}
\end{align}
because conditional expectation is the $L_2$ projection onto the
$\mathcal A_b$-measurable functions. The second term is
non-negative, so taking the infimum over
$F_b\in\mathcal H_b(\mathcal S_b)\subseteq L_2(\mathcal A_b)$ gives
\begin{equation}
\inf_{F_b\in\mathcal H_b(\mathcal S_b)}
\mathbb E\|F_b-T_b\|_2^2
\ge
\mathbb E\|T_b-\mu_b\|_2^2.
\label{eq:app_support_lower_bound}
\end{equation}

\paragraph{Nested information sets.}
At depth $L$, set
$\mathcal S_b^{(L)}:=\mathcal R^{(L)}(b)$ and define
\begin{equation}
\mathcal A_b^{(L)}
:=\sigma(X_j:j\in\mathcal S_b^{(L)}),
\qquad
\mu_b^{(L)}
:=\mathbb E[T_b\mid\mathcal A_b^{(L)}],
\end{equation}
with support-induced floor
\begin{equation}
\mathcal E_L^{\mathrm{supp}}
:=\mathbb E\|T_b-\mu_b^{(L)}\|_2^2.
\end{equation}
If $\mathcal S_b^{(L)}\subseteq\mathcal S_b^{(L+1)}$, then
$\mathcal A_b^{(L)}\subseteq\mathcal A_b^{(L+1)}$ and the tower
property gives
\[
\mu_b^{(L)}
=\mathbb E[\mu_b^{(L+1)}\mid\mathcal A_b^{(L)}].
\]
Conditional orthogonality therefore yields
\begin{equation}
\mathcal E_L^{\mathrm{supp}}
=
\mathcal E_{L+1}^{\mathrm{supp}}
+
\mathbb E\|\mu_b^{(L+1)}-\mu_b^{(L)}\|_2^2,
\label{eq:app_nested_information_bound}
\end{equation}
and hence
$\mathcal E_{L+1}^{\mathrm{supp}}\le\mathcal E_L^{\mathrm{supp}}$.
The inequality is strict exactly when the newly accessible
information changes the conditional mean on a set of nonzero
probability.

\paragraph{Connection to FC-VQC routing.}
Parallel routing leaves each output block restricted to its
corresponding input block. For the evaluated sliding-window mixer,
each destination block retains a self-routing edge. Hence
\begin{equation}
\mathcal R^{(L)}(b)\subseteq\mathcal R^{(L+1)}(b),
\end{equation}
so its structural information sets are nested with depth and the
preceding result applies. Theorem~\ref{thm:rf_sliding_window_main}
bounds how quickly this support can expand. Fully connected routing
provides global structural access after its first fully connected
routing step.

These statements concern information availability only. They do not
establish nesting of the finite FC-VQC hypothesis classes, attainment
of the conditional-mean lower bound, or monotonic improvement of
trained prediction error.

\clearpage
\clearpage
\section{Gradient Dynamics Analysis}
\label{app:gradients_concrete}

This appendix reports gradient dynamics on Concrete Strength. Each figure has rows $L\in\{1,3,5,7,9\}$ (top to bottom) and columns $K\in\{3,5,7,9\}$ (left to right).

At epoch $e$, let $g_i^{(e)}=\partial\mathcal{L}_{\mathrm{train}}(\boldsymbol{\theta}^{(e)})/\partial\theta_i$, where $\mathcal{L}_{\mathrm{train}}$ is the full-batch training loss and $P$ is the number of trainable parameters. Flattening all parameter gradients into one vector, we compute
\begin{equation}
V^{(e)}=\frac{1}{P-1}\sum_{i=1}^{P}\left(g_i^{(e)}-\bar g^{(e)}\right)^2,
\qquad
\bar g^{(e)}=\frac{1}{P}\sum_{i=1}^{P}g_i^{(e)}.
\label{eq:gradient_variance_concrete}
\end{equation}
Each curve plots $V^{(e)}$ against epoch on a logarithmic vertical scale. This statistic measures dispersion across parameter components, not variability across batches, samples, or seeds. A decrease can also reflect convergence; these empirical diagnostics alone establish neither a barren plateau nor its absence.

\begin{figure}[!htbp]
    \centering
    \includegraphics[width=0.245\linewidth]{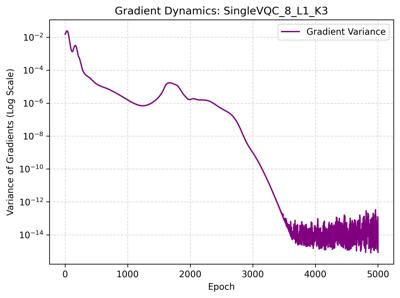}\hfill%
    \includegraphics[width=0.245\linewidth]{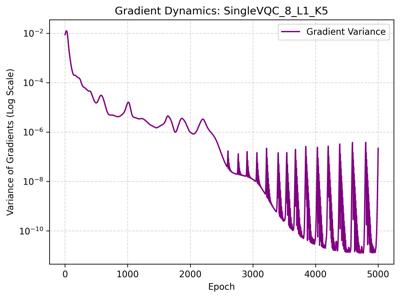}\hfill%
    \includegraphics[width=0.245\linewidth]{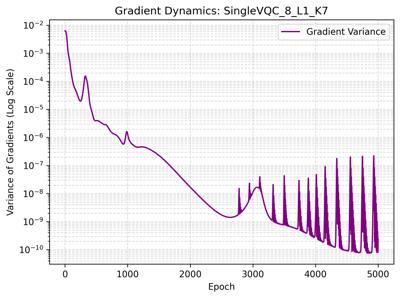}\hfill%
    \includegraphics[width=0.245\linewidth]{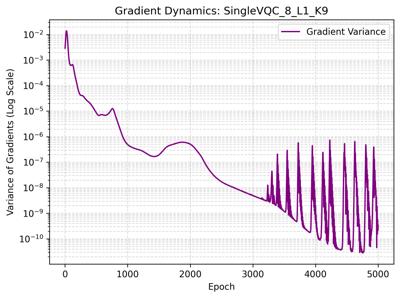}%
    \par\smallskip

    \includegraphics[width=0.245\linewidth]{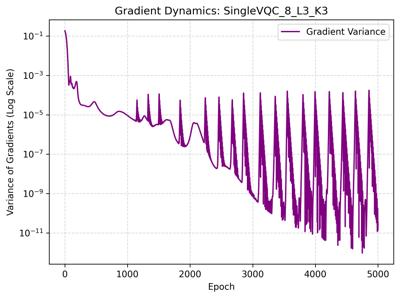}\hfill%
    \includegraphics[width=0.245\linewidth]{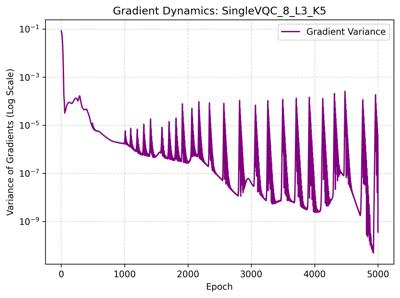}\hfill%
    \includegraphics[width=0.245\linewidth]{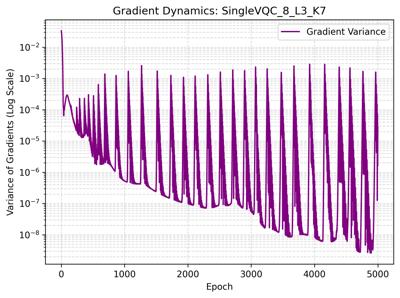}\hfill%
    \includegraphics[width=0.245\linewidth]{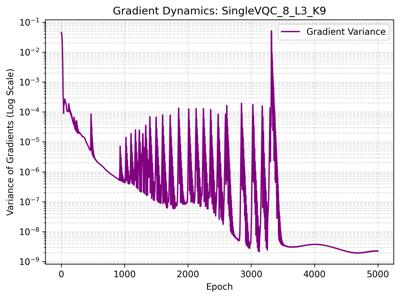}%
    \par\smallskip

    \includegraphics[width=0.245\linewidth]{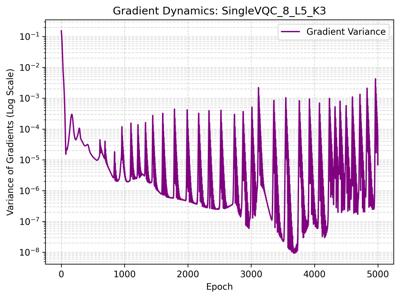}\hfill%
    \includegraphics[width=0.245\linewidth]{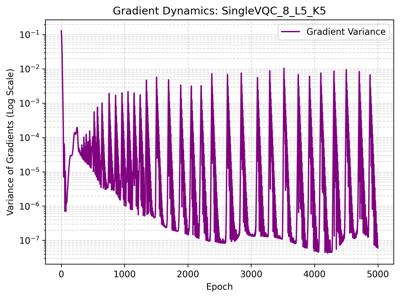}\hfill%
    \includegraphics[width=0.245\linewidth]{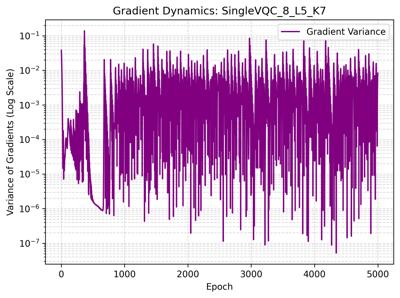}\hfill%
    \includegraphics[width=0.245\linewidth]{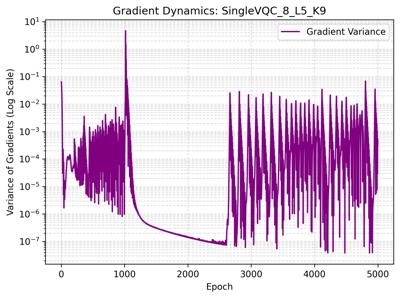}%
    \par\smallskip

    \includegraphics[width=0.245\linewidth]{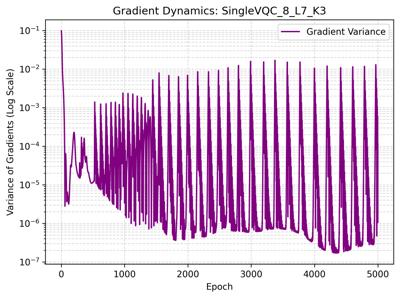}\hfill%
    \includegraphics[width=0.245\linewidth]{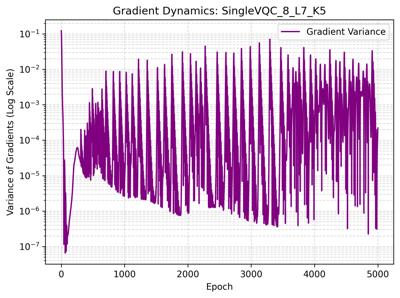}\hfill%
    \includegraphics[width=0.245\linewidth]{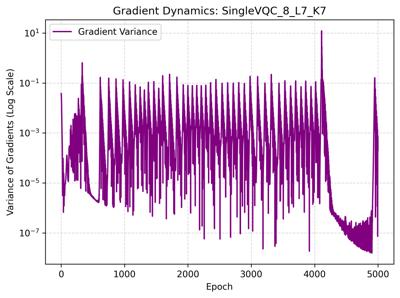}\hfill%
    \includegraphics[width=0.245\linewidth]{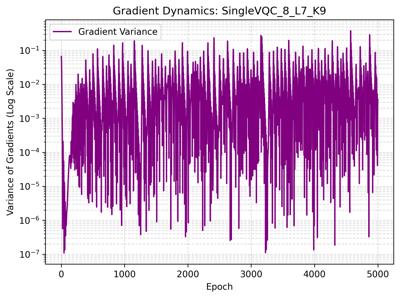}%
    \par\smallskip

    \includegraphics[width=0.245\linewidth]{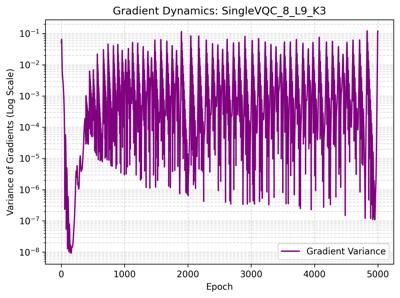}\hfill%
    \includegraphics[width=0.245\linewidth]{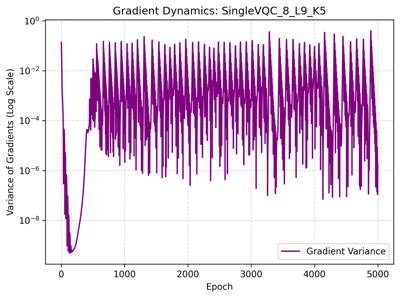}\hfill%
    \includegraphics[width=0.245\linewidth]{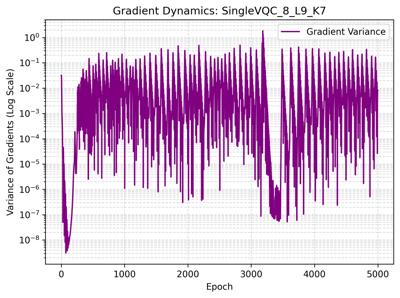}\hfill%
    \includegraphics[width=0.245\linewidth]{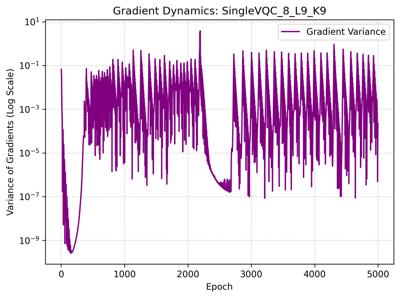}%

    \caption{\textbf{Gradient dynamics for the $8t1$ VQC.} Rows: $L=1,3,5,7,9$; columns: $K=3,5,7,9$. The first row is the Type~1 monolithic VQC; subsequent rows are Type~2 stacked full-width VQCs connected through measure-and-re-encode interfaces.}
    \label{fig:gradients_8t1}
\end{figure}

\clearpage

\begin{figure}[!htbp]
    \centering
    \includegraphics[width=0.245\linewidth]{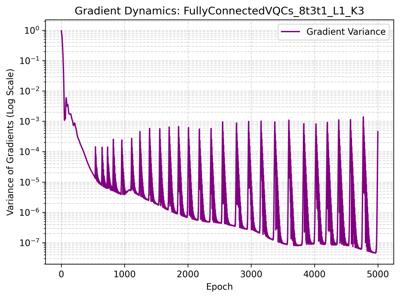}\hfill%
    \includegraphics[width=0.245\linewidth]{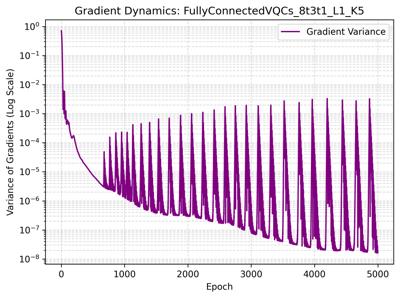}\hfill%
    \includegraphics[width=0.245\linewidth]{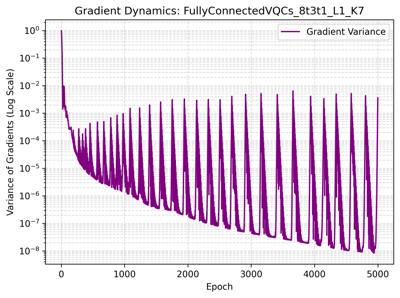}\hfill%
    \includegraphics[width=0.245\linewidth]{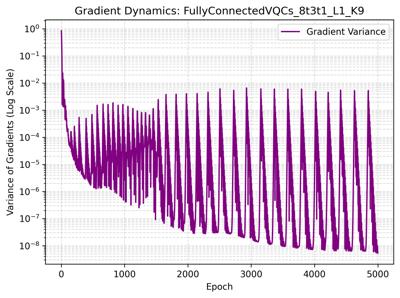}%
    \par\smallskip

    \includegraphics[width=0.245\linewidth]{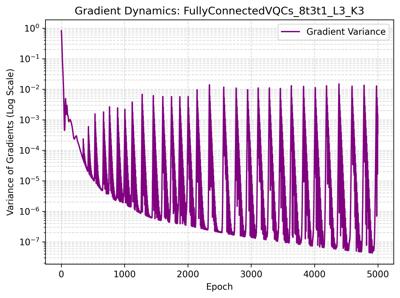}\hfill%
    \includegraphics[width=0.245\linewidth]{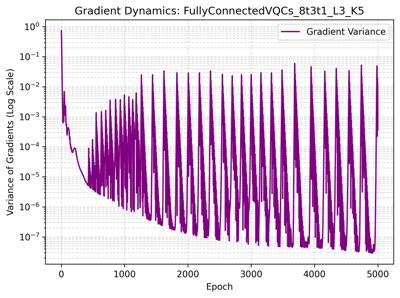}\hfill%
    \includegraphics[width=0.245\linewidth]{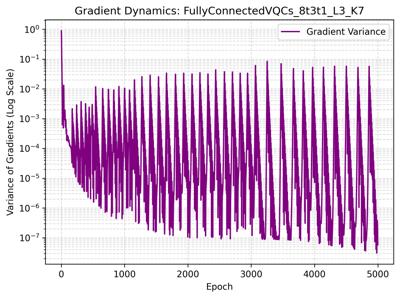}\hfill%
    \includegraphics[width=0.245\linewidth]{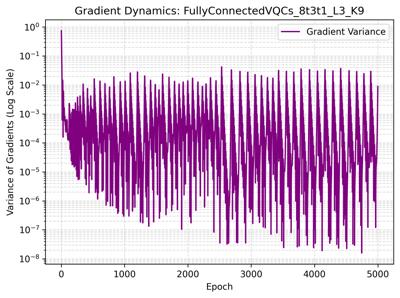}%
    \par\smallskip

    \includegraphics[width=0.245\linewidth]{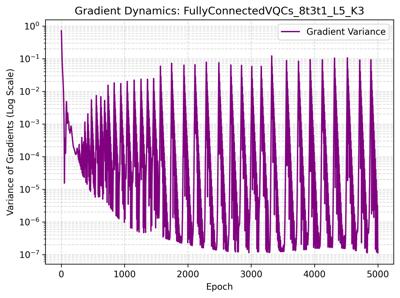}\hfill%
    \includegraphics[width=0.245\linewidth]{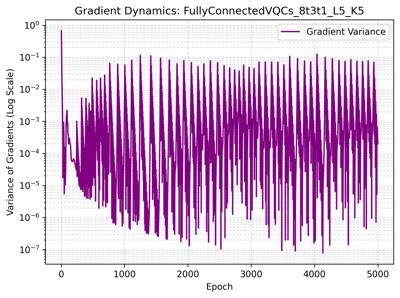}\hfill%
    \includegraphics[width=0.245\linewidth]{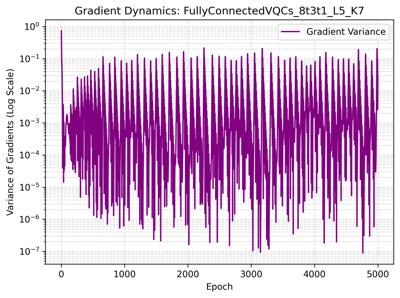}\hfill%
    \includegraphics[width=0.245\linewidth]{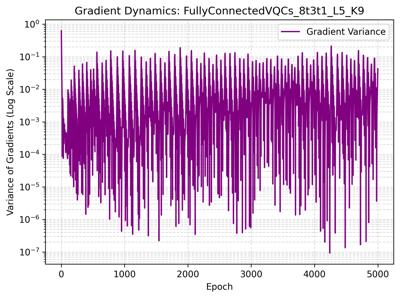}%
    \par\smallskip

    \includegraphics[width=0.245\linewidth]{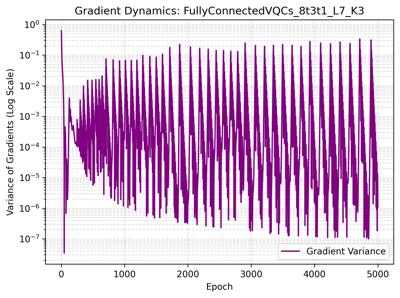}\hfill%
    \includegraphics[width=0.245\linewidth]{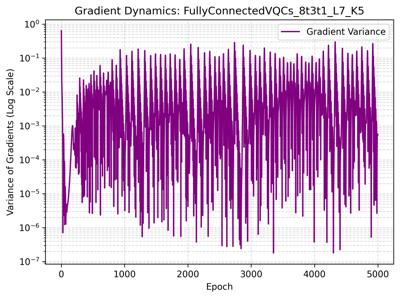}\hfill%
    \includegraphics[width=0.245\linewidth]{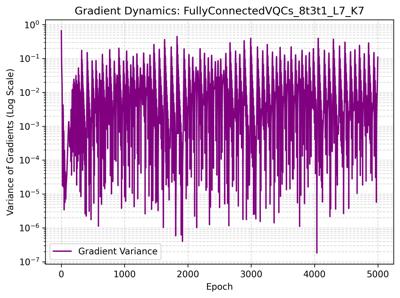}\hfill%
    \includegraphics[width=0.245\linewidth]{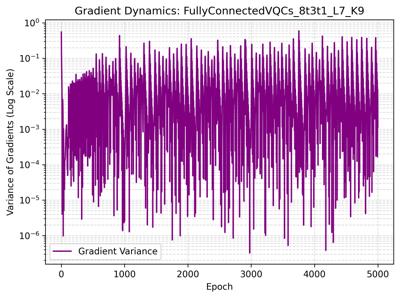}%
    \par\smallskip

    \includegraphics[width=0.245\linewidth]{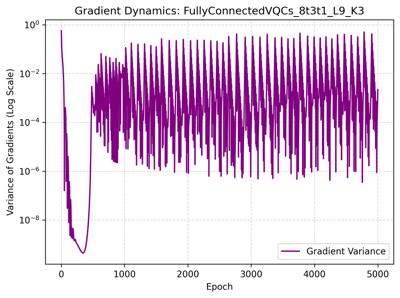}\hfill%
    \includegraphics[width=0.245\linewidth]{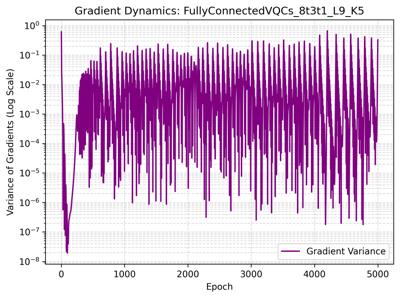}\hfill%
    \includegraphics[width=0.245\linewidth]{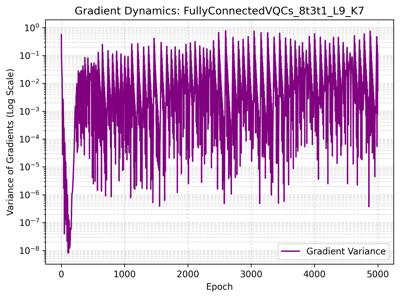}\hfill%
    \includegraphics[width=0.245\linewidth]{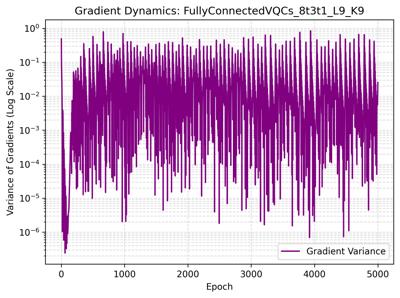}%

    \caption{\textbf{Gradient dynamics on Concrete Strength for the Type~3 $8t3t1$ FC-VQC architecture.} Rows: $L=1,3,5,7,9$ (top to bottom); columns: $K=3,5,7,9$ (left to right). The horizontal axis is training epoch; the vertical axis is gradient variance on a logarithmic scale.}
    \label{fig:gradients_8t3t1}
\end{figure}

\clearpage

\begin{figure}[!htbp]
    \centering
    \includegraphics[width=0.245\linewidth]{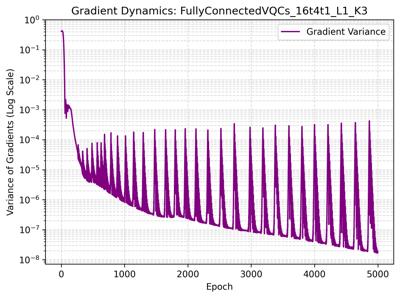}\hfill%
    \includegraphics[width=0.245\linewidth]{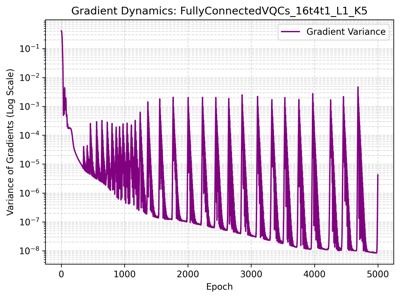}\hfill%
    \includegraphics[width=0.245\linewidth]{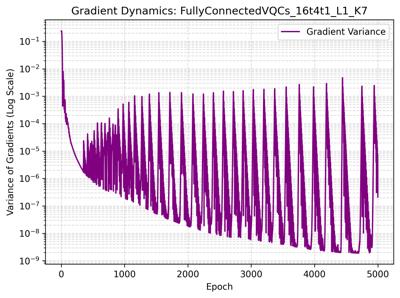}\hfill%
    \includegraphics[width=0.245\linewidth]{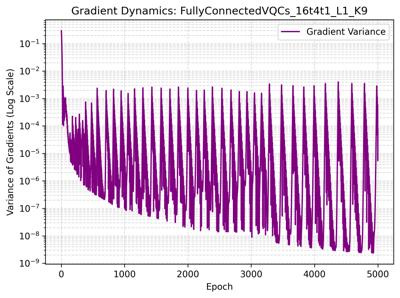}%
    \par\smallskip

    \includegraphics[width=0.245\linewidth]{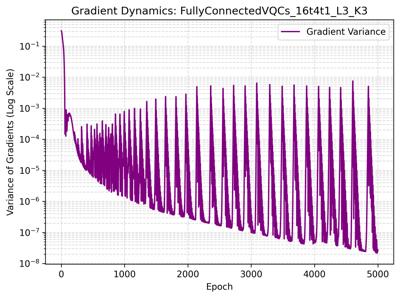}\hfill%
    \includegraphics[width=0.245\linewidth]{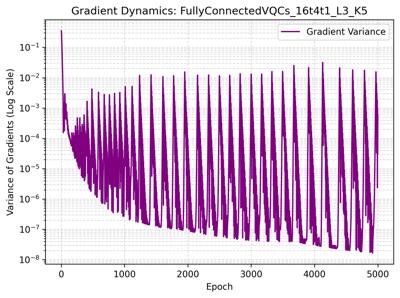}\hfill%
    \includegraphics[width=0.245\linewidth]{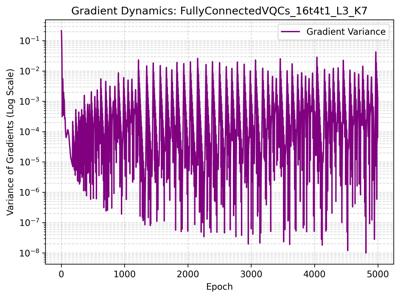}\hfill%
    \includegraphics[width=0.245\linewidth]{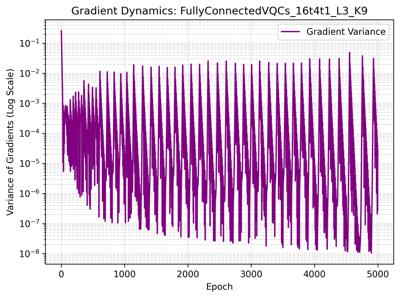}%
    \par\smallskip

    \includegraphics[width=0.245\linewidth]{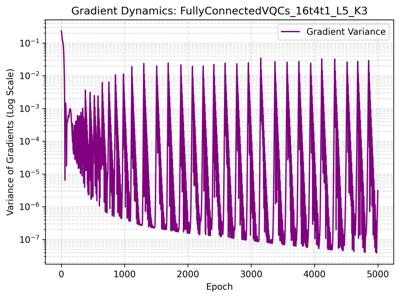}\hfill%
    \includegraphics[width=0.245\linewidth]{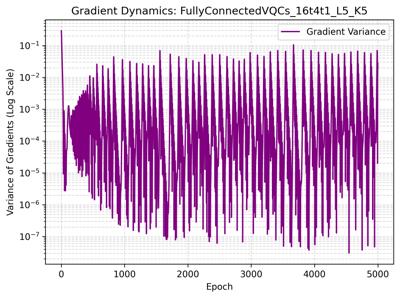}\hfill%
    \includegraphics[width=0.245\linewidth]{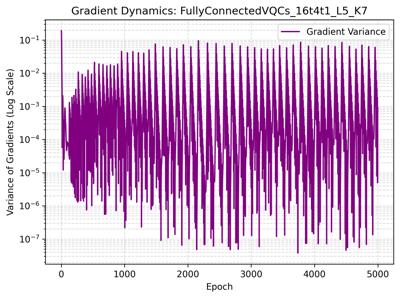}\hfill%
    \includegraphics[width=0.245\linewidth]{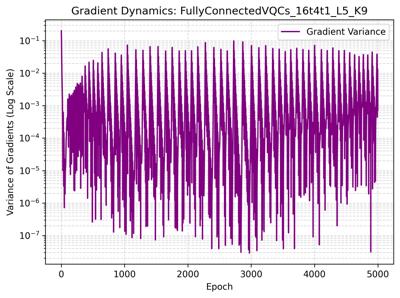}%
    \par\smallskip

    \includegraphics[width=0.245\linewidth]{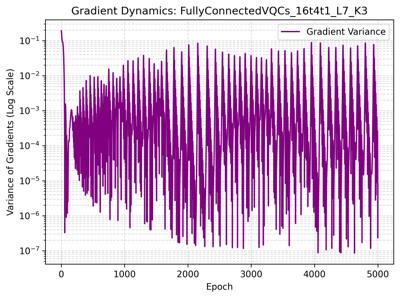}\hfill%
    \includegraphics[width=0.245\linewidth]{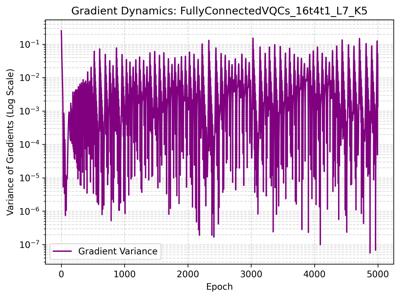}\hfill%
    \includegraphics[width=0.245\linewidth]{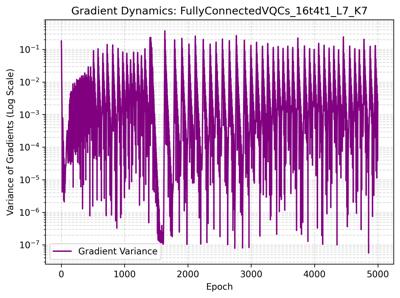}\hfill%
    \includegraphics[width=0.245\linewidth]{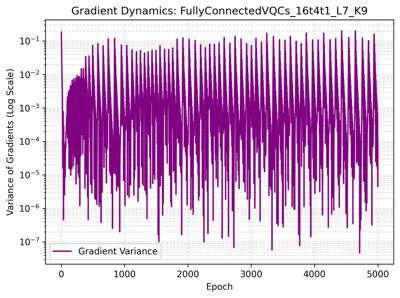}%
    \par\smallskip

    \includegraphics[width=0.245\linewidth]{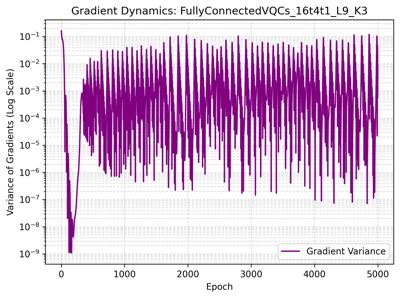}\hfill%
    \includegraphics[width=0.245\linewidth]{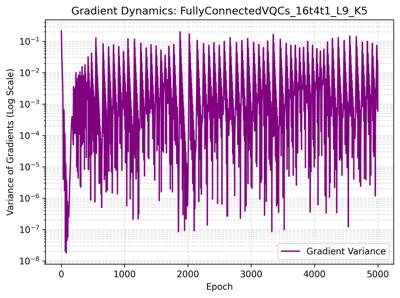}\hfill%
    \includegraphics[width=0.245\linewidth]{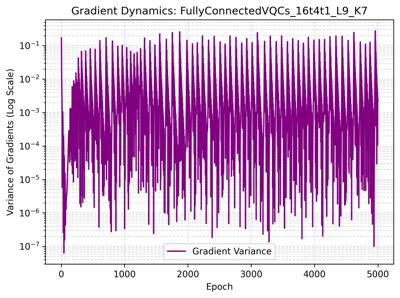}\hfill%
    \includegraphics[width=0.245\linewidth]{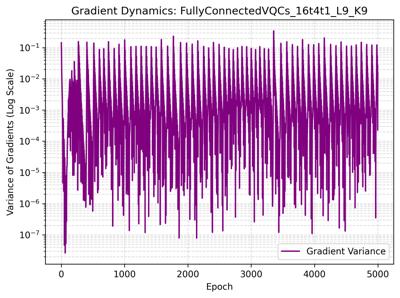}%

    \caption{\textbf{Gradient dynamics on Concrete Strength for the Type~4 $16t4t1$ FC-VQC architecture.} Rows: $L=1,3,5,7,9$ (top to bottom); columns: $K=3,5,7,9$ (left to right). The horizontal axis is training epoch; the vertical axis is gradient variance on a logarithmic scale.}
    \label{fig:gradients_16t4t1}
\end{figure}

\clearpage

\begin{figure}[!htbp]
    \centering
    \includegraphics[width=0.245\linewidth]{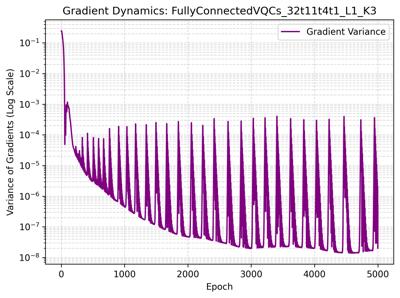}\hfill%
    \includegraphics[width=0.245\linewidth]{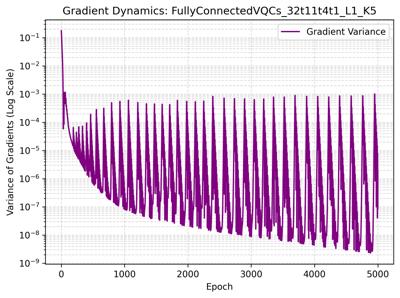}\hfill%
    \includegraphics[width=0.245\linewidth]{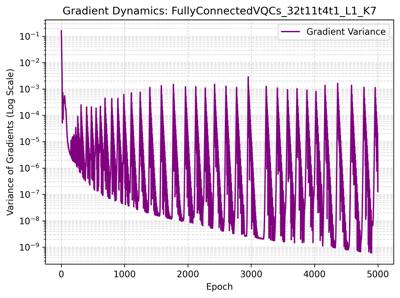}\hfill%
    \includegraphics[width=0.245\linewidth]{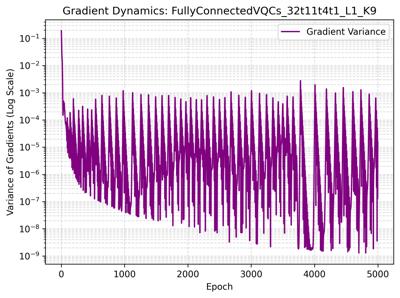}%
    \par\smallskip

    \includegraphics[width=0.245\linewidth]{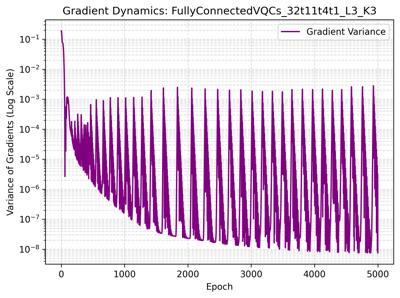}\hfill%
    \includegraphics[width=0.245\linewidth]{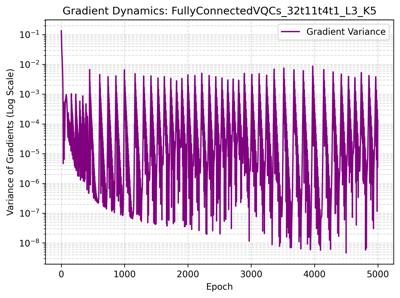}\hfill%
    \includegraphics[width=0.245\linewidth]{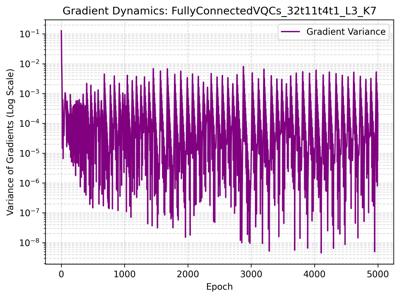}\hfill%
    \includegraphics[width=0.245\linewidth]{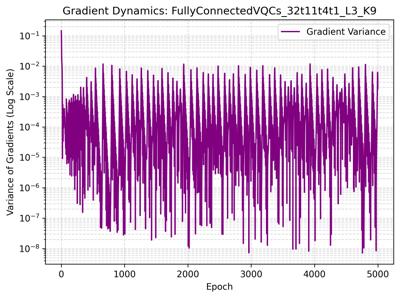}%
    \par\smallskip

    \includegraphics[width=0.245\linewidth]{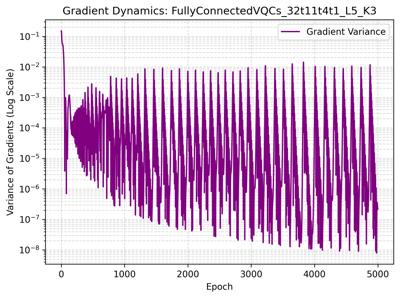}\hfill%
    \includegraphics[width=0.245\linewidth]{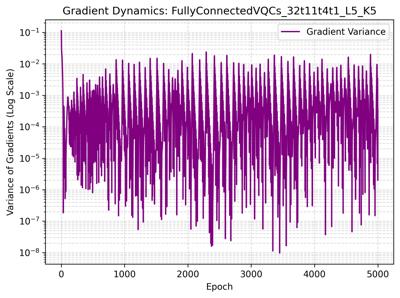}\hfill%
    \includegraphics[width=0.245\linewidth]{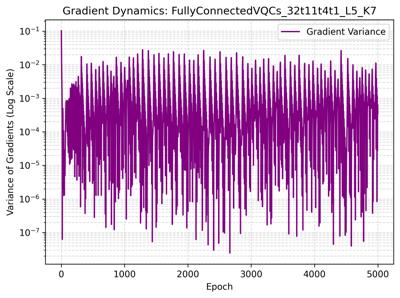}\hfill%
    \includegraphics[width=0.245\linewidth]{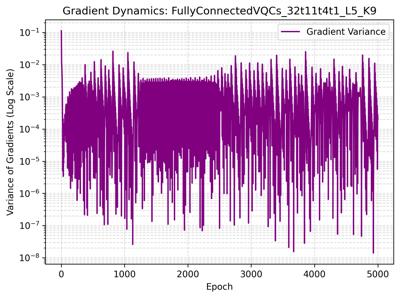}%
    \par\smallskip

    \includegraphics[width=0.245\linewidth]{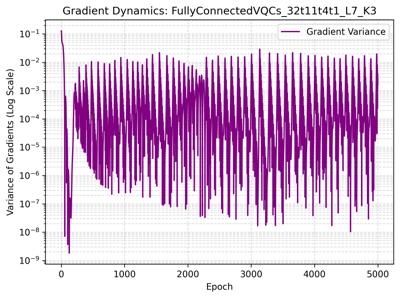}\hfill%
    \includegraphics[width=0.245\linewidth]{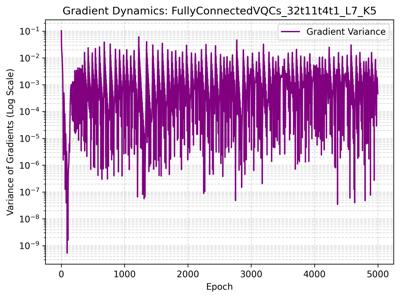}\hfill%
    \includegraphics[width=0.245\linewidth]{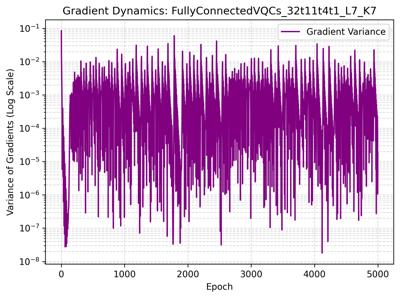}\hfill%
    \includegraphics[width=0.245\linewidth]{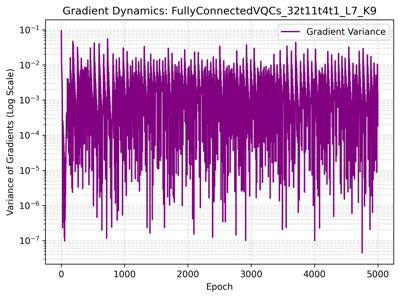}%
    \par\smallskip

    \includegraphics[width=0.245\linewidth]{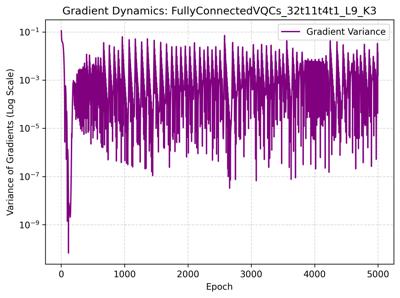}\hfill%
    \includegraphics[width=0.245\linewidth]{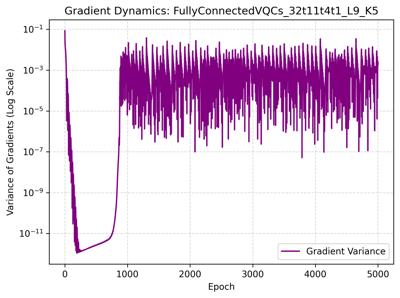}\hfill%
    \includegraphics[width=0.245\linewidth]{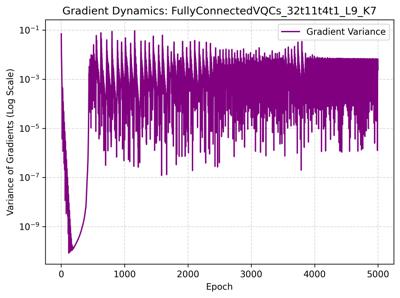}\hfill%
    \includegraphics[width=0.245\linewidth]{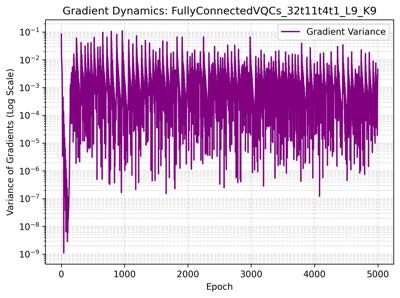}%

    \caption{\textbf{Gradient dynamics on Concrete Strength for the Type~4 $32t11t4t1$ FC-VQC architecture.} Rows: $L=1,3,5,7,9$ (top to bottom); columns: $K=3,5,7,9$ (left to right). The horizontal axis is training epoch; the vertical axis is gradient variance on a logarithmic scale.}
    \label{fig:gradients_32t11t4t1}
\end{figure}

\clearpage

\begin{figure}[!htbp]
    \centering
    \includegraphics[width=0.245\linewidth]{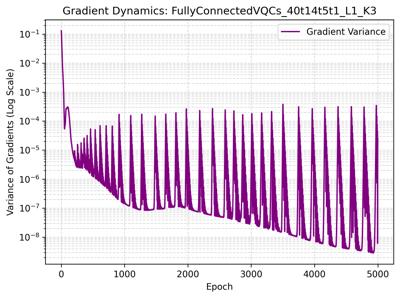}\hfill%
    \includegraphics[width=0.245\linewidth]{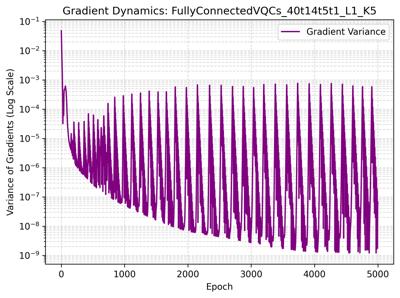}\hfill%
    \includegraphics[width=0.245\linewidth]{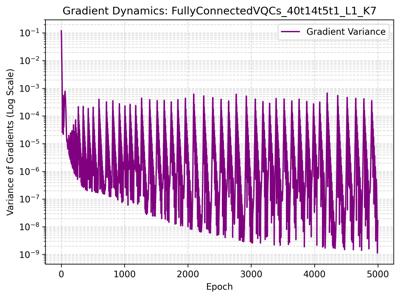}\hfill%
    \includegraphics[width=0.245\linewidth]{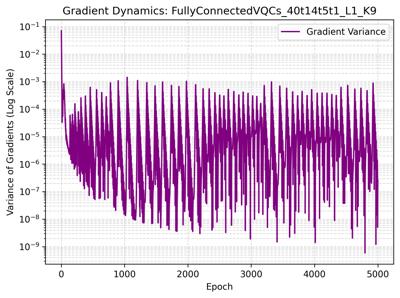}%
    \par\smallskip

    \includegraphics[width=0.245\linewidth]{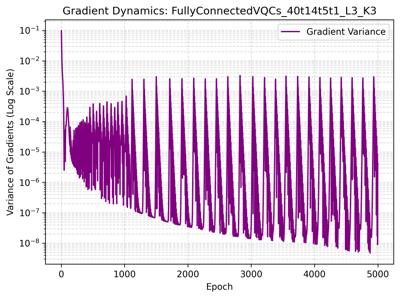}\hfill%
    \includegraphics[width=0.245\linewidth]{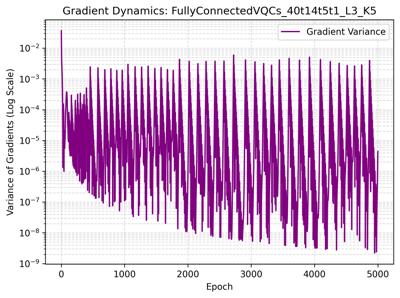}\hfill%
    \includegraphics[width=0.245\linewidth]{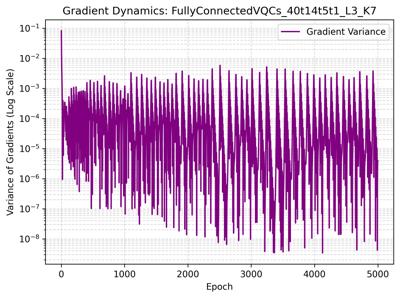}\hfill%
    \includegraphics[width=0.245\linewidth]{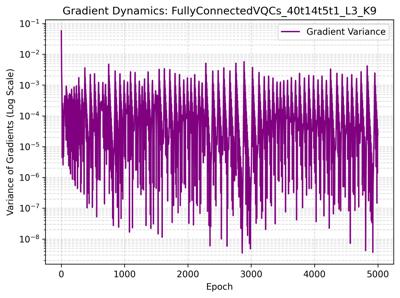}%
    \par\smallskip

    \includegraphics[width=0.245\linewidth]{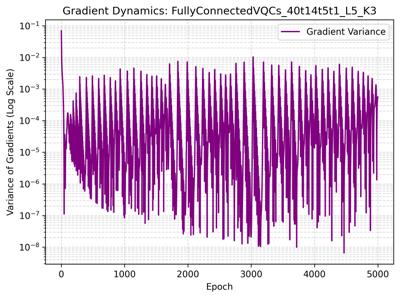}\hfill%
    \includegraphics[width=0.245\linewidth]{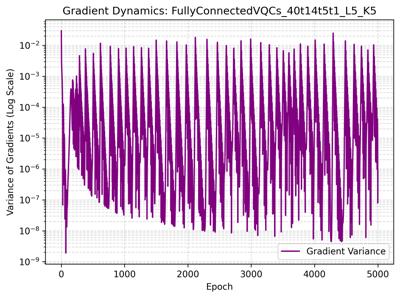}\hfill%
    \includegraphics[width=0.245\linewidth]{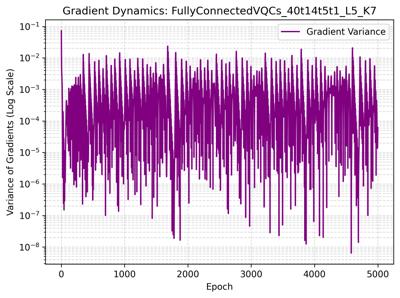}\hfill%
    \includegraphics[width=0.245\linewidth]{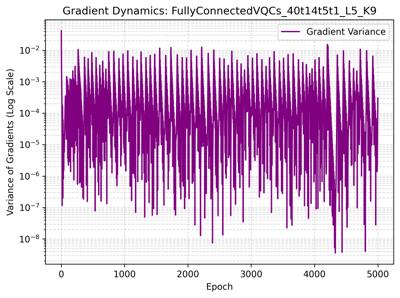}%
    \par\smallskip

    \includegraphics[width=0.245\linewidth]{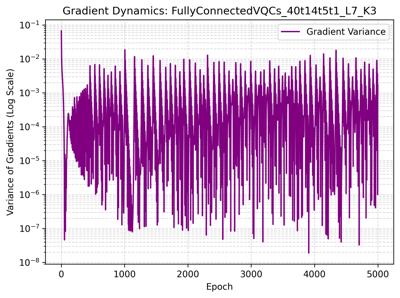}\hfill%
    \includegraphics[width=0.245\linewidth]{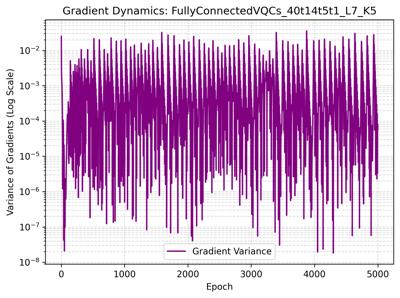}\hfill%
    \includegraphics[width=0.245\linewidth]{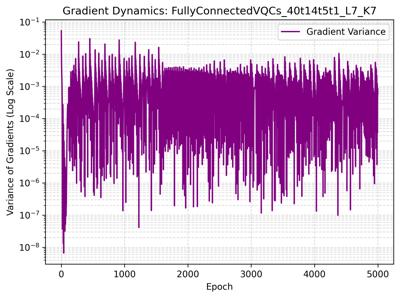}\hfill%
    \includegraphics[width=0.245\linewidth]{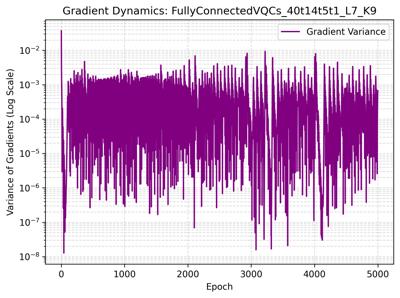}%
    \par\smallskip

    \includegraphics[width=0.245\linewidth]{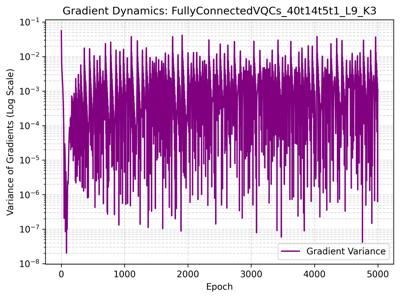}\hfill%
    \includegraphics[width=0.245\linewidth]{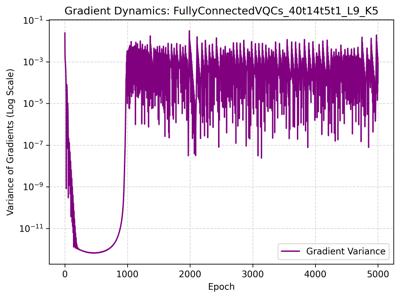}\hfill%
    \includegraphics[width=0.245\linewidth]{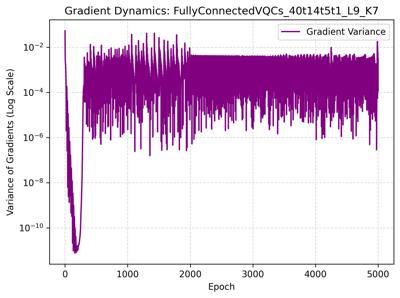}\hfill%
    \includegraphics[width=0.245\linewidth]{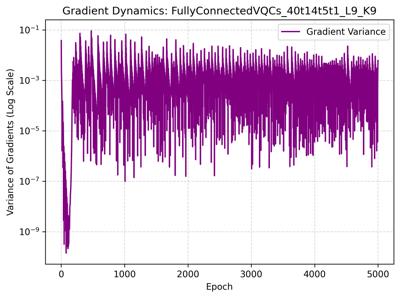}%

    \caption{\textbf{Gradient dynamics on Concrete Strength for the Type~4 $40t14t5t1$ FC-VQC architecture.} Rows: $L=1,3,5,7,9$ (top to bottom); columns: $K=3,5,7,9$ (left to right). The horizontal axis is training epoch; the vertical axis is gradient variance on a logarithmic scale.}
    \label{fig:gradients_40t14t5t1}
\end{figure}

\clearpage

\end{document}